\documentclass[runningheads]{cl2emult}

\def\R{\bf R}
\def\d{\partial}

\begin{document}


\title*{The Cauchy Problem for the Einstein Equations}
%
%
%
%
\titlerunning{The Cauchy Problem}
%
\author{Helmut Friedrich
\and Alan Rendall}
%
%
%
\institute{Max-Planck-Institut f\"ur Gravitationsphysik, Am M\"uhlenberg 1,\\
14476 Golm, Germany}

\maketitle              

\makeatletter                  
\@addtoreset{equation}{section}
\renewcommand{\theequation}{\arabic{section}.\arabic{equation}}%
\renewcommand{\thesubequation}{\arabic{section}.%
                               \arabic{equation}\alph{eqsubcnt}}%
\makeatother                   

\begin{abstract}
Various aspects of the Cauchy problem for the Einstein equations are
surveyed, with the emphasis on local solutions of the evolution equations.
Particular attention is payed to giving a clear explanation of
conceptual issues which arise in this context. The question of producing
reduced systems of equations which are hyperbolic is examined in detail
and some new results on that subject are presented. Relevant background
from the theory of partial differential equations is also explained at
some length.

\end{abstract}

\section{Introduction}

One of the most striking differences between the Newtonian theory of
gravity and its successor, general relativity, is that in the
latter the gravitational field acquires its own dynamical properties.
Its time evolution is complicated even in the absence of matter. This
contrasts with the fact that in the Newtonian theory the field vanishes
when no matter is present. The field equation, namely the Poisson
equation, together with the boundary condition that the field vanishes
at infinity, which is an essential part of the theory, combine to give
this result. The Einstein equations, the field equations of general
relativity, allow idealized situations which represent gravitational waves
in an otherwise empty universe, without any material sources. This
reflects the different mathematical nature of the equations involved in
these two cases. The Poisson equation is elliptic while the Einstein
equations are essentially hyperbolic in nature. The meaning of the term
'essential' in this context is not simple and explaining it is a major
theme in the following.

In order to understand the special theoretical difficulties connected
with the Einstein equations and what mathematical approaches may be
appropriate to overcome them, it is useful to compare gravitation
with electromagnetism. The motion of charged matter can be described
within the full Maxwell theory. However, there is also another
possibility, which is used when relativistic effects are small, for
instance in many situations in plasma physics. Here dynamical matter
is coupled to the electrostatic field generated by this matter at
any given time. In this quasi-static model the electric field follows
the sources in a passive way while in the full theory there are propagating
degrees of freedom. As in the case of gravitation, the elliptic equation in
the non-relativistic theory (the Poisson equation again) is replaced by a
system of hyperbolic equations, the Maxwell equations.

The fact that the Maxwell equations are so tractable is due to their
linearity, a convenient feature not shared by the Einstein equations. The
theory of linear partial differential equations in general, and of linear
hyperbolic equations in particular, are much better developed than
the corresponding nonlinear theories. As a side remark it may be noted
that the combined equations describing electromagnetic fields together
with their sources are nonlinear and that in that context
serious theoretical problems, such as that of describing radiation damping,
do appear.

Solutions of hyperbolic equations can be uniquely determined by their
values on a suitable initial hypersurface. The Cauchy problem is the
task of establishing a one to one correspondence between solutions
and initial data, and studying further properties of this
correspondence. The solution determined by a particular initial datum
may be {\it global}, i.e. defined on the whole space where the
equations are defined, or {\it local}, i.e. only defined on a neighbourhood
of the initial hypersurface. 'Local' and 'global' could be called local
and global {\it in time} since in the case where a preferred time
coordinate is present that is exactly what they mean.

{}From what has been said so far we see that in studying the Einstein
equations we are faced with a system of nonlinear hyperbolic equations.
Among nonlinear hyperbolic equations in physics, those which have been
studied most extensively are the Euler equations, and so we may hope to get
some insights from that direction. At the same time, it is wise to be
careful not to treat the analogy too uncritically, since the status of the
Euler and Einstein equations is very different. The Euler equations are
phenomenological in nature and much is understood about how they arise
from models on a more fundamental level. The Einstein equations have
been thought of as representing fundamental physics for most of their
history and the recent idea that they arise as a formal limiting case in
string theory will require, at the very least, a lot more work before it
can offer a solid alternative to this.
To return to the Euler equations, one of their well-known features
is the formation of shocks. While there is no indication of a directly
analogous phenomenon for the Einstein equations, it does draw attention
to a fundamental fact. For linear hyperbolic equations it is in general
possible to solve the Cauchy problem globally, i.e. to show the
existence of a global solution corresponding to each initial datum.
For nonlinear hyperbolic equations this is much more difficult and whether
it can be done or not must be decided on a case by case basis. The
formation of shocks in solutions of the Euler equations is an example of
the difficulties which can occur. A general theory for general equations can
only be hoped for in the case of the local (in time) Cauchy problem.

For the Einstein equations we must expect to encounter the problem that
solutions of the Cauchy problem for nonlinear hyperbolic equations do
not exist globally. In the case of the Einstein equations this issue
is clouded by the fact that the distinction between local and global
solutions made above does not apply. To define the notions
of local and global we used the concept of the space where the equations
are defined. In other words we used a background space. As we will see
in more detail later, in the case of the Einstein equations there is no
background space; the space-time manifold is part of the solution. It
is better in this case to talk only about local and global {\it properties}
of solutions and not about local and global {\it solutions}. We may then
loosely use the words 'local' and 'global' to refer to all aspects of
the Cauchy problem which refer to local and global properties of
solutions, respectively. Solutions of the Einstein equations present
global features such as the formation of black holes which are peculiar to
this system and which are made possible by the lack of a background
space.

In this article we are not concerned with general systems of nonlinear
hyperbolic equations, but with a particular one, which is given to us
by general relativity. Actually, when the coupling to matter fields
is taken into account, we do get a variety of hyperbolic systems.
Nevertheless, we might hope that in at least some situations of
interest, such as gravitational collapse, the dynamics of the
gravitational field would dominate the qualitative behaviour and let
the effects of the particular matter model fade into the background.
In any case, it is useful to retain the distinction between the local and
global Cauchy problems. The global problem is what we want to solve, but
the local problem is a natural first step. Our original plan was to cover
both topics, but along the way we discovered that the first step is already
so rich that on grounds of time and space we have relegated global
questions to passing remarks. Along the way we stumbled over a variety of
\lq well-known\rq\ things which turned out not to be known at all,
or even to be false.

The theory of the Cauchy problem allows us to formulate and establish
relativistic causality within general relativity. Another basic function
of the solution of the Cauchy problem is to parametrize solutions of the
field equations in a useful way. To single out the class of solutions
relevant to the description of a given physical situation, we can single
out an appropriate subclass of initial data, which is often simpler
to do. This does not mean that by identifying this class of solutions
we have solved the problem. Rather it means that we have found a
class of problems which may be of physical relevance and which it is
therefore desirable to investigate mathematically. In the end the central
mathematical problem is to discover some of the global properties of the
solutions being studied. As has already been said, this article is almost
entirely restricted to local questions, so that we will say little more
about this point. However we will return to it in the last section.

The point
of view of the Cauchy problem can also be used to throw light on various
other issues. For instance, it provides a framework in which it can be
shown that certain approximation methods used in general relativity
really provide approximations to solutions of the Einstein equations. It
can help to provide an analytical basis for the development of
efficient and reliable numerical schemes to solve the Einstein equations.
The Cauchy problem can furnish examples which throw light on general
conjectures about solutions of the Einstein equations. It can be used to
investigate whether certain properties of explicit solutions of the
Einstein equations of importance in general relativity are stable to
perturbations.

The structure of the article will now be outlined. In the second section
a number of fundamental concepts are introduced and critically reviewed.
In particular, the splitting of the field equations into evolution
equations and constraints is described. The rest of the article
concentrates on the evolution equations. The question of how to solve
the constraints on the initial hypersurface is not considered further.
The second section presents the basic elements which go into solving
the local Cauchy problem. It discusses in particular the question of
gauge freedom and how to show that the constraints are satisfied
everywhere, given that they are satisfied on the initial hypersurface
(propagation of the constraints). These are aspects of the question
of {\it hyperbolic reduction}, i.e. how questions about the Cauchy
problem for the Einstein equations can be reduced to questions about
the Cauchy problem for hyperbolic equations.

The solution of the Cauchy problem relies on the use of techniques from
the theory of partial differential equations (PDE). The third section
presents some of the relevant techniques and attempts to explain
some of the important concepts which play a role in the theory of
the Cauchy problem for hyperbolic equations. We have chosen to
concentrate on symmetric hyperbolic systems rather than on other
kinds of hyperbolic equations such as nonlinear wave equations, which
could also be used as a basis for studying the Cauchy problem for the
Einstein equations. It will be seen that symmetric hyperbolic systems
provide a very flexible tool. A comparative discussion of various notions
of hyperbolicity is also given.

In the fourth section we return to the question of hyperbolic reduction.
Different ways of reducing the Einstein equations in 3+1 form are
presented and compared. The ADM equations, which were already
introduced in Sect. \ref{basics}, are discussed further. A form
of these equations which has proved successful in numerical
calculations is analysed. A number of other illustrative examples are
treated in detail. The first example is that of the Einstein--Euler system
for a self-gravitating perfect fluid. The case of dust, which is
significantly different from that of a fluid with non-vanishing
pressure, and which leads to serious problems in some approaches,
is successfully handled. The second example is a variant of the pure
Cauchy problem, namely the initial boundary value problem. The third
example is the Einstein--Dirac system, which gives rise to particular
problems of its own.

The fifth section starts with a discussion of the proofs of the
basic statements of local existence, uniqueness and stability
for the Einstein equations, based on the PDE theory reviewed in
the third section and a particular hyperbolic reduction already
introduced in Sect. \ref{basics}. Then the extension of these
results to the case where different kinds of matter are present is
sketched. Existence, uniqueness and stability constitute together the
statement that the Cauchy problem is well-posed. The significance
of this is underlined by an example of a situation where the Cauchy
problem is ill-posed. Finally it is shown how existence and uniqueness
imply the inheritance of symmetries of the initial data by the
corresponding solutions.

The choice of topics covered in this article is heavily dependent on
the interests of the authors. In the last section we list
a number of the important topics which were not discussed. While
this list is also influenced by personal taste, we hope that it will
provide the interested reader with the opportunity to form a
balanced view of the subject.


\section{Basic Observations and Concepts}\label{basics}

In this section we give an introductory survey of various aspects of the
field equations. Most of them will be discussed again, in greater detail, in
later sections.

Gravitational fields represented by isometric space-times must be considered as
physically equivalent and therefore the field equations for the metric must
have the property that they determine isometry classes of solutions, not
specific coordinate representations. This is achieved by the covariant
Einstein equations
\[
R_{\mu \nu} - \frac{1}{2}\,R\,g_{\mu \nu} + \lambda\,g_{\mu \nu} =
\kappa\,T_{\mu \nu}.
\]
It is often said that many of the specific features of the Einstein equations
are related to this covariance. One may wonder, however, what is so special
about it, since the wave equation, the Yang--Mills equations, the Euler
equations,
just to name a few examples, are also covariant. The difference lies in the
fundamental nature of the metric field. While the examples just quoted are
defined
with respect to some background structure, namely a given Lorentz space (in
the case
of the Yang--Mills equations in four dimensions a conformal structure suffices),
the Einstein equations are designed to determine Lorentz manifolds without
introducing any extraneous structures. Consequently, the solutions of the
equations
themselves provide the background on which the equations are to be solved.

This makes it evident that Einstein's equations can at best be quasi-linear.
That
they are in fact quasi-linear can be seen from their explicit expression.
Writing the
equation in the form
\begin{equation}
\label{einst}
R_{\mu \nu} - \lambda\,g_{\mu \nu} = \kappa\,(T_{\mu \nu}
- \frac{1}{2}\,T\,g_{\mu
\nu}),
\end{equation}
the principal part of the differential operator of second order which acts on
the
metric coefficients is given by the left hand side. In arbitrary coordinates we
have
\begin{equation}
\label{ricexpl}
R_{\mu \nu} = - \frac{1}{2}\,g^{\lambda \rho}\,\left\{
\frac{\partial^2 g_{\mu \nu}}{\partial x^{\lambda}\,\partial x^{\rho}}
+ \frac{\partial^2 g_{\lambda \rho}}{\partial x^{\mu}\,\partial x^{\nu}}
- \frac{\partial^2 g_{\mu \rho}}{\partial x^{\lambda}\,\partial x^{\nu}}
- \frac{\partial^2 g_{\rho \nu}}{\partial x^{\mu}\,\partial x^{\lambda}}
\right \}
+ Q_{\mu \nu}(g, \partial\,g),
\end{equation}
where $Q$ denotes a rational function of the metric coefficients and their
first order derivatives. We see that the equations are quasi-linear but not
better,
i.e. they are linear in the derivatives of highest order but these
derivatives come
with coefficients which are given by the unknown itself. The Euler equations
are also
quasi-linear in this sense. However, they are studied in general on an
independent
background space-time in terms of which their solutions may be analysed.

\subsection{The Principal Symbol}\label{principal}

Consider  a system of $k$ partial differential equations of order $m$ for an
${\R}^k$-valued unknown $u$  defined on an open subset $U$ of
${\R}^n$. Suppose that in coordinates
$x^{\mu}$, $\mu = 1, \ldots, n$ on $U$ it takes the form
\begin{equation}\label{system}
\label{dop}
P\,u \equiv \sum_{|\alpha| \le m} A^{\alpha}\,D^{\alpha}\,u = f.
\end{equation}
Here the $A^{\alpha}$ denote smooth real $k \times k$-matrix-valued functions,
$f(x)$ is a smooth ${\R}^k$-valued function, the
$\alpha = (\alpha_1, \ldots, \alpha_n)$, with non-negative integers $\alpha_j$,
are multi-indices, $|\alpha| = \alpha_1 + \ldots + \alpha_n$, and
$D^{\alpha} = \partial^{\alpha_1}_{x^1} \ldots \partial^{\alpha_n}_{x^n}$.
Assume, first, that the equations are linear, so that
$A^{\alpha} = A^{\alpha}(x)$.

The question of the formal solvability of the equation leads to the important
notion of a characteristic. Suppose that  $H = \{ \Phi = {\rm const.},
d\,\Phi \neq 0\}$ is a hypersurface of $U$, defined in terms of a smooth
function $\Phi$. A {\it Cauchy data set on} $H$ for equation (\ref{system})
consists of a set of functions $u_0, u_1, \ldots , u_{(m-1)}$ on $H$. Let
the coordinates $x^{\mu}$ be chosen such that $\Phi = x^1$ near $H$.
Interpreting the functions $u_i$ as the derivatives $\partial^i_{x^1}u$ of a
solution $u$ to our equation, defined in a neighbourhood
of $H$, we ask whether equation (\ref{dop}) allows the function
$\partial^m_{x^1}u$ to be determined uniquely on $H$ from the Cauchy data.
Since all functions $D^{\alpha}\,u$ with $\alpha_1 \le m - 1$, can obviously
be derived from
the data, it follows that we can solve the equation on $H$ for
$\partial^m_{x^1}\,u$ if and only if the matrix $A^{(m, 0, \ldots, 0)}$ is
invertible at all points of $H$. Using for any covector $\xi_{\mu}$ and
multi-index
$\alpha$ the notation $\xi^{\alpha} = \xi^{\alpha_1}_1 \ldots
\xi^{\alpha_n}_n$, we
observe that with our assumptions
$A^{(m, 0, \ldots, 0)} = \sum_{|\alpha| = m} A^{\alpha}\,D^{\alpha}\,\Phi$.
The invertibility of the matrix on the right hand side is independent of the
coordinates chosen to represent equation (\ref{dop}) and the function $\Phi$
used to
represent $H$. This follows from the transformation law of the coefficients
of the
differential equation and of covectors under coordinate transformations.

We are thus led to introduce the following covariant notions.
For a covector $\xi$ at a given point $x \in M$, we define the
{\it principal symbol} of (\ref{dop}) at $(x, \xi)$ as the matrix
$\sigma(x, \xi) = \sum_{|\alpha| = m} A^{\alpha}(x)\,\xi^{\alpha}$ (or the
associated linear map). The reader should be warned that a definition
with $\xi^i$ replaced by $i\xi^i$, where $i$ is the imaginary unit,
is often used, since it fits well with the Fourier transform. The
latter convention will not be used here.
A hypersurface $H$, represented by a function $\Phi$
as above, is called {\it nowhere characteristic} for (\ref{dop}), if
$\det(\sigma(x,d\,\Phi)) \neq 0$ for $x \in H$ and we say that $H$ is  a
{\it characteristic
hypersurface}, or simply a  {\it characteristic} for (\ref{dop}), if
$\det(\sigma(x,
d\,\Phi(x))) = 0$ for $x \in H$.

Given Cauchy data on a hypersurface $H$ which is nowhere characteristic, we
can
determine a formal expansion of a possible solution $u$ in terms of $x^1$ on
$H$ by
taking formal derivatives of (\ref{dop}) and solving for $\partial^i_{x^1}\,u$
on $H$,
$i = m, m +1, \ldots$. Conversely, if $H$ is characteristic, Cauchy data
cannot be
prescribed freely on $H$, because the equation implies relations among the
Cauchy
data on $H$. We refer to these relations as the {\it inner equations}
induced
by the equations on the characteristic.

If the rank of the matrix $\sigma(x, d\,\Phi(x))$ is $k - j$ with
some positive integer $j$, there are $j$ such relations. The principal
symbol is not sufficient to describe the precise nature of these relations.
The complete information of (\ref{dop}) is needed for this.

If equation (\ref{dop}) is quasi-linear, so that the coefficients of the
equation do
not only depend on the points of $M$ but also on the unknown and its
derivatives of
order less than $m$, we have to proceed slightly differently. Suppose $u$ is a
solution of the quasi-linear equation (\ref{dop}). For given covector $\xi$ at
the
point $x \in M$ we define the {\it principal symbol} of (\ref{dop}) {\it with
respect
to} $u$ as the matrix $\sigma(x, \xi) = \sum_{|\alpha| = m}
A^{\alpha}(x, u(x), \ldots, D^{\beta}\,u(x)|_{|\beta|
= m - 1})\,\xi^{\alpha}$ and
use it to define, by the condition above, the {\it characteristics of}
(\ref{dop}) {\it with respect to} $u$. Thus for quasi-linear equations the
characteristics depend on the solution.

We return to the Einstein equations. Assuming that $(M, g)$ is a solution of
(\ref{einst}) of dimension $n$, we find that for given covector $\xi$ at
$x \in M$ the principal symbol of the operator in
(\ref{ricexpl}) defines the linear map
\begin{equation}
\label{riccisymbol}
k_{\mu \nu} \rightarrow (\sigma \cdot k)_{\mu \nu} =
- \frac{1}{2}\,g^{\lambda \rho}(x)\,\left\{
k_{\mu \nu}\,\xi_{\lambda}\,\xi_{\rho}
+ k_{\lambda \rho}\,\xi_{\mu}\,\xi_{\nu}
- k_{\mu \rho}\,\xi_{\lambda}\,\xi_{\nu}
- k_{\lambda \nu}\,\xi_{\mu}\,\xi_{\rho}
\right \},
\end{equation}
of the set of symmetric covariant tensors at $x$ into itself.
If $k_{\mu \nu}$ is in its kernel, we have
\begin{equation}
\label{kernelequ}
0 = k_{\mu \nu}\,\xi_{\rho}\,\xi^{\rho}
+ g^{\rho \lambda}\,k_{\rho \lambda}\,\xi_{\mu}\,\xi_{\nu}
- k_{\mu \rho}\,\xi^{\rho}\,\xi_{\nu}
- k_{\nu \rho}\,\xi^{\rho}\,\xi_{\mu},
\end{equation}
from which we see that tensors of the form
$k_{\mu \nu} = \xi_{\mu}\,\eta_{\nu} + \xi_{\nu}\,\eta_{\mu}$, with arbitrary
covector $\eta$, generate an $n$-dimensional subspace of the kernel. This
subspace
coincides with the kernel if $\xi_{\rho}\,\xi^{\rho} \neq 0$. If
$\xi_{\rho}\,\xi^{\rho} = 0$, $\xi_{\rho} \neq 0$, equation (\ref{kernelequ})
takes
the form $\xi_{\mu}\,\eta_{\nu} + \xi_{\nu}\,\eta_{\mu} = 0$, with some
covector
$\eta_{\mu}$. Since $\xi_{\mu} \neq 0$ we must have $\eta_{\mu} = 0$ or,
equivalently, $\frac{1}{2}\,g^{\rho \lambda}\,k_{\rho \lambda}\,\xi_{\nu}
= k_{\rho \nu}\,\xi^{\rho}$. The solutions, given by
$k_{\mu \nu} = h_{\mu \nu} + a\,(\xi_{\mu}\,\eta^{*}_{\nu} +
\xi_{\nu}\,\eta^{*}_{\mu})$ with $a \in {\R}$, $\eta^{*}_{\mu}$ a fixed
covector with $\xi^{\mu}\,\eta^{*}_{\mu} \neq 0$, and $h_{\mu \nu}$ satisfying
$\xi^{\mu}\,h_{\mu \nu} = 0$, $g^{\mu \nu}\,h_{\mu \nu} = 0$, span a space of
dimension $\frac{n\,(n - 1)}{2}$, which is strictly larger than $n$ if
$n \ge 4$.

It follows that any hypersurface is characteristic for the Einstein
equations. In
the case of spacelike or timelike hypersurfaces n relations are implied on
Cauchy
data sets, in the case of null hypersurfaces there are $\frac{n\,(n - 3)}{2}$
additional relations. The fact that the rank of the principal symbol drops
further
for covectors $\xi$ which are null has the immediate consequence that the
inner
equations induced on null hypersurfaces are completely different from the
inner
equations induced on nowhere characteristic hypersurfaces. This is related to
the
fact that null hypersurfaces (and appropriate generalizations admitting
caustics)
may represent wave fronts, swept out by high frequency perturbations of the
field.
The propagation of the latter is governed by the inner equations induced
on these hypersurfaces (cf. \cite{courant62} and, for a modern discussion, also
\cite{hormander}, \cite{taylor81} for the mathematical aspects of this
statement).

The fact that on any hypersurface the tensors of the form
$k_{\mu \nu} = \xi_{\mu}\,\eta_{\nu} + \xi_{\nu}\,\eta_{\mu}$ are in the
kernel of
the principal symbol map at $(x, \xi)$ can be deduced by an abstract argument
which
uses only the covariance and the quasi-linearity of the equations
(cf. \cite{besse},
\cite{kazdan:1981}). This emphasizes again the special role of null
hypersurfaces.
Assume that $\phi_{\tau}$ is the flow of a vector field $X$. If
we denote by $\phi^{*}_{\tau}$ the associated pull-back operation, we have
\[
\phi^{*}_{\tau}({\rm Ric}[g]) = {\rm Ric}[\phi^{*}_{\tau}(g)].
\]
Taking derivatives with respect to $\tau$ we obtain
\[
{\cal L}_X\,{\rm Ric}[g] = {\rm Ric}'_g[{\cal L}_X\,g],
\]
where the right hand side denotes the (at $g$) linearized Ricci operator
applied to
the Lie derivative
${\cal L}_X\,g_{\mu \nu} = \nabla_{\mu}\,X_{\nu} + \nabla_{\nu}\,X_{\mu}$ of
the
metric. Considering this equation as a differential equation for $X$, the
principal part is given by the terms of third order appearing on the right
hand side.
Since the equation is an identity, holding for all vector fields, these terms
must cancel. It follows that the expressions
$\xi_{\mu}\,X_{\nu} + \xi_{\nu}\,X_{\mu}$ are in the kernel of the principal
symbol
of the linearized and thus also of the non-linear Ricci operator.

For the further discussion it will be convenient to use coordinates related
to a
given hypersurface. For simplicity and since it is the case of most interest
to us,
we will assume the latter to be spacelike and denote it by $S$. Let $T$ be a
non-vanishing {\it time flow vector field} transverse to $S$ (but not
necesssarily
timelike) and $t = x^0$ a function with
$\{t = 0\} = S$ and $<dt, T>\,\,= 1$. We choose local coordinates $x^a$,
$a = 1, 2,
3$, on $S$ and extend them to a neighbourhood of $S$ such that
$<d\,x^a,T>\,\,= 0$.
In these coordinates $T = \partial_t$ and the metric has the
{\it ADM
representation}
\begin{equation}
\label{ADMmetric}
g = - (\alpha\,d\,t)^2
+ h_{ab}\,(\beta^a\,d\,t + d\,x^a)\,(\beta^b\,d\,t + d\,x^b),
\end{equation}
where $h =  h_{ab}\,d\,x^a\,d\,x^b$ represents the induced Riemannian metric on
the slice $S$. To ensure the non-degeneracy of the metric we assume the
{\it lapse function} $\alpha$ to be positive. We write
$\beta = \beta^{\mu}\,\partial_{\mu} = \beta^a\,\partial_a$ for the
{\it shift vector field}, which is tangent to the time slices. We assume
further
that the metric $h$ induced on the time slices is Riemannian and we denote its
contravariant form by $h^{ab}$, so that $h^{ab}\,h_{bc} = \delta^a\,_c$.

The (future-directed) unit normal to the time slices is given by
$n^{\mu} = \frac{1}{\alpha}\,(\delta^{\mu}\,_0 - \beta^{\mu})$, whence
$n_{\mu} = - \alpha\,\delta^0\,_{\mu}$. We write
$h^{\mu}\,_{\nu} = g^{\mu}\,_{\nu} +n^{\mu}\,n_{\nu}$ for the orthogonal
projector
onto the time slices, $h_{\mu \nu} = g_{\mu \nu} + n_{\mu}\,n_{\nu}$ for the
4-dimensional representation of the interior metric on the time slices, and
use $g$
to perform index shifts. The second fundamental form (extrinsic curvature) of
the
time slices is given by
\[
\chi_{\mu \nu} \equiv
h_{\mu}\,^{\lambda}\,h_{\nu}\,^{\rho}\,\nabla_{\lambda}\,n_{\rho} =
\frac{1}{2}\,h_{\mu}\,^{\lambda}\,h_{\nu}\,^{\rho}\,{\cal L}_n
\,g_{\lambda \rho}
= \frac{1}{2}\,{\cal L}_n\,h_{\mu \nu}.
\]
For tensor fields $t_{\mu} \ldots ^{\nu}$ intrinsic to the time slices, i.e.
having vanishing contractions with $n$, the
$h$-covariant derivative $D$ on a fixed time slice is given by
\[
D_{\rho}\,t_{\mu} \ldots ^{\nu} =
h^{\lambda}\,_{\rho}\,h^{\phi}\,_{\mu} \ldots h_{\psi}\,^{\nu}\,
\nabla_{\lambda}\,t_{\phi} \ldots ^{\psi}.
\]

Finally, we write $n(f) = n^{\mu}\,\partial_{\mu}\,f$, note that
$a_{\mu} = n^{\nu}\,\nabla_{\nu}\,n_{\mu} = D_{\mu}\,\log (\alpha)$,
and use the coefficients $\gamma_{\mu}\,^{\nu}\,_{\rho} =
\Gamma_{\lambda}\,^{\eta}\,_{\pi}\,h^{\lambda}\,_{\mu}\,
h^{\nu}\,_{\eta}\,h^{\pi}\,_{\rho}$ which represent the covariant derivative
$D$
in the sense that the components $\gamma_a\,^b\,_c$ are the Christoffel
symbols of
the metric $h_{ab}$. The connection coefficients of $g_{\mu \nu}$ can then be
written
\begin{equation}
\label{gammadecomp}
\Gamma_{\mu}\,^{\nu}\,_{\rho} =
n_{\mu}\,n^{\nu}\,n_{\rho}\,n(\log\,\alpha)
+ n_{\mu}\,a^{\nu}\,n_{\rho}
- a_{\mu}\,n^{\nu}\,n_{\rho}
- n_{\mu}\,n^{\nu}\,a_{\rho}
\end{equation}
\[
+ n_{\mu}\,n_{\rho}\,\frac{1}{\alpha}\,\beta^{\nu}\,_{,\pi}\,n^{\pi}
- n_{\mu}\,\frac{1}{\alpha}\,\beta^{\nu}\,_{,\pi}\,h^{\pi}\,_{\rho}
- n_{\rho}\,\frac{1}{\alpha}\,\beta^{\nu}\,_{,\pi}\,h^{\pi}\,_{\mu}
\]
\[
+ n^{\nu}\,\chi_{\mu \rho}
- n_{\mu}\,\chi^{\nu}\,_{\rho}
- \chi_{\mu}\,^{\nu}\,n_{\rho}
+ \gamma_{\mu}\,^{\nu}\,_{\rho}.
\]

\subsection{The Constraints}\label{constraints}
To isolate the geometrically relevant information contained in the Cauchy
data, we
reduce the coordinate freedom tranverse to $S$ by assuming $T$ to be a geodesic
unit vector field normal to $S$. The metric then takes the form
\begin{equation}
\label{gauss1fufo}
g = - d\,t^2 + h_{ab}\,d\,x^a\,d\,x^b,
\end{equation}
and the pull-back of the second fundamental to the time slices is given by
\begin{equation}
\label{gauss2fufo}
\chi_{ab} = \frac{1}{2}\,\partial_t\,h_{ab}.
\end{equation}
Expressions (\ref{gauss1fufo}) and (\ref{gauss2fufo}) suggest that the
{\it essential Cauchy data} for the metric field are given by the induced
metric
$h_{ab}$ and the second fundamental form $\chi_{ab}$ on the spacelike
hypersurface $S$. Since coordinate transformations are unrestricted, the $n$
inner relations induced on general Cauchy data on a spacelike hypersurface
$S$ must
be conditions on the essential Cauchy data.

If we contract the covector $\xi$, which corresponds in our context to the
normal $n$
of $S$, with the principal symbol of the Einstein tensor $G_{\mu \nu}
= R_{\mu \nu}
- \frac{1}{2}\,R\,g_{\mu \nu}$ and evaluated on $k$, we get
\[
\xi^{\mu}\,\left((\sigma \cdot k)_{\mu \nu} -
\frac{1}{2}\,g_{\mu \nu}\,g^{\lambda \rho}\,(\sigma \cdot k)_{\lambda \rho}
\right)
= 0.
\]
This identity indicates the combinations of the equations which
contain only derivatives of first order in directions transverse to $S$.
Indeed, if we express the equations
\[
0 = Z^H \equiv
n^{\mu}\,n^{\nu}\,(G_{\mu \nu} + \lambda\,g_{\mu \nu} - \kappa\,T_{\mu
\nu}),
\]
\[
0 = Z^M_{\nu} \equiv n^{\mu}\,h_{\nu}\,^{\rho}\,(G_{\mu \rho}
+ \lambda\,g_{\mu \rho}
- \kappa\,T_{\mu \rho}),
\]
on $S$ in terms of $h$ and $\chi$, write
$\rho = n^{\mu}\,n^{\nu}\,T_{\mu \nu}$,
$j_{\nu} = - n^{\mu}\,h^{\rho}\,_{\nu}\,T_{\mu \rho}$, and pull-back to $S$,
we get
the {\it constraint equations on spacelike hypersurfaces},
i.e. the {\it Hamiltonian constraint}
\begin{equation}
\label{hamconstr}
0 = 2\,Z^H = r - \chi_{ab}\,\chi^{ab} + (\chi_a\,^a)^2 - 2\,\lambda
- 2\,\kappa\,\rho,
\end{equation}
and the {\it momentum constraint}
\begin{equation}
\label{momconstr}
0 = Z^M_b = D_a\,\chi_b\,^a - D_b\,\chi_a\,^a + \kappa\,j_b,
\end{equation}
where $r$ denotes the Ricci scalar and $D$ the connection of $h$.

These equations have the following geometric meaning. On a spacelike
hypersurface
$S$ the curvature tensors $R^{\mu}\,_{\nu \rho \eta}$ and
$r^{\mu}\,_{\nu \rho \eta}$ of $g$ and $h$ respectively and the second
fundamental
form $\chi_{\mu \nu}$ are related, irrespective of any field equation,
by the {\it Gauss equation}
\begin{equation}
\label{Gausseq}
r^{\mu}\,_{\nu \rho \eta} =
h^{\mu}\,_{\lambda}\,R^{\lambda}\,_{\pi \phi \psi}\,
\,h^{\pi}\,_{\nu}\,h^{\phi}\,_{\rho}\,h^{\psi}\,_{\eta}
- \chi^{\mu}\,_{\rho}\,\chi_{\nu \eta}
+ \chi^{\mu}\,_{\eta}\,\chi_{\nu \rho}
\end{equation}
and the {\it Codazzi equation}
\begin{equation}
\label{Codazzieq}
D_{\mu}\,\chi_{\nu \eta} - D_{\nu}\,\chi_{\mu \eta}
= - n_{\lambda}\,
R^{\lambda}\,_{\pi \phi \psi}\,
\,h^{\pi}\,_{\eta}\,h^{\phi}\,_{\mu}\,h^{\psi}\,_{\nu}.
\end{equation}
The constraint equations follow from (\ref{Gausseq}) and (\ref{Codazzieq}) by
contractions, the use of the field equations, and pull-back to $S$. Thus the
constraints represent the covariant condition for the isometric
embeddibility of an
{\it initial data set} $(S, h_{ab}, \chi_{ab}, \rho, j_a)$ into a solution
of the
Einstein equations.

We note here that the constraints (\ref{hamconstr}) and (\ref{momconstr}),
which are analogues of the constraints of Maxwell's equations, have
important physical consequences. One of the most important of these is
the positivity of the mass which can be associated with an asymptotically
flat initial data set (subject to reasonable conditions) \cite{schoen:yau},
\cite{witten}.

\subsection{The Bianchi Identities}\label{bianchiid}

Before analysing the structure of the field equations further, we note some
important identities. The Riemann tensor ${\rm Riem}[g]$ of the metric $g$,
given by
\begin{equation}
\label{riccid}
R^{\mu}\,_{\nu \lambda \rho} =
\partial_{\lambda}\,\Gamma_{\rho}\,^{\mu}\,_{\nu}
- \partial_{\rho}\,\Gamma_{\lambda}\,^{\mu}\,_{\nu}
+ \Gamma_{\lambda}\,^{\mu}\,_{\delta}\,\Gamma_{\rho}\,^{\delta}\,_{\nu}
- \Gamma_{\rho}\,^{\mu}\,_{\delta}\,\Gamma_{\lambda}\,^{\delta}\,_{\nu},
\end{equation}
where $\Gamma_{\rho}\,^{\mu}\,_{\nu}$ denotes the Christoffel symbols of
$g_{\mu \nu}$, has the covariance property
\[
{\rm Riem}[\phi^*(g)] = \phi^*({\rm Riem}[g]),
\]
where $\phi$ denotes a diffeomorphism of $M$ into itself.
Two important identities are a direct (cf. \cite{kazdan:1981}) consequence of
this,
the {\it first Bianchi identity}
\begin{equation}
\label{1biid}
R^{\mu}\,_{\lambda \nu \rho}
+ R^{\mu}\,_{\rho \lambda \nu}
+ R^{\mu}\,_{\nu \rho \lambda} = 0,
\end{equation}
and the {\it second Bianchi identity}
\begin{equation}
\label{2biid}
\nabla_{\mu}\, R^{\gamma}\,_{\lambda \nu \rho}
+ \nabla_{\rho}\, R^{\gamma}\,_{\lambda \mu \nu}
+ \nabla_{\nu}\, R^{\gamma}\,_{\lambda \rho \mu} = 0.
\end{equation}
The latter implies the further identities
\begin{equation}
\label{cbiid}
\nabla_{\mu}\, R^{\mu}\,_{\nu \lambda \rho} =
\nabla_{\lambda}\, R_{\nu \rho} - \nabla_{\rho}\, R_{\nu \lambda},
\end{equation}
\begin{equation}
\label{ccbiid}
\nabla^{\mu}\,R_{\mu \nu} - \frac{1}{2}\,\nabla_{\nu}R = 0.
\end{equation}

The second Bianchi identity will serve us two quite different purposes.
Firstly, it will allow us to resolve certain problems arising from the
degeneracy of
the principal symbol considered above (it is the integrability condition which
allows us to show the propagation of suitably chosen gauge conditions and the
preservation of the constraints). Secondly, it will provide us with alternative
representations of the field equations.

\subsection{The Evolution Equations}\label{evole}

In this section we shall discuss a few basic ideas about the evolution
problem.
Our observations about the constraints and the decomposition of
\[
Z_{\mu \nu} \equiv G_{\mu \nu} + \lambda\,g_{\mu\nu}
- \kappa\,T_{\mu \nu},
\]
given by
\begin{equation}
\label{Zdecomp}
Z_{\mu \nu} = Z^S_{\mu \nu} - n_{\nu}\,Z^M_{\mu} - n_{\mu}\,Z^M_{\nu}
+ n_{\mu}\,n_{\nu}\,Z^H,
\end{equation}
with
$Z^S_{\mu \nu} = h_{\mu}\,^{\lambda}\,h_{\nu}^{\rho}\,Z_{\lambda \rho}$,
suggest that the basic information on the evolution equations should be
contained
in
\[
Z^S_{\mu \nu} = 0,
\]
or any combination of it with the constraints. To obtain simple expressions
in terms
of the field
$h_{\mu
\nu}$ and the second fundamental form $\chi_{\mu \nu}$, defined by the
generalization
\begin{equation}
\label{2fufo}
{\cal L}_T\,h_{\mu \nu} = 2\,\alpha\,\chi_{\mu \nu} + {\cal L}_{\beta}\,h_{\mu
\nu},
\end{equation}
of (\ref{gauss2fufo}), we consider the equation
\begin{equation}
\label{ADMeinst}
 0 = Z^S_{\mu \nu}  - \frac{1}{2}\,h_{\mu \nu}\,
(h^{\lambda \rho}\,Z^S_{\lambda \rho} - Z^H)
\end{equation}
\[
= \frac{1}{\alpha}\,({\cal L}_T\,\chi_{\mu \nu}
- {\cal L}_{\beta}\,\chi_{\mu \nu}
- D_{\mu}\,D_{\nu}\,\alpha) + \,r_{\mu \nu} +
\chi_{\rho}\,^{\rho}\,\chi_{\mu \nu} - 2\,\chi_{\rho \mu}\,\chi_{\nu}\,^{\rho}
\]
\[
- \lambda\,h_{\mu\nu}
- \kappa\,h_{\mu}\,^{\rho}\,h_{\nu}\,^{\lambda}
\,(T_{\rho \lambda} - \frac{1}{2}\,T\,g_{\rho \lambda}).
\]
Together with (\ref{2fufo}) it should be regarded as an evolution equation for
the
fields $h_{\mu \nu}$, $\chi_{\mu \nu}$.

It then appears natural to analyse the general solution of the
Einstein equations by the following procedure: Find initial data, i e. a
solution $h_{ab}$,
$\chi_{ab}$ of the constraints, on the slice $S = \{t = 0\}$. Then find the
solution
$h_{ab}$, $\chi_{ab}$ of the equations
\begin{equation}
\label{pbad}
\partial_t\,h_{ab} = 2\,\alpha\,\chi_{ab} + {\cal L}_{\beta}\,h_{ab},
\end{equation}
\begin{equation}
\label{pbag}
\partial_t\,\chi_{ab} =
- \alpha\,(r_{ab} + \chi_c\,^c\,\chi_{ab} - 2\,\chi_{ac}\,\chi_b\,^c)
\end{equation}
\[
+ D_a\,D_b\,\alpha + {\cal L}_{\beta}\,\chi_{ab}
+ \alpha\,(\lambda\,h_{ab} + \kappa\,(T_{ab} - \frac{1}{2}\,T\,h_{ab})),
\]
equivalent to (\ref{2fufo}) and (\ref{ADMeinst}), which induces these data
on $S$. The first step will not be considered further in this article;
we shall give some relevant references on the problem of
solving the constraint equations in Sect. \ref{outlook}.
Here we want to comment on the second step, which raises several questions.

i) What determines the functions $\alpha$, $\beta^a$? Is it possible
to prescribe them, at least locally near $S$, as arbitrary functions
$\alpha = \alpha(t, x^c)$, $\beta^a = \beta^a(t, x^c)$ of the coordinates,
possibly with the restriction $\alpha^2 - \beta_c\,\beta^c > 0$, which would
make
$\partial_t$ timelike?  We could give a positive answer to this question,
if,
starting from a representation  of the metric (\ref{ADMmetric}) in terms of
some coordinate system
$x^{\mu'}$ with $t' = x^{0'} = 0$ on $S$, we could always find a coordinate
transformation
$t = t(x^{\mu'})$, $x^a = x^a(x^{\mu'})$ with $t(0, x^{a'}) = 0$, which casts
the metric into the desired form, i.e. achieves
\begin{equation}
\label{1neik}
- \frac{1}{\alpha^2(t,x^c)} = g^{00} =
g^{\mu' \nu'}(x^{\lambda'})\,t_{,\mu'}\,t_{,\nu'},
\end{equation}
\begin{equation}
\label{2neik}
\frac{1}{\alpha^2(t,x^c)}\,\beta^a(t,x^c) = g^{0a} =
g^{\mu' \nu'}(x^{\lambda'})\,t_{,\mu'}\,x^a\,_{,\nu'}.
\end{equation}
If the left hand side of the first equation only depended on $t$ and
$x^{\mu'}$,
the standard theory of first order PDE's for a single unknown could be
applied to
this equation and second equation would essentially reduce to an ODE.
However, in
general this theory does not apply, because the dependence of the function
$\alpha$
on $x^a$ introduces a coupling to the second equation.

ii) Suppose we could prescribe lapse and shift arbitrarily. Could we then
show the existence of a (unique) solution $h_{ab}$, $\chi_{ab}$ of the
initial value problem for the equations (\ref{pbad}) and  (\ref{pbag})
(possibly coupled to some matter equations)?

iii) Suppose we could answer the last question positively, would the
resulting solution to (\ref{pbad}) and (\ref{pbag}) then satisfy the
constraints (\ref{hamconstr}) and (\ref{momconstr}) on the slices
$\{t = {\rm const.}\}$? Only
then would we know that the metric $g_{\mu \nu}$, obtained from our fields
$h_{ab}$, $\chi_{ab}$, $\alpha$ $\beta^c$ by (\ref{ADMmetric}), is a solution
of the Einstein equations.

We can answer question (iii) as follows. Suppose equation (\ref{ADMeinst}) is
satisfied on a set $M = ]- a, c[ \times S$, with $a,\,c > 0$, and the solution
induces the given data on $\{0\} \times S$, which we assume to be identified
with $S$ in the obvious way. Since (\ref{ADMeinst}) is equivalent to
$Z^S_{\mu \nu} - h_{\mu\nu}\,Z^H = 0$, we can write on $M$ by (\ref{Zdecomp})
\[
G_{\mu \nu} + \lambda\,g_{\mu \nu} - \kappa\,T_{\mu \nu}
= - n_{\nu}\,Z^M_{\mu} - n_{\mu}\,Z^M_{\nu}
+ \{2\,n_{\mu}\,n_{\nu} + g_{\mu \nu}\}\,Z^H.
\]
Taking the divergence, using the contracted Bianchi identity
(\ref{ccbiid}),
assuming that the matter field equations
have been given such as to ensure $\nabla^{\mu}\,T_{\mu \nu} = 0$, and
splitting
into normal and tangential parts, we get the equations
\[
n^{\mu}\,\nabla_{\mu}\,Z^H - h^{\mu \nu}\,D_{\mu}\,Z^M_{\nu}
= 2\,Z^M_{\nu}\,n^{\mu}\,\nabla_{\mu}\,n^{\nu}
- 2\,Z^H\,\nabla_{\mu}\,n^{\mu},
\]
\[
n^{\mu}\,\nabla_{\mu}\,Z^M_{\nu} - D_{\nu}\,Z^H
= - Z^M_{\mu}\,\nabla^{\mu}\,n_{\nu}
- Z^M_{\nu}\,\nabla_{\mu}\,n^{\mu}
\]
\[
+ Z^M_{\rho}\,n_{\nu}\,n^{\mu}\,\nabla_{\mu}\,n^{\rho}
+ 2\,Z^H\,n^{\mu}\,\nabla_{\mu}\,n_{\nu},
\]
which imply {\it subsidiary equations}, satisfied by $Z^H$ and $Z^M_a$,
\begin{equation}
\label{ADMsubs}
\left\{ \left( \begin{array}{cc}
\frac{1}{\alpha} & 0 \\
0 & \frac{1}{\alpha}\,h^{ac}\\
\end{array}
\right)
\partial_0 +
\left( \begin{array}{cc}
- \frac{1}{\alpha}\,\beta^d & - h^{cd}\\
- h^{ad} & \,\,- \frac{1}{\alpha}\,h^{ac}\,\beta^d \\
\end{array}
\right)
\partial_d \right\}
\left( \begin{array}{c}
Z^H\\
Z^M_c\\
\end{array}
\right)
=
\left( \begin{array}{c}
h\\
h^a\\
\end{array}
\right),
\end{equation}
where $h$, $h^a$ denote linear functions of $Z^H$, $Z^M_a$. Since it is a
system for
$v =\,^t(Z^H, Z^M_c)$ of the form
\[
A^{\mu}\,\partial_{\mu}\,v+B\,v=0,
\]
with symmetric matrices $A^{\mu}$ and a positive definite matrix $A^0$ , it
is {\it symmetric hyperbolic} (cf. Sect. \ref{symhyp}). Moreover, it has
{\it characteristic polynomial}
\begin{equation}
\label{auxpol}
\det(A^{\mu}\,\xi_{\mu}) = - \det(h^{ab})\,
(n^{\rho}\,\xi_{\rho})^2\,g^{\mu \nu}\xi_{\mu}\,\xi_{\nu},
\end{equation}
which implies that its characteristics are hypersurfaces which are timelike or
null with respect to the metric $g_{\mu \nu}$.

Consequently, if $S' = \{\phi = 0, \phi_{,\mu} \neq 0\} \subset M$, with
$\phi \in C^{\infty}(M)$, is a spacelike hypersurface, the matrix
$A^{\mu}\,\phi_{,\mu}$ is positive definite on $S'$.
Suppose $S'$, $S''$ are two spacelike hypersurfaces which intersect at
their common $2-$dimensional boundary $Z$ and bound a compact
\lq lens-shaped region\rq\ in $M$. Then it follows from the discussion of
symmetric hyperbolic systems in Sect. \ref{symhyp} that the fields $Z^H$,
$Z^M_c$ must vanish on $S''$ if they vanish on $S'$.

To make a precise statement about the consequences of this property, we need
the
important notion of the domain of dependence. Let us assume that there is
given a
time orientation on $(M, g)$. If $U$ is a closed subset of $M$ we define the
{\it future (past) domain of dependence of $U$ in $M$} as the set of points
$x \in M$ such that any $g-$non-spacelike curve in $M$ through $x$ which is
inextendible in the past (future) intersects $U$. We denote this set by
$D^{+}(U)$ (resp. $D^{-}(U)$).

It can be shown that the result about lens-shaped regions referred to above
and the fact that the fields $Z^H$, $Z^M_c$ vanish on $S$ imply that $Z^H$
and $Z^M_c$, whence also $Z^M_{\mu}$, vanish on the {\it domain of dependence}
$D(S) = D^{+}(S) \cup D^{-}(S)$ of $S$ in $M$. This shows the
{\it preservation of the constraints} under the evolution defined by
equations (\ref{pbad}) and (\ref{pbag}) and the prescribed lapse and shift.

Questions (i), (ii) are more delicate. Let us assume that the coefficients
$g^{\mu'\nu'}(x^{\lambda'})$ are real analytic functions for $x^{\lambda'}$
in an
open subset $V$ of ${\R}^4$ with $V \cap \{x^{0'} = 0\} \neq \emptyset$,
and that
$\alpha > 0$ and $\beta^c$ are real analytic functions of $t$ and $x^a$. Then
equations (\ref{1neik}) and (\ref{2neik}) can be written in the form
\[
t_{,0'} = F_1(t, x^a, t_{,c'}, x^a\,_{,c'}, x^{\mu'}),\,\,\,\,\,\,
x^a\,_{,0'} = F_2(t, x^a, t_{,c'}, x^a\,_{,c'}, x^{\mu'}),
\]
with functions $F_1$, $F_2$ which are real analytic for
$(t, x^a, t_{,c'}, x^a\,_{,c'}, x^{\mu'}) \in {\R}^{16} \times V$. Thus,
by the
theorem of Cauchy--Kowalewskaya (cf. \cite{dieudonne:IV}), the
differential problem
considered in question (i) can be solved in a neighbourhood of the set
$\{x^{0'} = 0\} \subset V$. Using the covariance of equations
(\ref{1neik}), (\ref{2neik}), it follows that given a real analytic Lorentz
space $(M, g)$, an analytic spacelike hypersurface $S$ in $M$ with coordinates
$x^a$ on $S$, and analytic functions $\alpha = \alpha(t, x^c) > 0$,
$\beta^a = \beta^a(t, x^c)$ there exist unique real analytic coordinates
$t$, $x^a$ on some neighbourhood of $S$ in $M$ such that $t = 0$ on $S$ and
the lapse and shift in the expression of $g$ in these coordinates are given by
$\alpha$ and $\beta^a$.

Assume now that the $3-$manifold $S$, the initial data $h_{ab}$, $\chi_{ab}$
solving the constraints on $S$, as well as the functions $\alpha(t, x^c) > 0$,
$\beta^a(t, x^c)$ are real analytic and assume for simplicity that
$T_{\mu \nu} = 0$. Then we can derive from (\ref{pbad}) and (\ref{pbag}) a
differential
system for $u = (h_{ab}, k_{abc} \equiv h_{ab,c}, \chi_{ab})$ of the form
$\partial_t\,u = H(u, t, x^a)$ with a function $H$ which is real analytic
where $\det(h_{ab}) \neq 0$. Again the theorem of Cauchy--Kowalewskaya tells
us that
this system, whence also (\ref{pbad}) and (\ref{pbag}), has a unique real
analytic
solution on ${\R} \times S$ near $\{0\} \times S$ for the data which are
given on
$\{0\} \times S$ after identifying the latter in the obvious way with $S$.
By our discussion of (iii) we know that we thus obtain a unique analytic
solution to
the full Einstein equations.

It should be noted that the solution obtained in this way depends a priori
not only on
the data but also on the chosen lapse and shift. That the latter do in fact
only affect the coordinate representation of the solution can be seen as
follows. Given another set of analytic functions $\alpha'(t', x^{'c}) > 0$,
$\beta^{'a}(t', x^{'c})$, we can either find, as remarked above, a coordinate
transformation $t \rightarrow t'$, $x^{a} \rightarrow x^{'a}$ such that the
given metric has the new values of lapse and shift in the new
coordinates , or we can deal with the initial value problem for (\ref{pbad})
and (\ref{pbag}) with the new lapse and shift.
However, due to the uniqueness of this solution it must coincide with the
first solution in its new coordinate representation.

It is a remarkable fact that in the course of solving the Einstein equations
we can prescribe rather arbitrarily four functions $\alpha$, $\beta^a$
(in the analytic case) which are considered first as functions on some
abstract ${\R}^4$ but which, once the solution has been constructed,
acquire the meaning of lapse and shift for the coordinate expression of the
metric. Since the coordinates in which they have
this meaning are defined by $\alpha$, $\beta^a$ implicitly (via the field
equations), we refer to these functions as the
{\it gauge source functions} of
our procedure (we shall see below that, depending on the chosen equations,
quite
different objects can play the role of gauge source functions) and to the act
of prescribing these functions as imposing a {\it gauge condition}. The
considerations above show also that {\it the manifold on which the solution is
constructed must be regarded as part of the solution}. The transition
functions relating
the different coordinates we have considered, as well that the domains of
definition
of these coordinate systems themselves, are determined by the gauge source
functions,
the field equations, and the initial data.

We have seen that in the case where the data and the given lapse and shift
are real
analytic, we can answer our questions in a satisfactory way. However, the
assumption is not satisfactory. This is not meant to say anything against
analytic solutions. In fact, most of the \lq exact solutions\rq\ which
are the
source of our intuition for general relativistic phenomena are (piecewise)
analytic. However, we should not restrict to analyticity in principle. One
reason is that it would be in conflict with one of the basic tenets
of general relativity. Given two non-empty open subsets $U$, $V$ of a connected
space-time $(M, g)$ such that no point of $U$ can be connected by a causal
curve
with a point of $V$, any process in $U$ should be independent of what happens
in
$V$. However, if the space-time is analytic, the field in $V$ is essentially
fixed
by the behaviour of $g$ in $U$. For instance, we would not be able to study the
evolution of data $h_{ab}$, $\chi_{ab}$ on a $3-$manifold $S$ where $h_{ab}$ is
conformally flat in some open subset of $S$ but not in another one.

Therefore, Einstein's equations should allow us to discuss the existence and
uniqueness of solutions, and also the continuous dependence of the latter
on the
data (stability), in classes of functions which are $C^{\infty}$ or of even
lower
smoothness. In other words, Einstein's equations should imply evolution
equations for
which the Cauchy problem is {\it well posed} (cf. \cite{garabedian},
\cite{hadamard52},
\cite{treves}). Whether an initial value problem is well posed cannot be
decided on
the level of analytic solutions and with the methods used to prove the
Cauchy--Kowalewskaya theorem. On this level there is no basic distinction
between
initial value problems based on spacelike and those based on timelike
hypersurfaces,
though in the latter case the stability property is known not to be
satisfied. Thus
we are led to search for evolution equations which satisfy some
\lq hyperbolicity condition\rq, i.e. a condition (essentially) on the
algebraic structure of the equations
which entails the well-posedness of the Cauchy problem.

A number of different hyperbolicity conditions are known, all of them
having in
common that they require the equations to admit at each point a maximal
number of \lq real\rq\ characteristics: if the equations have a local
expression of the form
(\ref{dop}), then the operator $P$ is hyperbolic at $x \in U$ only if
there is a covector $\zeta \in T^*_x\,M \setminus \{0\}$ such that every
$2-$dimensional plane in the cotangent space $T^*_x\,M$ containing $\zeta$
intersects the {\it conormal cone}
$\{det(\sum_{|\alpha| = m} A^{\alpha}\,\xi^{\alpha}) = 0\}$ in
$k \times m$ real lines (counting multiplicities) (cf.
\cite{courant62}).
But notice that this condition alone does not ensure the well-posedness
of the Cauchy problem. Some further remarks about different notions of
hyperbolicity can be found in Sect. \ref{otherhyp}.

To analyse the situation in the case of (\ref{pbad}) and (\ref{pbag}), we
solve (\ref{pbad}) for $\chi_{ab}$ and insert into (\ref{pbag}) to obtain
a system of second order for $h_{ab}$ which takes the form
\begin{equation}
\label{2ndADM}
\frac{1}{\alpha^2}\,
\{ \partial^2_t\,h_{ab}
- \beta^c\,\partial_c\,\partial_t\,h_{ab}
- \beta^c\,\partial_t\,\partial_c\,h_{ab}
+ \beta^c\,\beta^d\,\partial_c\,\partial_d\,h_{ab} \}
\end{equation}
\[
- h^{cd}\,\left(
\partial_c\,\partial_d\,h_{ab} + \partial_a\,\partial_b\,h_{cd}
- \partial_a\,\partial_c\,h_{bd} - \partial_b\,\partial_c\,h_{ad}
\right)
\]
\[
- \frac{2}{\alpha}\,\partial_a\,\partial_b\,\alpha
- \frac{2}{\alpha}\,\left(h_{c(a}\,\partial_{b)}\,\partial_t\,\beta^c
- \beta^d\,h_{c(a}\,\partial_{b)}\,\partial_d\,\beta^c \right)
\]
\[
= \quad\mbox{terms of lower order in}\quad h_{ab},\,\alpha,\,\beta^c.
\]
To analyse the characteristic polynomial, we have to know how $\alpha$ and
$\beta^c$ are related to the solution $h_{ab}$. Suppose, for simplicity,
that
$\alpha$ and $\beta^c$ are given functions.. Then we have to calculate for
given
covector $\xi_{\mu} \neq 0$ the determinant of the linear map
\[
k_{ab} \mapsto \bar{k}_{ab} =
- g^{\mu \nu}\,\xi_{\mu}\,\xi_{\nu}\,k_{ab}
- h^{cd}\,k_{cd}\,\xi_a\,\xi_b + 2\,\xi^c\,k_{c(a}\,\xi_{b)},
\]
of symmetric tensors, where we set $\xi^a = h^{ab}\,\xi_b$. Denoting by
$A'(\xi)$
the linear transformation which maps the independent components $k_{ab}$,
$a \le b$,
onto the $\bar{k}_{ab}$, $a \le b$, we find
\[
\det\,A'(\xi) = - (n^{\mu}\,\xi_{\mu})^6\,(g^{\mu
\nu}\,\xi_{\mu}\,\xi_{\nu})^3.
\]
Thus, if we consider lapse and shift as given, the conormal cone of system
(\ref{2ndADM}) satisfies the condition required by hyperbolicity. Moreoever,
the characteristics are timelike or null as one would expect for the evolution
equations
of a theory which is founded on the idea that physical processes propagate on
or inside the light cone of the metric field.

{\it However, in spite of the fact that these equations have been used for a
long time and in various contexts, and in spite of the naturality of equation
(\ref{pbag}) which appears to be indicated by the hyperbolicity of the
subsidiary
equations, it is not known whether the Cauchy problem for equations
(\ref{2ndADM}) is
well posed. It appears that we need to make use of the constraint equations to
obtain suitable evolution equations.}

Therefore we proceed along a different route. While the form of the constraint
equations induced on a given hypersurface is unique, there is a huge freedom
to
modify our evolution equations. We can try to bring the principal part of
(\ref{2ndADM}) into a suitable form  by using the constraints, by assuming
lapse and
shift to be functionals of the metric, or by subjecting them to some
equations. Then
gauge source functions of quite a different nature may appear. Our choice in
this section is motivated by the following two observations (\cite{lanczos},
\cite{bruhat52}, cf. also \cite{friedrich:1985}):

(i) Suppose $S$ is some spacelike hypersurface of some Lorentz space
$(M, g)$ and
$x^a$ are coordinates on some open subset $U$ of $S$. Let
$F^{\mu'} = F^{\mu'}(x^{\nu'})$ be four smooth real functions defined on
${\R}^4$.
Then there exist coordinates $x^{\nu'}$ on some neighbourhood of $U$ in $M$
with
$x^{0'} = 0$, $x^{a'} = x^a$ on $U$ and such that the Christoffel
coefficients of
$g$ in these coordinates satisfy the relations
$\Gamma^{\mu'}(x^{\nu'}) = F^{\mu'}(x^{\nu'})$, where
$\Gamma^{\mu'} = g^{\lambda' \rho'}\,\Gamma_{\lambda'}\,^{\mu'}\,_{\rho'}$.

Obviously, one can construct coordinates $x^{\nu'}$ on some neighbourhood of
$U$ by
solving Cauchy problems for the semi-linear wave equations
$\nabla_{\mu}\,\nabla^{\mu}\,x^{\nu'} = - F^{\mu'}(x^{\rho'})$ with
Cauchy data on
$U$ which are consistent with our requirements. When the wave equations are
expressed in these coordinates the relations above result.

(ii) The $4-$dimensional Ricci tensor can be written in the form
\begin{equation}
\label{einlanc}
R_{\mu \nu} = - \frac{1}{2}\,g^{\lambda \rho}\,g_{\mu \nu,\lambda \rho} +
\nabla_{(\mu}\Gamma_{\nu)}
\end{equation}
\[
+ \Gamma_{\lambda}\,^{\eta}\,_\mu\,g_{\eta \delta}
\,g^{\lambda \rho}\,\Gamma_{\rho}\,^{\delta}\,_\nu
+ 2\,\Gamma_{\delta}\,^{\lambda}\,_{\eta}\,
g^{\delta \rho}\,g_{\lambda(\mu}\,\Gamma_{\nu)}\,^{\eta}\,_{\rho},
\]
where we set
\[
\Gamma_{\nu} = g_{\nu \mu}\,\Gamma^{\mu},\,\,\,\,\,\,\,
\nabla_{\mu}\Gamma_{\nu} =
\partial_{\mu}\Gamma_{\nu}
- \Gamma_{\mu}\,^{\lambda}\,_{\nu}\,\Gamma_{\lambda}.
\]
Thus, if we consider the $\Gamma_{\nu}$ as given functions, the Einstein
equations take the form of a system of wave equations for the metric
coefficients.

Before we show that these observations lead to a short and elegant argument
for the
well-posedness of the initial value problem for Einstein's equations
(assuming well
behaved matter equations), we indicate how the form of equations
(\ref{einlanc})
relates to our previous considerations.

{}From the expressions (\ref{gammadecomp}) we get the relations
\begin{equation}
\label{atransp}
\partial_t\,\alpha - \alpha_{,a}\,\beta^a
= \alpha^2\,(\chi - n_{\nu}\Gamma^{\nu}),
\end{equation}
\begin{equation}
\label{btransp}
\partial_t\,\beta^a - \beta^a\,_{,b}\,\beta^b
= \alpha^2\,(\gamma^a - D^a\log\,\alpha
- h^a\,_{\nu}\,\Gamma^{\nu}),
\end{equation}
which indicate that prescribing the functions $\Gamma^{\nu}$ may
fix the evolution of lapse and shift. Writing the $3-$Ricci tensor similarly to
(\ref{einlanc})
\[
r_{ab} = - \frac{1}{2}\,h^{cd}\,h_{ab,cd} + D_{(a}\gamma_{b)}
+ \gamma_c\,^d\,_a\,h_{fd}\,h^{ce}\,\gamma_e\,^f\,_b
+ 2\,\gamma_c\,^d\,_e\,h^{cf}\,h_{d(a}\,\gamma_{b)}\,^e\,_f,
\]
with
\begin{equation}
\label{gaexpr}
\gamma_a = h_{ab}\,\gamma^b = h_{ab}\,h^{cd}\,\gamma_c\,^b\,_d
= h^{cd}\,(h_{ac,d} - \frac{1}{2}\,h_{cd,a}),
\end{equation}
we obtain (\ref{2ndADM}) in the form
\[
\frac{1}{\alpha^2}\,(\partial^2_t\,h_{ab} - 2\,h_{ab,tc}\,\beta^c
+ h_{ab,cd}\,\beta^c\,\beta^d)
- h^{cd}\,h_{ab,cd}
\]
\[
- \frac{2}{\alpha^2}\,D_{(a}\,\left[h_{b)c}\,(\partial_t\,\beta^c -
\beta^d\,\partial_d\,\beta^c)\right]
+ 2\,D_{(a}\gamma_{b)}
- 2\,D_a\,D_b\,\log\,\alpha
\]
\[
= \quad\mbox{terms of lower order in}\quad h_{ab},\,\alpha,\,\beta^c.
\]
Using (\ref{btransp}) to replace the terms of second order in the second line,
we get
\begin{equation}
\label{hwave}
- g^{\mu \nu}\,h_{ab,\mu \nu}
= - 2\,D_{(a}\{h_{b)c}\,h^c\,_{\mu}\,\Gamma^{\mu}\}
\end{equation}
\[
+ \frac{2}{\alpha^2}\left\{D_a\,\alpha\,D_b\,\alpha
+ \chi_{ab}\,(\partial_t\,\alpha - {\cal L}_{\beta}\,\alpha)
+ 2\,D_{(a}\alpha\,h_{b)c}\,(\partial_t\beta^c - \beta^c\,_{,d}\,\beta^d)
\right.
\]
\[
\left. + 2\,\beta^c\,_{,(a}\,h_{b)c,t}
+ h_{c(a}\,\beta^d\,_{,b)}\,\beta^c\,_{,d}
- \beta^d\,_{,(a}\,{\cal L}_{\beta}\,h_{b)d}
\right\}
\]
\[
+ 2\,\left\{2\,\chi_{ac}\,\chi_b\,^c - \,\chi_c\,^c\,\chi_{ab}
- \gamma_c\,^d\,_a\,h_{fd}\,h^{ce}\,\gamma_e\,^f\,_b\right.
\]
\[
\left.- 2\,\gamma_c\,^d\,_e\,h^{cf}\,h_{d(a}\,\gamma_{b)}\,^e\,_f
+ \lambda\,h_{ab} + \kappa\,(T_{ab} - \frac{1}{2}\,T\,h_{ab})\right\},
\]
which can be read as a wave equation for $h_{ab}$.
{}From (\ref{atransp}) and (\ref{pbad}) we get
\begin{equation}
\label{1awave}
\frac{1}{\alpha^2}\,(\partial_t - \beta^c\,\partial_c)^2\,\alpha
= (\partial_t - \beta^c\,\partial_c)\,\chi
+ 2\,\alpha\,\chi^2 - 5\,\alpha^2\,\chi\,n_{\nu}\Gamma^{\nu}
\end{equation}
\[
+ 3\,\alpha^3\,(n_{\nu}\Gamma^{\nu})^2
- \alpha\,(\partial_t - \beta^c\,\partial_c)\,
(n_{\nu}\Gamma^{\nu}).
\]
Taking the trace of (\ref{pbag}) and using (\ref{pbad}) and the
Hamiltonian constraint (\ref{hamconstr}), we get
\begin{equation}
\label{chievol}
(\partial_t - \beta^c\,\partial_c)\,\chi = D_a\,D^a\,\alpha
- \alpha\left\{\chi_{ab}\,\chi^{ab} - \lambda
+ \kappa\,\frac{1}{2}\,(\rho + h^{ab}\,T_{ab})
\right\},
\end{equation}
whence, using (\ref{1awave}), the wave equation
\begin{equation}
\label{awave}
\frac{1}{\alpha^2}\,(\partial_t - \beta^c\,\partial_c)^2\,\alpha
- D_a\,D^a\,\alpha
= - \alpha\left\{\chi_{ab}\,\chi^{ab} - \lambda
\right.
\end{equation}
\[
\left. + \kappa\,\frac{1}{2}\,(\rho + h^{ab}\,T_{ab})
+ 5\,\alpha\,\chi\,n_{\nu}\Gamma^{\nu}
- 3\,\alpha^2\,(n_{\nu}\Gamma^{\nu})^2
+ (\partial_t - \beta^c\,\partial_c)\,
(n_{\nu}\Gamma^{\nu})\right\}.
\]
{}From (\ref{gaexpr}), (\ref{pbad}), and the momentum constraint
(\ref{momconstr}) follows
\begin{equation}
 \label{bellipt}
(\partial_t - \beta^c\,\partial_c)\,\gamma^a = h^{cd}\,\beta^a\,_{,cd}
+ \alpha\,D^a\chi + 2\,\alpha\,\kappa\,j^a
\end{equation}
\[
+ 2\,\chi^{ac}\,D_c\alpha
- \chi\,D^a\alpha - 2\,\alpha\,\chi^{cd}\,\gamma_c\,^a\,_d -
\beta^a\,_{,c}\,\gamma^c,
\]
which implies together with (\ref{atransp}), (\ref{btransp}),
and (\ref{pbad}) the wave equation
\begin{equation}
\label{bwave}
\frac{1}{\alpha^2}\,(\partial_t - \beta^c\,\partial_c)^2\,\beta^a
- h^{cd}\,\beta^a\,_{,cd} =  2\,\alpha\,\kappa\,j^a
\end{equation}
\[
+ 4\,(\chi^{ac} - \chi\,h^{ac})\,D_c\alpha
- 2\,\alpha\,(\chi^{bc} - \chi\,h^{bc})\,\gamma_b\,^a\,_c
- \beta^a\,_{,c}\,\gamma^c
+ \beta^a\,_{,c}\,D^c\log\,\alpha
\]
\[
- 2\,\alpha\,n_{\nu}\,\Gamma^{\nu}\,(\gamma^a - D^a\log\,\alpha -
h^a\,_{\nu}\,\Gamma^{\nu})
- 2\,\alpha\,\chi\,h^a\,_{\nu}\,\Gamma^{\nu}
\]
\[
+ D^a(\alpha\,n_{\nu}\,\Gamma^{\nu})
- (\partial_t - \beta^c\,\partial_c)(h^a\,_{\nu}\,\Gamma^{\nu}).
\]
Equations (\ref{hwave}), (\ref{awave}), and (\ref{bwave}) form a hyperbolic
system for the
fields $h_{ab}$, $\alpha$, $\beta^a$ if we consider the functions
$\Gamma^{\mu}$ as
given. We note that besides (\ref{pbag}) we used the constraints as well as
(\ref{gammadecomp}) to derive this system.

We return to the evolution problem. Suppose we are given smooth data $h_{ab}$,
$\chi_{ab}$, $\rho$, $j_a$, i. e. a solution of the constraints, on some
$3-$dimensional manifold $S$, which, for simplicity, we assume to be
diffeomorphic to ${\R}^3$ and endowed with global coordinates $x^a$.
Following our previous considerations, we set $M = {\R} \times S$, denote
by $t = x^0$ the natural coordinate on ${\R}$ and extend the $x^a$ in
the obvious way to $M$. We embed our initial data set into $M$ by identifying
$S$ diffeomorphically with $\{0\} \times S$ (the need for this embedding
shows that it is in general not useful to restrict the choice of the
coordinates $x^a$).

We now choose four smooth real functions $F_{\mu}(x^{\lambda})$ on $M$ which
will be assigned the role of gauge source functions in the following
procedure. As suggested by (\ref{einlanc}) and the preceeding discussion, we
study the Cauchy problem for the {\it reduced equations}
\begin{equation}
\label{2redequ}
- \frac{1}{2}\,g^{\lambda \rho}\,g_{\mu \nu,\lambda \rho} +
\nabla_{(\mu}\,F_{\nu)}
+ \Gamma_{\lambda}\,^{\eta}\,_\mu\,g_{\eta \delta}
\,g^{\lambda \rho}\,\Gamma_{\rho}\,^{\delta}\,_\nu
\end{equation}
\[
+ 2\,\Gamma_{\delta}\,^{\lambda}\,_{\eta}\,
g^{\delta \rho}\,g_{\lambda(\mu}\,\Gamma_{\nu)}\,^{\eta}\,_{\rho}
= - \lambda\,g_{\mu \nu} + \kappa\,(T_{\mu \nu} - \frac{1}{2}\,T\,g_{\mu \nu}).
\]
Since we do not want to get involved in this section with details of the matter
equations, we assume the Cauchy problem for them to be well posed and the
energy-momentum tensor to be divergence free as a consequence of these
equations (one of the remarks following formula (\ref{qksusseq}) shows that
the situation can be occasionally somewhat more subtle).

To prepare Cauchy data for (\ref{2redequ}), we choose on $S$ a smooth positive
lapse
function  $\alpha$ and a smooth shift vector field $\beta^a$, which determine
with
the datum $h_{ab}$ the ADM representation (\ref{ADMmetric}) for $g_{\mu \nu}$
on $S$.
Using now $h_{ab}$, $\chi_{ab}$, $\alpha$, $\beta^a$ in equations
(\ref{atransp}),
(\ref{btransp}) (with $\Gamma^{\nu}$ replaced by $g^{\nu \mu}\,F_{\mu}$), and
(\ref{pbad}), we obtain the corresponding datum $\partial_t\,g_{\mu \nu}$ on
$S$.

It will be shown in Sect. \ref{locevol}
that the Cauchy problem for (\ref{2redequ}) and
the
initial data above is well posed. Thus there exists a smooth solution
$g_{\mu \nu}$ to it on some neighbourhood $M'$ of $S$. To assure uniqueness, we
assume that $M'$ coincides with the domain of dependence of $S$ in $M'$ with
respect to $g_{\mu \nu}$.

With the functions $\Gamma_{\mu}$ calculated from $g_{\mu \nu}$, the reduced
equation (\ref{2redequ}) takes the form
\[
G_{\mu \nu} + \lambda\,g_{\mu \nu} - \kappa\,T_{\mu \nu}
= \nabla_{(\mu}\,D_{\nu)} - \frac{1}{2}\,g_{\mu \nu}\,\nabla_{\rho}\,D^{\rho}
\]
with $D_{\mu} = \Gamma_{\mu} - F_{\mu}$. To see that our {\it gauge
conditions are
preserved} under the evolution defined by (\ref{2redequ}) and our gauge source
functions, we show that $D_{\mu} = 0$ on $M'$.

Using the Bianchi identity (\ref{ccbiid}) and our assumptions on the matter
equations, we get, by taking the divergence of the equation above, the
subsidiary
equation
\[
\nabla_{\mu}\,\nabla^{\mu}\,D_{\nu} + R^{\mu}\,_{\nu}\,D_{\mu} = 0.
\]
Since we used (\ref{atransp}) and (\ref{btransp}) to determine the initial
data, we know
that $D_{\mu} = 0$ on $S$. Moreover, equations (\ref{awave})and (\ref{bwave})
may be used,
on the one hand, to calculate $\partial_t\,\Gamma_{\mu}$ from $g_{\mu \nu}$,
but
they are, on the other hand, contained in (\ref{2redequ}), with $\Gamma_{\mu}$
replaced by $F_{\mu}$. This implies that $\partial_t\,D_{\mu} = 0$ on $S$.
Using the uniqueness property for systems of wave equations discussed in
Sect. \ref{locexist}, we
conclude that $D_{\mu}$ vanishes on $M'$ and $g_{\mu \nu}$ solves indeed
Einstein's equations (\ref{einst}) on $M'$.

We refer to the process of reducing the initial value problem for Einstein's
equations to an initial value problem of a hyperbolic system as a
{\it hyperbolic reduction}. The argument given here was developed for the first
time in \cite{bruhat52} with the \lq harmonicity assumption\lq\
$\Gamma_{\mu} = 0$. We prefer to keep the complete freedom to specify the gauge
source functions, because some important and complicated problems are related
to
this. Our derivation of the system (\ref{hwave}), (\ref{awave}),
(\ref{bwave})
illustrates the intricate relations between equation (\ref{pbag}), the
constraints,
and the conservation of the gauge condition in our case. Though other such
arguments
may differ in details, the overall structure of all hyperbolic reduction
procedures
are similar.

\subsection{Assumptions and Consequences}\label{asscon}

If one wants to investigate single solutions or general classes of solutions
to the
Einstein equations by an abstract analysis of the hyperbolic reduced
equations, there
are certain properties which need to be \lq put in by hand\rq\ and others
which are determined by the field equations.

The properties in the first class depend largely on the type of physical
systems
which are to be modelled and on the structure of the hypersurface on which
data are
to be prescribed. The latter could be spacelike everywhere as in the Cauchy
problem
considered above. One may wish in this case to consider cosmological models
with
compact time slices, one of which is the initial hypersurface, one may
wish to
model the field of an isolated system, in which case the initial hypersurface
should
include a domain extending to infinity with the field satisfying certain
fall-off
conditions, or one may wish to consider initial hypersurfaces with \lq inner
boundaries\rq\ whose nature may depend on various geometrical and physical
considerations. Other cases of interest are the \lq initial-boundary value
problem\rq (cf. Sect. \ref{ibvp}),
where data are given on a spacelike and a timelike hypersurface which
intersect in
some spacelike
2-surface, the \lq characteristic initial value problem\rq\ where data are
given on
two null hypersurfaces which intersect in some spacelike two-surface, or
various
other combinations of hypersurfaces. Finally, we have to make a choice of
matter
model which may introduce specific problems which are independent of the
features
of Einstein's equations considered above. All these considerations will
affect the
problem of finding appropriate solutions to the corresponding constraint
equations
and the nature of this problem depends to a large extent on the nature of the
bounding hypersurface. We note that the fall-off behaviour of the data on
spacelike
hypersurfaces can only partly be specified freely, other parts being
determined by
the constraint equations.

There are conditions which we just assume because we cannot do better. For
instance,
one may wish to analyse solutions which violate some basic causality
conditions (cf.
\cite{hawking:ellis}) or which admit various identifications in the future
and past
of the initial hypersurface. Such violations or identifications introduce
compatibility conditions on the data which cannot be controlled locally. At
present there are no techniques available to analyse such situations in any
generality. In the context of the Cauchy problem we shall only consider
solutions
$(M, g)$ of the Einstein equations which are {\it globally hyperbolic} and such
that the initial hypersurface $S$ is a {\it Cauchy hypersurface} of the
solution,
i.e. every inextendible non-spacelike curve in $M$ intersects $S$ exactly
once
\cite{hawking:ellis}.

The class of properties which need to be inferred from the structure of the
data and the field equations includes in general everything which has to do
with the long time evolution of the field: the development of singularities
and horizons or of asymptotic regimes where the field becomes in any sense
weak and possibly approximates the Minkowski field. This does not, of course,
preclude the possibility of making assumptions on the singular or asymptotic
behaviour of the field and analysing the consistency of these assumptions
with the field equations. But in the end we would like to characterize the
long time behaviour of the field in terms of the initial data.

\section{PDE Techniques}\label{pde}

\subsection{Symmetric Hyperbolic Systems}\label{symhyp}

In order to obtain results on the existence, uniqueness and stability
of solutions of the Einstein equations it is necessary to make contact
with the theory of partial differential equations. A good recent textbook
on this theory is \cite{evans98}. One part of the theory
which can usefully be applied to the Einstein equations is the theory of
symmetric hyperbolic systems. This will be discussed in some detail in
the following. The aim is not to give complete proofs of the results of
interest in general relativity, but to present some arguments which illustrate
the essential techniques of the subject. It is important to note that
the use of symmetric hyperbolic systems is not the only way of proving
local existence and uniqueness theorems for the Einstein equations. The
original proof \cite{bruhat52} used second order hyperbolic equations.
Other approaches use other types of equations such as mixed
hyperbolic-elliptic systems (see, e.g. \cite{christodoulou93},
Chap. 10).
The basic tools used to prove existence are the same in all cases. First a
family of approximating problems is set up and solved. Of course, in order
that this be helpful, the approximating problems should be easier to solve
than the original one. Then energy estimates, whose definition is discussed
below, are used to show that the solutions of the approximating problems
converge to a solution of the original problem in a certain limit.

Now some aspects of the theory of symmetric hyperbolic systems will be
discussed. We consider a system of equations for $k$ real variables which are
collected into a vector-valued function $u$. The solution will be defined on
an appropriate subset of $\R\times\R^n$. (The case needed for the Einstein
equations is $n=3$.) A point of $\R\times\R^n$ will be denoted by $(t,x)$.
The equations are of the form:
$$A^0(t,x,u)\d_t u+A^i(t,x,u)\d_i u+B(t,x,u)=0$$
This system of equations is called symmetric hyperbolic if the matrices
$A^0$ and $A^i$ are symmetric and $A^0$ is positive definite. This system
is quasi-linear, which means that it is linear in its dependence on the first
derivatives. The notion of symmetric hyperbolicity can be defined more
generally, but here we restrict to the quasi-linear case without further
comment. It is relatively
easy to prove a uniqueness theorem for solutions of a symmetric hyperbolic
system and so we will do this before coming to the more complicated existence
proofs. A hypersurface $S$ is called spacelike with respect to a solution $u$
of the equation if for any 1-form $n_\alpha$ conormal to $S$ (i.e. vanishing
on vectors tangent to $S$) the expression $A^\alpha (t,x,u) n_\alpha$ is
positive definite. This definition has a priori nothing to do with the usual
sense of the word spacelike in general relativity. However, as will be seen
below, the two concepts are closely related in some cases.
Define a {\it lens-shaped region} to be an open subset $G$ of
$\R\times\R^n$ with compact closure whose boundary is the union of a subset
$S_0$ of the hypersurface $t=0$ with smooth boundary $\d S_0$ and a spacelike
hypersurface $S_1$ with a boundary which coincides with $\d S_0$. It will be
shown that if $G$ is a lens-shaped region with respect to solutions $u_1$
and $u_2$ of a symmetric hyperbolic system and if the two solutions agree on
$S_0$ then they agree on all of $G$. This statement is subject to
differentiability requirements.

Consider a symmetric hyperbolic system where $A^\alpha$ and $B$ are $C^1$
functions of their arguments. Let $u_1$ and $u_2$ be two solutions of class
$C^1$. Let $G$ be a lens-shaped region with respect to $u_1$ and $u_2$ whose
boundary is the union of hypersurfaces $S_0$ and $S_1$ as before. There
exist continuous functions $M^\alpha$ and $N$ such that:
\begin{eqnarray*}
A^\alpha(t,x,u_1)-A^\alpha(t,x,u_2)&=&M^\alpha(t,x,u_1,u_2)(u_1-u_2)\\
B(t,x,u_1)-B(t,x,u_2)&=&N(t,x,u_1,u_2)(u_1-u_2)
\end{eqnarray*}
This sharp form of the mean value theorem is proved for instance in
\cite{hamilton82}. It follows that:
$$A^\alpha(u_1)\d_\alpha(u_1-u_2)+[M^\alpha(u_1,u_2)(\d_\alpha u_2)
+N(u_1,u_2)](u_1-u_2)=0$$
Here the dependence of functions on $t$ and $x$ has not been written out
explicitly. This equation for $u_1-u_2$ can be abbreviated to
$$A^\alpha(u_1)\d_\alpha(u_1-u_2)=Q(u_1-u_2)$$
where $Q$ is a continuous function of $t$ and $x$. It follows that
\begin{eqnarray*}
&&\d_\alpha (\langle e^{-kt}A^\alpha (u_1)(u_1-u_2),u_1-u_2\rangle)\\
&&=e^{-kt}\langle [-kA^0(u_1)+(\d_\alpha A^\alpha)(u_1)+2Q](u_1-u_2),
u_1-u_2\rangle
\end{eqnarray*}
for any constant $k$. Now apply Stokes' theorem to this on the region $G$.
This gives an equation of the form $I_1=I_0+I_G$ where $I_0$, $I_1$ and
$I_G$ are integrals over $S_0$, $S_1$ and $G$ respectively. Because $S_1$
is spacelike, $I_1$ is non-negative. If $k$ is chosen large enough the matrix
$P=kA^0(u_1)-(\d_\alpha A^\alpha)(u_1)-2Q$ is uniformly positive definite on
$G$. This means that there exists a constant $C>0$ such that
$\langle v,P(t,x,u_1) v\rangle \ge C\langle v,v\rangle$ for all $v\in\R^k$ and
$(t,x)$ in $G$. Thus if $u_1-u_2$ is not identically zero on $G$, then the
volume integral $I_G$ can be made negative by an appropriate choice of $k$.
If $u_1$ and $u_2$ have the same initial data then $I_0=0$. In this case
$I_0+I_G$ is negative and we get a contradiction. Thus in fact $u_1-u_2=0$.

This argument shows that locally in a neighbourhood of the initial
hypersurface $S$ the solution of a symmetric hyperbolic system at
a given point is determined by initial data on a compact subset of
$S$. For any point near enough to $S$ is contained in a lens-shaped region.
In this context we would like to introduce the concept of domain of
dependence for a symmetric hyperbolic system. There are problems with
conflicting terminology here, which we will attempt to explain now.
If a given subset $E$ of $S$ is chosen, then the {\it domain of dependence}
$\tilde D(E)$ of $E$ is the set of all points of $\R\times\R^n$
such that the value of a solution of the
symmetric hyperbolic system at that point is determined by the restriction
of the initial data to $E$. Comparing with the definition of the domain
of dependence for the Einstein equations given in Sect. \ref{evole}
we find a situation similar to that seen already for the term \lq spacelike\rq.
The applications of the term to the Einstein equations and to symmetric
hyperbolic systems might seem at first to be completely unrelated, but
in fact there is a close relation. This fact was already alluded to in
Sect. \ref{evole}.
Even for hyperbolic equations there is another ambiguity in terminology.
The definition for symmetric hyperbolic systems conflicts with
terminology used in some places in the literature \cite{courant62}. Often
the set which is called domain of dependence here is referred to as the
domain of determinacy of $E$ while $E$ is said to be a domain of dependence
for a point $p$ if $p$ lies in what we here call the domain of
dependence. In the following we will never use the terminology of
\cite{courant62}. However, we felt it better to warn the reader of the
problems than to pass over them in silence.

The relationship of the general definition of the domain of dependence for
symmetric hyperbolic systems adopted in this article with the domain of
dependence in general relativity will now be discussed in some more detail.
As will be explained later, the harmonically reduced Einstein equations can
be transformed to a symmetric hyperbolic system by introducing the first
derivatives of the metric as new variables. The comparison we wish to make
is between the domain of dependence $\tilde D(S)$
defined by this particular symmetric
hyperbolic system and the domain of dependence $D(S)$ as defined in
Sect.~\ref{evole}. The first important point is that a hypersurface is
spacelike with respect to the symmetric hyperbolic system if and only if it
is spacelike in the usual sense of general relativity, i.e. if the metric
induced on it by the space-time metric is positive definite. Given this fact,
we see that the notion of lens-shaped region used in this section coincides
in the case of this particular symmetric hyperbolic system derived from the
Einstein equations with the concept introduced in Sect. \ref{evole}.

The above uniqueness statement will now be compared for illustrative purposes
with the well-known uniquess properties of solutions of the wave equation
in Minkowski space. Here we see a simplified form of the situation which
has just been presented in the case of the Einstein equations. The wave
equation, being second order, is not symmetric hyperbolic. However, it can
be reduced to a symmetric hyperbolic system by introducing first derivatives
of the unknown as new variables in a suitable way. It then turns out that
once again the concept of a spacelike hypersurface, as introduced for
symmetric hyperbolic systems, coincides with the usual (metric) notion for
hypersurfaces in Minkowski space. It is well known that if initial data for
the wave equation is given on the hypersurface $t=0$ in Minkowski space, and
if $p$ is a point in the region $t>0$, then the solution at $p$ is determined
uniquely by the data within the intersection of the past light cone of $p$
with the hypersurface $t=0$. Once the wave equation has been reduced to
symmetric hyperbolic form this statement can be deduced from the above
uniqueness theorem for symmetric hyperbolic systems. It suffices to note the
simple geometric fact that any point in the region $t>0$ strictly inside the
past light cone of $p$ is contained in a lens-shaped region with the same
property. Thus the solution is uniquely determined inside the past light cone
of $p$ and hence, by continuity, at $p$.

The existence of the domain of dependence has the following consequence.
If two sets $u^0_1$ and $u^0_2$ of initial data on the same initial
hypersurface $S$ coincide on an open subset $U$ of $S$, then the
corresponding domains of dependence $\tilde D_1(U)$ and $\tilde D_2(U)$,
as well as the corresponding solutions $u_1$ and $u_2$ on them, coincide.
In other words, on $\tilde D_1(U)$ the solution $u_1$ is independent of
the extension of $u^0_1$ outside $U$. It follows in particular that there
is no need to impose boundary conditions or fall-off conditions
on $u_1$ 'on the edge of $S$' or some periodicity condition to determine
the solution locally near a point of $S$. This is referred to as the
localization property of symmetric hyperbolic systems since it means that
the theory is
not dependent on knowing how the initial data behave globally in space. In
the following data on $\R^n$ will be considered which are periodic in the
spatial coordinates. This is equivalent to replacing $\R^n$ as domain of
definition of the unknown by the torus $T^n$ obtained by identifying the
Cartesian coordinates in $\R^n$ modulo $2\pi$. In order to prove statements
about solutions of a symmetric hyperbolic system, something must be assumed
about the regularity of the coefficients. This will in particular imply that
$A^0$ is continuous. The continuity of $A^0$ and the compactness of the torus
then show together that $A^0(t,x,u)$ is uniformly positive definite on any
finite closed time interval for any given continuous function $u$ defined on
this interval.

The general strategy of the existence proof we will present here will now be
discussed in more detail. This proof is essentially that given
in \cite{taylor96}. The first detailed proof of an existence theorem for
general quasi-linear symmetric hyperbolic systems was given by Kato
\cite{kato75} using the theory of semigroups. The proof here uses less
sophisticated functional analysis, but the basic pattern of approximations
controlled by energy estimates is the same in both cases.
For simplicity we will restrict to the special case where
$A^0$ is the identity matrix. Once the proof has been carried out in that
case the differences which arise in the general case will be pointed out.
Assuming now that $A^0$ is the identity, the equation can be written in the
form:
$$\d_t u=-A^i(t,x,u)\d_i u-B(t,x,u)$$
This looks superficially like an ordinary differential equation in an
infinite dimensional space of functions of $x$. Unfortunately, this point of
view is not directly helpful in proving local existence. The essential
point is that if $A^i(t,x,u)\d_i$ is thought of as an operator on a space
of functions of finite differentiability, this operator is unbounded. The
strategy is now to replace the unbounded operator by a bounded one. In fact
we even go further and replace the infinite dimensional space by a finite
dimensional one. This is done by the introduction of a mollifier (smoothing
operator) at appropriate places in the equation. The mollifier contains a
parameter $\epsilon$.
The resulting family of equations depending on $\epsilon$
defines the family of approximating problems referred to above in this
particular case. The approximating problems can be solved using the standard
theory of ordinary differential equations. It then remains to show that the
solutions to the approximating problems converge to a solution of the
original problem as $\epsilon$ tends to zero.

A convenient mollifier on the torus can be constructed using the Fourier
transform. We now recall some facts concerning the Fourier transform on
the torus. If $u$ is a continuous complex-valued function on $T^n$ then its
Fourier coefficients are defined by:
$$\hat u(\xi)=\int_{T^n} u(x)e^{-i\langle x,\xi\rangle} dx$$
Here $\xi$ is an element of $Z^n$, i.e. a sequence of $n$ integers. In fact
this formula makes sense for any square integrable function on the torus.
It defines a linear mapping from the space $L^2(T^n)$ of complex
square integrable functions on the torus to the space of all complex-valued
functions on $Z^n$. Its image is the space of square summable functions
$L^2(Z^n)$ and the $L^2$ norm of $u$ is equal to that of $\hat u$ up to a
constant factor. In particular the Fourier transform defines an isomorphism
of $L^2(T^n)$ onto $L^2(Z^n)$. Thus to define an operator on $L^2(T^n)$ it
suffices to define the operator on $L^2(Z^n)$ which corresponds to it under
the Fourier transform. These statements follow from the elementary theory
of Hilbert spaces once it is known that trigonometric polynomials (i.e.
finite linear combinations of functions of the form $e^{i\langle\xi,x\rangle}$)
can be used to approximate all continuous functions on the torus in the
sense of uniform convergence. This is worked out in detail for the case $n=1$
in \cite{rudin87}, Chap. 4. The case of general $n$ is not much
different,
once the approximation property of trigonometric polynomials is known in that
context. This follows, for instance, from the Stone--Weierstrass theorem
(\cite{rudin91}, p. 122).

Let $\phi$ be a smooth real-valued function with compact support on $\R^n$
which is identically one in a neighbourhood of the origin and satisfies $0\le
\phi(\xi)\le 1$ for all $\xi$ and $\phi(-\xi)=\phi(\xi)$. For a positive
real number $\epsilon$ let $\phi_\epsilon(\xi)
=\epsilon^{-n}\phi(\xi/\epsilon)$.
If we identify $Z^n$ with the set of points in $\R^n$ with integer
coordinates it is possible to define functions $\phi_\epsilon$ on $Z^n$
by restriction. The mollifier $J_\epsilon$ is defined by
$$\widehat{J_\epsilon u}=\phi_\epsilon\hat u$$
for any square integrable complex-valued function $u$ on the torus. It is a
bounded self-adjoint linear operator on $L^2(T^n)$. In fact
$\|J_\epsilon\|=1$ for all $\epsilon$. The
symmetry condition imposed on $\phi$ ensures that if $u$ is real,
$J_\epsilon u$ is also real. Note that, for given
$\epsilon$, $\widehat{J_\epsilon u}$ is only non-zero at a finite number of
points, the number of which depends on $\epsilon$. It follows that
$J_\epsilon u$ is a trigonometric polynomial and that the space of
trigonometric polynomials which occurs for fixed $\epsilon$ and all possible
functions $u$ is finite dimensional. As a consequence the image of
$L^2_{{\R}}(T^n)$, the subspace of $L^2(T^n)$ consisting of real-valued
functions, under the operator $J_\epsilon$ is a finite dimensional space
$V_\epsilon$. As $\epsilon$ tends to zero the dimension of $V_\epsilon$ tends
to infinity and this is why these spaces can in a certain sense approximate
the infinite-dimensional space $L^2_{{\R}}(T^n)$.

The original problem will now be approximated by problems involving ordinary
differential equations on the finite-dimensional spaces $V_\epsilon$. If
the initial datum for the original problem is $u^0$, the initial datum
for the approximating problem will be $u^0_\epsilon=J_\epsilon u^0$. The
equation to be solved in the approximating problem is:
\begin{equation}\label{approxu}
\d_t u_\epsilon=-J_\epsilon [A^i(t,x,u_\epsilon)J_\epsilon\d_i u_\epsilon
+B(t,x,u_\epsilon)]=F(u_\epsilon)
\end{equation}
Here the unknown $u_\epsilon$ takes values in the vector space $V_\epsilon$.
In order to apply the standard existence and uniqueness theory for ordinary
differential equations to this it suffices to check that the right hand
side is a smooth function of $u_\epsilon$ provided $A^i$ and $B$ are smooth
functions of their arguments. Let us first check that $F$ is continuous.
If $u_n$ is a sequence of elements of $V_\epsilon$ which converges to $u$
then $J_\epsilon \d_i u_n$ converges to $J_\epsilon\d_i u$. The sequence of
functions $A^i(t,x,u_n)J_\epsilon\d_i u_n+B(t,x,u_n)$ converges uniformly to
$A^i(t,x,u)J_\epsilon\d_i u+B(t,x,u)$.
To see that $F$ is continous it remains to show that if $v_n\to v$ uniformly
$J_\epsilon v_n\to J_\epsilon v$. If $v_n\to v$ uniformly then the convergence
also holds in the $L^2$ norm. This implies that $\hat v_n\to\hat v$ in
$L^2(Z^n)$.  Hence $\phi_\epsilon \hat v_n\to \phi_\epsilon \hat v$ pointwise.
{}From this it can be concluded that $J_\epsilon v_n\to J_\epsilon v$.
This completes the proof of the continuity of $F$. Differentiability can be
shown in a similar way. The main step is to show that if $u$ and $v$ are
elements of $V_\epsilon$ then
$\lim_{s\to 0}s^{-1}[A^i(t,x,u+sv)-A^i(t,x,u)]$ exists (in
the sense of uniform convergence). By smoothness of $A^i$ it does exist and is
equal to $D_3 A(t,x,u)v$, where $D_3$ denotes the derivative with respect to
the third argument. In this way the existence of the first derivative of $F$
can be demonstrated and the explicit expression for the derivative shows that
it is continuous. Higher derivatives can be handled similarly. Hence $F$ is a
smooth function from $V_\epsilon$ to $V_\epsilon$. The standard theory of
ordinary differential equations now gives the existence of a unique solution
of the approximating problem for each positive value of $\epsilon$. Existence
is only guaranteed for a short time $T_\epsilon$ and at this point in the
argument it is not excluded that $T_\epsilon$ could tend to zero as
$\epsilon\to 0$. It will now be seen that this can be ruled out by the
use of energy estimates.

The basic energy estimate involves computing the time derivative of the
energy functional defined by
\begin{equation}\label{energy}
{\cal E}=\int_{T^n} |u_\epsilon|^2 dx
\end{equation}
Since the functions $u_\epsilon$ are smooth and defined on a compact manifold
differentiation under the integral is justified.
\begin{eqnarray*}
d{\cal E}/dt&=&2\int\langle u_\epsilon,\d_t u_\epsilon\rangle\\
            &=&-2\int\langle u_\epsilon, J_\epsilon [A^iJ_\epsilon\d_i
            u_\epsilon+B]\rangle
\end{eqnarray*}
Using the fact that $J_\epsilon$ commutes with the operators $\d_i$ and is
self-adjoint, gives
$$\int\langle u_\epsilon, J_\epsilon [A^iJ_\epsilon\d_i u_\epsilon]\rangle=
\int\langle J_\epsilon u_\epsilon, A^i\d_iJ_\epsilon u_\epsilon\rangle$$
Using Stokes theorem and the symmetry of $A^i$ gives
$$\int\langle J_\epsilon u_\epsilon, A^i\d_iJ_\epsilon u_\epsilon\rangle
=-\int\langle \d_iA^iJ_\epsilon u_\epsilon, J_\epsilon u_\epsilon\rangle
-\int\langle A^iJ_\epsilon\d_i u_\epsilon, J_\epsilon u_\epsilon\rangle$$
It follows that
$$\int\langle u_\epsilon, J_\epsilon [A^iJ_\epsilon \d_i u_\epsilon]\rangle=
-\frac{1}{2}\int\langle \d_i A^i J_\epsilon u_\epsilon, J_\epsilon
u_\epsilon\rangle$$
Substituting this in equation (\ref{energy}) gives:
$$\d_t{\cal E}=\int\langle (\d_i A^i)J_\epsilon u_\epsilon-2B
,J_\epsilon u_\epsilon\rangle$$
Now
$$\|(\d_i A^i)J_\epsilon u_\epsilon-2B \|_{L^2}
\le\|\d_i A^i\|_{L^\infty}\|u_\epsilon\|_{L^2}+2\|B\|_{L^2}$$
Hence it follows by the Cauchy-Schwarz inequality that
$$\d_t{\cal E}\le\|\d_i A^i\|_{L^\infty}{\cal E}+2\|B\|_{L^2}{\cal E}^{1/2}$$
This is the fundamental energy estimate. Note that this computation is
closely related to that used to prove uniqueness for solutions of symmetric
hyperbolic systems above.

The existence proof also requires higher order energy estimates. To
obtain these, first differentiate the equation one or more times with respect
to the spatial coordinates.
Higher order energy functionals are defined by
$${\cal E}_s=\sum_{|\alpha|\le s}\int_{T^n} (D^\alpha u_\epsilon)^2 dx$$
The square root of the energy functional ${\cal E}_s$ is a norm which
defines the Sobolev space $H^s$. It is because of the energy estimates that
Sobolev spaces play such an important role in the theory of hyperbolic
equations. Differentiating the equation for (\ref{approxu} gives
\begin{eqnarray}\label{highere}
\d_t(D^\alpha u_\epsilon)&=&-J_\epsilon [A^i(t,x,u_\epsilon)J_\epsilon\d_i
D^\alpha u_\epsilon]-J_\epsilon[D^\alpha(A^i(t,x,u_\epsilon)J_\epsilon\d_i u)
\nonumber\\
&&-A^i(t,x,u_\epsilon)J_\epsilon\d_i D^\alpha u_\epsilon]
-J_\epsilon D^\alpha(B(t,x,u_\epsilon))
\end{eqnarray}
Differentiating the expression for ${\cal E}_s$ with respect to time causes the
quantity $\d_t(D^\alpha u_\epsilon)$ to appear. This can be substituted for
using the equation (\ref{highere}). Stokes theorem can be used to eliminate
the highest
order derivatives, as in the basic energy estimate. To get an inequality
for ${\cal E}_s$ similar to that derived above for $\cal E$ it remains to
obtain $L^2$ estimates for the quantities
$$D^\alpha(A^i(t,x,u_\epsilon)J_\epsilon\d_i u_\epsilon)-A^i(t,x,u_\epsilon)
J_\epsilon\d_i D^\alpha u_\epsilon$$
and
$$D^\alpha B(t,x,u_\epsilon)$$
This can be done by means of the Moser inequalities, which will be stated
without proof. (For the proofs see \cite{taylor96}.)

The first Moser estimate says that there exists a constant $C>0$ such that
for all bounded functions $f,g$ on $T^3$ belonging to the Sobolev space
$H^s$ the following inequality holds:
$$\|D^{\alpha}(fg)\|_{L^2}\le C(\|f\|_{L^\infty}\|D^s g\|_{L^2}+
\|D^s f\|_{L^2}\|g\|_{L^\infty})$$
Here $s=|\alpha|$ and for a given norm $\|D^s g\|$ is shorthand for
the maximum of $\|D^\alpha g\|$ over all multiindices $\alpha$ with
$|\alpha|=s$. The second estimate says that there exists a constant $C>0$
such that for all bounded functions $f,g$ on $T^3$ such that the first
derivatives of $g$ are bounded, $f$ is in $H^s$ and $g$ is in $H^{s-1}$
the following inequality holds:
$$\|D^\alpha(fg)-fD^\alpha g\|_{L^2}\le C(\|D^s f\|_{L^2}
\|g\|_{L^\infty}+\|Df\|_{L^\infty}\|D^{s-1} g\|_{L^2})$$
The third estimate concerns composition with nonlinear functions. Let
$F$ be a smooth function defined on an open subset of $\R^k$ and taking
values in $\R^k$. There exists a constant $C>0$ such that for all functions
$f$ on $T^3$ taking values in a fixed compact subset $K$ of $U$ and belonging
to the Sobolev space $H^s$ and any multiindex $\alpha$ with $s=|\alpha|\ge 1$
the following inequality holds:
$$\|D^{\alpha}(F(f))\|_{L^2}\le C\|Df\|_{L^\infty}^{s-1}\|D^s f\|_{L^2}$$

The result of using the Moser inequalities in the expression for
$d{\cal E}_s/dt$ is an estimate of the form:
$$\d_t{\cal E}_s\le C{\cal E}_s$$
where the constant $C$ depends on a compact set $K$ in which $u_\epsilon$
takes values, as above, and the $L^\infty$ norm of the first derivatives of
$u_\epsilon$. This can be estimated by a $C^1$ function of the
$C^1$ norm of $u_\epsilon$. By the Sobolev embedding theorem (see e.g.
\cite{kreiss89}), there is a
constant $C$ such that $\|u_\epsilon\|_{C^1}\le C\|u_\epsilon\|_{H^s}$ for
any $s>n/2+1$. Thus we obtain a differential inequality of the form
$$\d_t{\cal E}_s\le f({\cal E}_s)$$
for a $C^1$ function $f$. It follows that the quantity ${\cal E}_s(t)$
satisfies the inequality
$${\cal E}_s(t)\le z(t)$$
where $z(t)$ is the solution of the differential equation $dz/dt=f(z)$ with
initial value ${\cal E}(0)$ at $t=0$. (A discussion of comparison arguments of
this type can be found in \cite{hartman82}, Chap. 3.) Since the
function
$z$ remains finite on some time interval $[0,T]$, the same must be true for
${\cal E}_s$ for any $s>n/2+1$ and the solutions $u_\epsilon$ exist on the
common interval $[0,T]$ for all $\epsilon$. Morover $\|u_\epsilon(t)\|_{H^s}$
is bounded independently of $\epsilon$ for all $t\in [0,T]$. Putting this in
the equation shows that $\|\d_t u_\epsilon (t)\|_{H^{s-1}}$ is also bounded.

The proof of existence of a solution with the given initial datum now
follows by some functional analysis. The main point is that the boundedness
of a sequence of functions in some norm often implies the existence of a
sequence which is convergent in another topology. The reader who does not
possess and does not wish to acquire a knowledge of functional analysis may
wish to skip the rest of this paragraph. The space defined by the $L^\infty$
norm of $\|u(t)\|_{H^s}$ is a Banach space and is also the dual of a
separable Banach space. The family $u_\epsilon$ is bounded in this space and
hence, by the Banach--Alaoglu theorem (see \cite{rudin91}, pp. 68-70) there
is a sequence
$\epsilon_k$ tending to zero such that $u_{\epsilon_k}(t)$ converges
uniformly in the weak topology of $H^s(T^n)$ to some function $u$ which is
continuous in $t$ with values in $H^s(T^n)$ with repect to the weak topology.
The same argument can be applied to the family $\d_t u_\epsilon$.
As a result, after possibly passing to a subsequence,
$\d_t u_{\epsilon_k}(t)$ converges uniformly in the weak topology of
$H^{s-1}(T^n)$. On the other hand, since $T^n$ is compact, the inclusion of
$H^s(T^n)$ in $H^{s-1}(T^n)$ is compact. (This is Rellich's Lemma. See e.g.
\cite{griffiths78}, p. 88) Thus, by the vector-valued Ascoli
theorem \cite{dieudonne69} it can be assumed, once again passing to a
subsequence if
necessary, that $u_{\epsilon_k}(t)$ converges uniformly in the norm topology
of $H^{s-1}(T^n)$. If $s>n/2+2$ then the Sobolev embedding theorem can be
used to deduce that it converges uniformly in the norm of $C^1(T^n)$. Once
this is known it follows directly that the expression on the right hand side
of the approximate equation converges uniformly to that on the right hand
side of the exact equation. On the other hand $\d_t u_k$ converges to $\d_t u$
in the sense of distributions and so $u$ satifies the original equation in
the sense of distributions. The equation then implies that the solution is
$C^1$ and is a classical solution. It also has the desired initial datum.

\medskip
\noindent
{\bf Remarks}

\noindent
1. The existence and uniqueness theorem whose proof has just been sketched
says that given an initial datum $u_0$ in $H^s(T^n)$ with $s>n/2+2$ there is
a solution for this datum which is a bounded measurable function $u(t)$
with values in $H^s(T^n)$ and is such that $\d_t u$ is a bounded measurable
function with values in  $H^{s-1}(T^n)$. With some further work it can be shown
that $n/2+2$ can be replaced by $n/2+1$ and that the bounded measurable
functions are in fact continuous with values in the given space
\cite{taylor96}.
\hfil\break\noindent
2. The time of existence of a solution which is continuous with values in
$H^s(T^n)$ depends only on a bound for the norm of the initial data in
$H^s(T^n)$.
\hfil\break\noindent
3. The analogues of the higher energy estimates whose derivation was
sketched here for the solutions of the approximating problems hold for the
solution itself \cite{taylor96}. This implies that as long as the $C^1$ norm
of $u(t)$ remains bounded the $H^s$ norm also remains bounded. This leads
to the following continuation criterion. If $u(t)$ is a solution on an
interval $[0,T)$ and if $\|u(t)\|_{C^1}$ is bounded independently of $t$
by a constant $C>0$ then the solution exists on a longer time
interval. For there is some $T_1$ such that a solution corresponding to any
data $u_0$ with $\|u_0\|_{H^s}\le C$ exists on $[0,T_1)$. Considering
setting data at times just preceding $T$, and using uniqueness to show that
the solutions obtained fit together to give a single solution shows the
existence of a solution of the original problem on the interval $[0,T+T_1)$.
\hfil\break\noindent
4. The previous remark implies the existence of $C^\infty$ solutions
corresponding to $C^\infty$ data. For suppose that a datum $u_0$ of class
$C^\infty$ is given. Then $u_0$ belongs to $H^s(T^n)$ for each $s$. Thus
there is a corresponding solution on an interval $[0,T_s)$ which is
continuous with values in $H^s$. Assume that $T_s$ has been chosen as large
as possible. It is not possible that $T_s\to 0$ as $s\to\infty$. For the
boundedness of the solution in $H^s$, $s>n/2+1$ implies its boundedness in
$C^1$ and this in turn implies that the solution can be extended to a longer
time interval. Thus in fact $T_s$ does not depend on $s$. Using the equation
then shows that the solution is $C^\infty$.
\hfil\break\noindent
5. Analogous statements to all of the above
can be obtained for the case where $A^0$ is not the
identity with similar techniques. The essential point is to modify the
expressions for the energy functionals using $A^0$. For instance the basic
energy for the approximating problems is given by
${\cal E}=\int\langle A^0u_\epsilon,u_\epsilon\rangle$.
\hfil\break\noindent
6. Our treatment here differs slightly from that of Taylor \cite{taylor96}
by the use of a reduction to finite-dimensional Banach spaces. Which
variant is used is a matter of taste.

\vskip 10pt\noindent
The statement made in the second remark is a part of what is referred to as
Cauchy stability. This says that when the initial data is varied the
corresponding solution changes in a way which depends continuously on the
data. This assertion applies to solutions defined on a fixed common time
interval $[0,T]$. The above remark shows that a common time interval of this
kind can be found for data which are close to a given datum in $H^s(T^n)$.
A related statement is that if the coefficients of the equation and the
data depend smoothly on a parameter, then for a compact parameter interval
the corresponding solutions for different parameter values exist on a
common time interval $[0,T]$ and their restrictions to this time interval
depend smoothly on the parameter.

\subsection{Symmetric Hyperbolic Systems on Manifolds}\label{symhypman}

In the last section it was shown how existence and uniqueness statements
can be obtained for solutions of symmetric hyperbolic systems with data
on a torus. It was also indicated that these imply results for
more general settings by means of the domain of dependence. In this
section some more details concerning these points will be provided. First
it is necessary to define what is meant by a symmetric hyperbolic system on
a manifold $M$. (In the context of the last section $M=T^3\times I$.) In
general this will be an equation for sections of a fibre bundle $E$ over
$M$ which, for simplicity, we will take to be a vector bundle. The
easiest definition is that a symmetric hyperbolic system is an equation
of the form $P(u)=0$, where $P$ is a nonlinear differential operator on
$M$ which in any local coordinate system satisfies the definition given
previously. For the general invariant definition of a differential
operator on sections of a bundle the reader is referred to \cite{palais68}.
The object defined in local coordinates by $\sigma(\xi)=A^\alpha\xi_\alpha$
is called the principal symbol of the differential operator
(cf. Sect. \ref{principal}). Here $A^\alpha$
is supposed to be evaluated on a given section $u$. The principal symbol is
defined invariantly by the differential operator. Let $L(V)$ be the bundle
of linear mappings from $V$ to itself. Then the invariantly defined principal
symbol is a section of the pullback of the bundle $L(V)$ to the cotangent
bundle $T^*M$. Given a local trivialization of $V$ it can be identified
locally with a $k\times k$ matrix-valued function on the cotangent bundle.
Here $k$ is the dimension of the fibre of $V$, i.e. the number of unknowns
in the local coordinate representation.

In order to define the condition of symmetry in the definition of symmetric
hyperbolic systems in an abstract context, it is necessary to introduce
a Riemannian metric on the bundle $V$. This metric is not visible in the
local coordinate representation, since in that context this role is played
by the flat metric whose components in the given coordinates are given by
the Kronecker delta. The symmetry condition is expressed in terms of
the principal symbol and the chosen inner product as:
$$\langle\sigma(\xi) v,w\rangle=\langle v,\sigma(\xi) w\rangle$$
for all sections $v,w$ of $V$ and all covectors $\xi$. The positivity
of $A^0$ is replaced by the condition that for some covector $\xi$ the
quadratic form associated to $\sigma(\xi)$ via the metric on $V$ (by
lowering the index) is positive definite. A hypersurface such that all
its non-vanishing conormal vectors have this property is called spacelike.

Suppose now we have a symmetric hyperbolic system on a manifold $M$, a
submanifold $S$ of $M$ and initial data such that $S$ is spacelike. The
aim is to show the existence of a solution on an open neighbourhood $U$ of
$S$ in $M$ with the prescribed data and that this is the unique solution
on $U$ with this property. Choose a covering of $S$ by open sets
$U_\alpha$ with the property that for each $\alpha$ the closure of this
set is contained
in a chart domain over which the bundle $V$ can be trivialized globally.
The problem of solving the equations on some open region in $M$ with
data the restriction of the original initial data to $U_\alpha$ can be
solved by means of the results already discussed. For the local
coordinates can be used to identify $U_\alpha$ with an open subset
$U'_\alpha$ of a torus $T^n$ and the initial data can be extended smoothly
to the whole torus. The equation itself can also be transferred and
extended. Corresponding to the extended data there exists a solution of
the equations on $T^n\times I$ for some interval $I$. The coordinates can
be used to transfer (a restriction of) the solution to an open subset
$W_\alpha$ of $M$ which is an open neighbourhood of $U_\alpha$. The
domain of dependence can then be used to show that there is an open
neighbourhood where the solutions on the different $W_\alpha$ agree on
the intersections. More specifically, on an intersection of this kind
the two in principle different solutions can be expressed in terms of
the same local coordinates and trivialization. By construction the
coordinate representation of the data is the same in both cases. Thus
the local uniqueness theorem for symmetric hyperbolic systems implies
that both solutions are equal in a neighbourhood of the initial hypersurface.
Thus the solutions on different local patches fit together to define the
desired solution. Moreover the domain of dependence argument shows that it
is unique on its domain of definition.

\subsection{Other Notions of Hyperbolicity}\label{otherhyp}

There are two rather different aspects of the concept of hyperbolicity.
The one is an intuitive idea of what good properties a system of equations
should have in order to qualify for the name hyperbolic. The archetypal
hyperbolic equation is the wave equation and so the desired properties
are generalizations of properties of solutions of the wave equation.
The first property is that the system should have a well-posed initial
value problem. This means that there should exist solutions corresponding
to appropriate initial data, that these should be uniquely determined by
the data, and that they should depend continuously on the data in a suitable
sense. The second property is the existence of a finite domain of
dependence. This means that the value of solutions at a given point close
to the initial hypersurface should depend only on data on a compact subset
of the initial hypersurface.

The first aspect of hyperbolicity which has just been presented is a list
of wishes. The second aspect is concerned with the question, how these
wishes can be fulfilled. The idea is to develop criteria for hyperbolicity.
For equations satisfying one of these criteria, the desired properties can
be proved once and for all. Then all the user who wants to check the
hyperbolicity of a given system has to do is to check the criteria, which
are generally more or less algebraic in nature. An example is symmetric
hyperbolicity, which we have already discussed in detail.

Symmetric hyperbolicity can be applied to very many problems in general
relativity, but there are also other notions of hyperbolicity which have
been applied and cases where no proof using symmetric hyperbolic systems is
available up to now. We will give an overview of the situation, without
going into details. The concepts which will be discussed are strict
hyperbolicity, strong hyperbolicity and Leray hyperbolicity. This
multiplicity of definitions is a consequence of the fact that there is no one
ideal criterion which covers all cases.

There is unfortunately to our knowledge no fully general treatment of these
matters in the mathematical literature. A general theory of this kind
cannot be given in the context of this article, and we will refrain from
stating any general theorems in this section. We will simply indicate
some of the relations between the different types of hyperbolicity
and give some references where more details can be found.

The discussion will use some ideas from the theory of pseudodifferential
operators. For introductions to this theory see \cite{taylor81},
\cite{taylor91}, \cite{alinhac91}. One important tool
is an operator $\Lambda$ which defines an isomorphism from functions in
the Sobolev space $H^s$ to functions in the Sobolev space $H^{s-1}$.
Powers of this operator can be used to associate functions with a given
degree of differentiability (in the sense of Sobolev spaces) with
functions with a different degree of differentiability. The operator
$\Lambda$ is non-local. On the torus it could be chosen to be the
operator $(1-\Delta)^{1/2}$, defined in an appropriate way.

The concept of strict hyperbolicity is easily defined. Suppose that we
have a system of $N$ equations of order $m$. Suppose there is a covector
$\tau$ such that for any covector $\xi$ not proportional to $\tau$
the equation $\det\sigma(x,\tau+\lambda\xi)$ has $Nm$ distinct real
solutions
$\lambda$. In this case we say that all characteristics are real and
distinct and that the system is strictly hyperbolic. The weakness of
this criterion is due to the fact that it is so often not satisfied for
systems of physical interest. For example, the system for two functions
$u$ and $v$ given by the wave equation for each of them is not strictly
hyperbolic. The wave equation itself is strictly hyperbolic but simply
writing it twice side by side leads to characteristics which are no longer
distinct, so that strict hyperbolicity is lost.

Given a strictly hyperbolic system of order $m$ with unknown $u$, let
$u_i=\Lambda^{-i}\d_t^i u$ for $i=1,2,\ldots,m-1$. Then the following
system of equations is obtained:
\begin{eqnarray*}
\d_t u_i=\Lambda u_{i+1};\ \ \ 0\le i\le m-1       \\
\d_t u_{m-1}=\Lambda^{-m+1}(\d_t^m u)
\end{eqnarray*}
where it is understood that $\d_t^m u$ should be expressed in terms
of $u_0,\ldots,u_{m-1}$ by means of the original system. This system is
first order in a sense which is hopefully obvious intuitively and which
can be made precise using the theory of pseudodifferential operators.
This is not a system of differential equations, due to the nonlocality
of $\Lambda$, but rather a system of pseudodifferential equations.
Suppose we write it as $\d_t v+A^i(v)\d_i v+B(v)=0$, where $v$ is an
abbreviation for $u_0,\ldots,u_{m-1}$. The idea is now to find a
pseudodifferential operator $A^0(v)$ of order zero depending on the
unknown so that $A^0(v)A^i(v)$ is symmetric for all $v$ and $i$ and
$A^0$ has suitable positivity properties. If an $A^0$ of this kind
can be found, then multiplying the equation by it gives something
which looks like a pseudodifferential analogue of a symmetric hyperbolic
system. This symmetrization of the reduced equation can in fact be
carried out in the case of strictly hyperbolic systems and the
resulting first order pseudodifferential equation be treated by a
generalization of the methods used for symmetric hyperbolic systems
\cite{taylor91}, \cite{taylor96}. This gives an existence theorem, but
does not directly give information about the domain of dependence. This
information can be obtained afterwards by a different method \cite{taylor81}.

The algebra involved in symmetrizing the reduction of a strictly hyperbolic
system has not been detailed. Instead we will present some general
information about symmetrization of first order systems using
pseudodifferential operators. Consider then the equation
$$
\d_t u+A^i(u)\d_i u=0
$$
If there exists a function $A^0(u)$ such that $A^0$ is positive definite
and $A^0A^i$ is symmetric for each $i$ then multiplying the equation by
$A^0(u)$ gives the equation:
$$
A^0(u)\d_t u+A^0(u)A^i(u)\d_i u=0
$$
which is symmetric hyperbolic. This can be generalized as follows. Each
differential operator has a symbol which is polynomial in $\xi$. In the
theory of pseudodifferential operators, operators are associated to symbols
which no longer have a polynomial dependence on $\xi$. Within this theory
the general statement (which must of course be limited in order to become
strictly true) is that any algebra which can be done with symbols can be
mimicked by operators. In terms of symbols the problem of symmetrization
which has been stated above can be formulated as follows. Given symbols
$\xi_i a^i(x,u)$, which are matrix-valued functions, find a positive definite
symmetric real matrix-valued function $a^0(x,u)$ such that
$a^0(x,u)\xi_ia^i(x,u)$ is symmetric. When the problem has been formulated in
this way a generalization becomes obvious. Why not allow $a^0$ to depend on
$\xi$? In the context of
pseudodifferential operators this can be done. In order for this to be
useful $a^0$ must be of order zero, which means that it must be bounded
in $\xi$ for each fixed $(x,u)$ and that its derivatives with respect to $\xi$
and $x$ satisfy similar conditions which will not be given here. In fact
it is enough to achieve the symmetrization of the symbol for $\xi$ of unit
length, since the $\xi$ dependence of the procedure is essentially
homogeneous in $\xi$ anyway. It is also important that it be possible to
choose $a^0$ in such a way that it depends smoothly on $\xi$, since the
theory of pseudodifferential operators requires sufficiently smooth symbols.
To prevent confusion in the notation, $a^0$ will from now on be denoted by
$r$ while we write $a(x,\xi,u)=a^i(x,u)\xi_i$.

A criterion which ensures that a first order system of PDE can be symmetrized
in the generalized sense just discussed is that of strong hyperbolicity
\cite{kreiss89}. It is supposed that all characteristics are real, that they
are of constant multiplicity, and that the symbol is everywhere
diagonalizable. Furthermore, it is assumed that this diagonalization can be
carried out in a way which depends smoothly on $x$, $\xi$ and $u$. This
means that there is a symbol $b(x,\xi,u)$ which satifies
$$
b^{-1}(x,\xi,u)a(x,\xi,u)b(x,\xi,u)=d(x,\xi,u)
$$
where $d$ is diagonal. Let $r(x,\xi,u)=[b(x,\xi,u)b^T(x,\xi,u)]^{-1}$ where
the superscript $T$ denotes the transpose. Then $r$ has the desired properties.
For $r=b^{-1}(b^{-1})^T$ is symmetric and positive definite while
$ra=(b^{-1})^T d b^{-1}$ is also symmetric. The one part of this criterion
which is not purely algebraic is the smoothness condition. It is, however,
often not hard to check once it is known how to verify the other conditions.

Another related notion of hyperbolicity is Leray hyperbolicity \cite{leray53}.
It allows operators with principal parts of different orders in the same
system. We will discuss this in terms of using powers of the operator
$\Lambda$ to adjust the orders of the equations. This is somewhat different
from the approach of Leray but is closely related to it. Suppose we have
a system of PDE with unknown $u$. Let us split the vector $u$ into a
sequence of vectors $u_1,\ldots,u_L$. Then the differential equation
which we may write schematically as $P(u)=0$ can be rewritten as
$P_j(u_i)=0$, where $i$ and $j$ run from $1$ to $L$.
Let $v_i=\Lambda^{s(i)} u_i$ multiply the equations $P_j(u_i)=0$ with
$\Lambda^{t(j)}$. The result is a system of equations of the form
$Q_j(v_i)=0$, where $Q_j(v_i)=\Lambda^{t(j)} P_j(\Lambda^{-s(i)}v_i)$.
The order of $Q_j$ in its dependence on $v_i$ is the order of $P_j$ in its
dependence on $u_i$ plus $t(j)-s(i)$. Now we would
like to choose $t(j)$ and $s(i)$ in such a way that all operators $Q_i$
have the same order (say one) in their dependence on the corresponding
variable $v_i$ and lower order (say zero) in their dependence on $v_j$
for $i\ne j$. Adding the same amount to all indices
simultaneously is irrelevant. The indices are only determined up to a
common additive constant. The system is Leray hyperbolic if this can be
achieved by suitable choices of the decomposition and the indices, if the
operators $P_i$ corresponding to the given decomposition, considered with
respect to their dependence on $u_i$, are strictly hyperbolic, and if the
characteristics of these strictly hyperbolic operators satisfy a certain
condition. (We say more on this condition later.) It is now plausible that
this decomposition can be combined with the symmetrization of strictly
hyperbolic systems to obtain a system of first order pseudodifferential
equations which admit a pseudodifferential symmetrization. We do however
stress that as far as we know the only place where the existence proof for
Leray hyperbolic systems is written down in the literature
is in the original lecture notes of
Leray \cite{leray53} and that the details of the
plausibility argument presented here have not been worked out. It has
the advantage that it gives an intuitive interpretation of the meaning
of the indices $s(i)$ and $t(j)$. All the components of $v$ in the solution
will have the same degree of differentiability. If this is $H^k$ then the
variables $u_i$ will have differentiability $H^{k-s(i)}$.

We now comment on the condition on the characteristics mentioned in the
discussion of Leray hyperbolicity. For a strictly hyperbolic system there
is a notion of spacelike covectors analogous to that for symmetric
hyperbolic systems. The position of these spacelike covectors is closely
related to the position of the characteristics. The required condition
is that there should be vectors which are simultaneously spacelike for
the $L$ strictly hyperbolic systems occurring in the definition.

These general ideas concerning Leray hyperbolic systems will now be
illustrated by the example of the Einstein-dust system. This example was
first treated by Choquet-Bruhat \cite{choquet58}. A discussion of this and
other examples can also be found in the books of Lichnerowicz
\cite{lichnerowicz67}, \cite{lichnerowicz94}. We will have more to say
about this system in Sects. \ref{euler} and \ref{matmod}.
The Einstein-dust equations are
\begin{eqnarray*}
&&G_{\alpha\beta}=8\pi\rho\,U_\alpha U_\beta    \\
&&\nabla_\alpha(\rho\,U^\alpha)=0               \\
&&U^\alpha\nabla_\alpha U^\beta=0
\end{eqnarray*}
The problem with these equations from the point of view of symmetric
hyperbolic systems is that while the equation for the evolution of $\rho$
contains derivatives of $U^\alpha$ the equation for the evolution of
$U^\alpha$ does not contain $\rho$ This could be got around if the
derivatives of $U^\alpha$ in the evolution equation for $\rho$ were
considered as lower order terms. This is only possible if the first
derivatives of $U^\alpha$ are considered on the same footing as $\rho$.
The adjustment of orders involved in the Leray theory, combined with an
extra device, allows this to be achieved. In order that everything be
consistent we expect the order of differentiability of the metric to
be one greater than that of $U^\alpha$. For the Christoffel symbols
occur in the evolution equations for $U^\alpha$. Combining these two
things means that the density should be two times less differentiable
than the metric. This creates problems with the Einstein equations.
For these are essentially (and the harmonically reduced equations are
precisely) non-linear wave equations for the metric. The solution of
a system of this kind is only one degree more differentiable than the
right hand side. This does not fit, since the density occurs on the right
hand side. The extra device consists is differentiating the Einstein
equations once more in the direction $U^\alpha$ and then substituting
the evolution equation for $\rho$ into the result. This gives the equation:
$$
U^\gamma\nabla_\gamma G_{\alpha\beta}=-8\pi\rho
U_\alpha U_\beta\nabla_\gamma U^\gamma
$$
Note that the right hand side of this contains no derivatives of $\rho$ and
so is not worse than the right hand side of the undifferentiated equation.
On the other hand, the left hand side is, in harmonic coordinates, a third
order hyperbolic equation in $g_{\alpha\beta}$ and the solution of an
equation of this type has (by the Leray theory) two more degrees of
differentiability than the right hand side.

After this intuitive discussion of the Einstein-dust system, let us show
how it is related to the choice of indices needed to make the Leray
hyperbolicity of the system manifest. The first thing which needs to be
done is to decide which equations should be considered as the evolution
equations for which variables. This has already been done in the above
discussion. The relative orders of differentiability we have discussed
suggest that we choose $s(g)=1$, $s(U)=2$ and $s(\rho)=3$. This
equalizes the expected differentiability of the different variables.
In the new variables the evolution equations for $g$, $U$ and $\rho$ are
of order four, three and four respectively. To equalize the orders we can
choose $t(g)=1$, $t(U)=2$ and $t(\rho)=1$. The blocks into which the
system must be split are not just three but fifteen, one for each
component. The indices $s(i)$ and $t(j)$ are here chosen the same for
each component of one geometrical object. (Here we have ignored the
difficulty that because of the normalization condition
$U_\alpha U^\alpha=-1$ the variables $U^\alpha$
are not all independent.)


\section{Reductions}\label{reductions}

In Sect. \ref{basics} we saw an example of a hyperbolic reduction for
Einstein's field
equations which is, at least in the vacuum case, sufficient to obtain
local existence
results and to demonstrate that the local evolution is dominated by the
light cone
structure and the associated concept of the domain of dependence. Thus,
  besides the
existence problem, it settles the conceptual question whether the
evolution process
determined by Einstein's equations is consistent with the basic tenets of
the theory.

Nevertheless, there are various reasons for considering other types of
reductions.
Different matter fields may require different treatments, various physical
or geometrical considerations may require reductions satisfying certain
side conditions, the desire to control the gauge for an arbitrarily long time
may motivate the search for new gauge conditions. In particular, in recent
years various systems of hyperbolic equations deduced from Einstein's
equations have been put forward with the aim of providing \lq good\rq\ systems
for numerical calculations (cf. \cite{abrahams et
al.}, \cite{anderson:york}, \cite{bona:masso},
\cite{choquet-bruhat:ruggeri},
\cite{friedrich:hypred}, \cite{fritelli:reula}, \cite{iriondo et al.}).
Since it is difficult to judge the relative efficiency of such systems by
a few abstract arguments, detailed numerical calculations are needed to
test them and it still remains to be seen which of these systems will
serve the intended purpose best.

A general discussion of the problem of finding reductions which are
useful in numerical or analytical studies should also include systems
combining hyperbolic with elliptic equations (cf. \cite{christodoulou93},
\cite{rendall:II}, \cite{choquet96} for analytic treatments of such
situations) or with systems of still other types. For simplicity we
will restrict ourselves to purely hyperbolic reductions. But even then a
general discussion does not yet appear feasible. Ideally, one would like to
exhibit a kind of \lq hyperbolic skeleton\rq\ of the Einstein equations and
a complete characterization of the freedom to fix the gauge from which
all hyperbolic reductions should be derivable. Instead, there are
at present various different methods available
which have been invented to serve specific needs.

Therefore, we will present some of these methods without striving for
completeness. There are various different boundary value problems of
interest for Einstein's equations and tomorrow a new question may lead
to a new solution of the reduction problem. Our aim is rather to
illustrate the enormous richness of possibilities to adapt the equations
to various geometrical and physical situations, and to comment on some of the
new features which may be observed. Apart from the
illustrative purpose it should not be forgotten, though, that each of
the reductions outlined here implies, if worked out in detail, the local
existence of solutions satisfying certain side conditions.

We begin by recalling in general terms the steps to be considered in a
reduction procedure.
It should be noted that these steps are not independent of each other,
and that the same is true of the following considerations.

As a first step we need to {\it choose a representation of the field
equations}. Above we used the standard representation in which Einstein's
equations are written as a system of second order for the metric
coefficients $g_{\mu \nu}$ and we also considered the ADM equations with
the first and second fundamental forms of the time slicing as the basic
variables. Later we shall also employ representations involving
equations of third order in the metric field, in which the metric is
expressed in terms of an orthonormal frame field.

Contrary to what has been claimed by too ambitious or suggested by
ambiguous formulations, there is no way to get hyperbolic evolution
equations without fixing a gauge. Hyperbolicity implies uniqueness in
the PDE sense (in contrast to the notion of geometric uniqueness used
for the Einstein equations, cf. Sect. \ref{unic}). It is
necessary to make
{\it a choice of precisely four gauge source functions}.

Although in the proof of an existence result one has to fix a coordinate
system, it is of interest to note that there are choices of gauge
conditions which render the reduced equations hyperbolic for any fixed
choice of coordinates. In Sect. \ref{basics} we considered the functions
$F^{\nu}(x^{\mu})$ as gauge source functions determining the coordinates.
Since these functions can be chosen arbitrarily and since the principal
part of the reduced equations (\ref{2redequ}) will not be changed if they
depend on the metric coefficients, they can be assumed to be of the
form $F^{\nu} = F^{\nu}(x^{\mu}, g_{\lambda \rho})$.

The following choice is particularly interesting. Let an affine connection
$\bar{\Gamma}_{\nu}\,^{\mu}\,_{\lambda}$ be given on the manifold
$M = {\R} \times S$ on which we want to construct our solution.
Since the difference $\Gamma_{\nu}\,^{\mu}\,_{\lambda} -
\bar{\Gamma}_{\nu}\,^{\mu}\,_{\lambda}$, where the
first term denotes the connection of the prospective solution
$g_{\mu \nu}$, defines a tensor field, the requirement that the equation
$\Gamma^{\mu} = g^{\nu\lambda}\,\bar{\Gamma}_{\nu}\,^{\mu}\,_{\lambda}$ be
satisfied, has an invariant meaning. This suggests a way to impose a
gauge condition which removes the freedom to perform diffeomorphisms
while leaving all the freedom to perform coordinate transformations. In
particular, if we assume the connection
$\bar{\Gamma}_{\nu}\,^{\mu}\,_{\lambda}$ to be the Levi--Civita connection
of a metric
$\bar{g}_{\mu \nu}$ on $M$, the condition
$F^{\mu} = g^{\nu\lambda}\,\bar{\Gamma}_{\nu}\,^{\mu}\,_{\lambda}$ in
equation (\ref{2redequ}) corresponds to the requirement that the identity
map of $M$ onto itself is a harmonic map from $(M,g)$ to $(\bar M,\bar g)$
(cf. also \cite{friedrich:global}, \cite{friedrich:hypred}).

Again, we do not have a good overview of all the possibilities to impose
gauge conditions. There exist conditions which work well with quite
different representations of the field equations. Examples are given by
the choice of gauge source functions $F^{\nu}(x^{\mu})$ considered above,
which work with suitably chosen gauge conditions for the frame field
also in the frame formalism (\cite{friedrich:1985}), or by the
gauge, considered also below, in which the shift $\beta^a$ and the
function $q = \log (\alpha\,h^{-\sigma})$, with
$h = \det (h_{ab})$ and $\sigma = {\rm const.} > 0$, are  prescribed as gauge
source functions (cf. \cite{anderson:york}, \cite{choquet-bruhat:ruggeri},
\cite{friedrich:hypred} for reductions based on $\sigma = \frac{1}{2}$ and
different representations of the Einstein equations). There are other
gauge conditions which only work for specific representations like some
of the ones we shall consider in the context of the frame formalism. For
some gauge conditions, like the ones employing the gauge source functions
$F^{\nu}(x^{\mu})$, the universal applicability has been shown (cf. our
argument in Sect. \ref{basics}) or is easy to see. For others, like the
ones using the gauge source function $q$ above, it has apparently never
been shown.

The gauge problem does not admit a \lq universal solution\rq\ which works in
all possible situations of interest. In fact, choosing the gauge is
related to some of the
most complicated questions of constructing general solutions to
Einstein's equations.
Often one would like to find a system of coordinates which covers the
complete domain of
existence of a solution arising from given initial data. If there existed
such
coordinates $x^{\mu}$, we could, in principle, characterize them in terms
of the
associated gauge source functions $F^{\nu}(x^{\mu})$ considered in
Sect. \ref{basics}. However,
the domain of validity, in particular the \lq lifetime\rq, of a coordinate
system depends
on the data, the equations, the type of gauge condition, as well as on
the given gauge
source functions. In practice,  gauge source functions which ensure that
the coordinates exist globally, if there exist any at all, have to be
identified in the course of constructing the solution.

Having implemented the gauge conditions, we have to {\it find a
hyperbolic system of
reduced equations} from our representation of the Einstein equations. As we
shall see,
there are often various possibilities and the final choice will depend on
the desired
application. With the reduced equations at hand we have to
{\it arrange initial data}
which satisfy the constraints and are consistent with the gauge
conditions. While the
second point usually poses no problem, the first point involves solving
elliptic
equations and requires a seperate discussion.

The reduced equations define together with the initial data a well-posed
initial value
problem and we can show, by using standard techniques of the theory of
partial
differential equations (cf. Sect. \ref{symhyp}), the {\it existence of
solutions to the
reduced problem}, work out differentiability properties of solutions etc.
or start
the numerical evolution.

As the final step one needs to show that the {\it constraints and gauge
conditions are
preserved} by the evolution defined by the reduced equations. One may
wonder whether
there exists a universal argument which tells us that this will be the
case for
any system of hyperbolic reduced equations deduced from the Einstein
equations.
The example of the spinor equations discussed below shows that this
cannot be true
without restrictions on the matter fields. However, even if we ask this
question
about reduced problems for the Einstein vacuum field equations, the
answer seems not to
be known. The standard method here is to use the
reduced equations, as in Sect. \ref{basics},
together with some differential identities, to show that the \lq constraint
quantities\rq\
satisfy a certain subsidiary system which allows us to argue that these
quantities vanish.

If the gauge condition used in our reduced equations has been shown
to be
universally applicable and if there already exists another hyperbolic
reduction using
these gauge conditions and applying to the same geometric and physical
situation (choice of matter model), it can be argued, invoking the uniqueness
property of initial
value problems for hyperbolic equations, that our reduced equations must
preserve the
constraints and the gauge conditions.


In the examples which follow we shall not always produce the complete
reduction argument.
Often we shall only exhibit a symmetric hyperbolic system of reduced
equations and remark
on some of its properties.

\subsection{Hyperbolic Systems from the ADM Equations}\label{bssn}

In the article \cite{baumgarte:shapiro} (cf. also
\cite{shibata:nakamura}) the authors derived a
system from equations (\ref{hamconstr}), (\ref{momconstr}),
(\ref{pbad}) and (\ref{pbag}) which seems to be numerically distinctively
better behaved than
any other system derived so far from the ADM equations. However, at
present there
appears to be no clear understanding as to why this should be so. The
symmetric
hyperbolic system to be discussed below was found in the course of
an attempt to understand whether
the system considered in \cite{baumgarte:shapiro} is in any sense related
to a hyperbolic system. Since other hyperbolic systems related to the
equations in
\cite{baumgarte:shapiro} have been discussed in \cite{alcubierre:etal},
\cite{fritelli:reula:II} there is now a large family of such systems
available.

In the following we shall consider the fields
\begin{equation}
\label{qgauge}
\beta^a,\,\,\,\,\,\,\,q = \log(\alpha\,h^{- \sigma}),
\end{equation}
with $h = \det(h_{ab})$, $\sigma = {\rm const.} > 0$, as the gauge source
functions.
The density $h$ will be used to rescale the $3$-metric and the trace-free
part of the extrinsic curvature to obtain the densities
$\tilde{h}_{ab} = h^{-\frac{1}{3}}\,h_{ab}$,
$\tilde{\chi}_{ab}
= h^{-\frac{1}{3}}\,(\chi_{ab} - \frac{1}{3}\,\chi\,h_{ab})$.
For simplicity
we shall refer to $\tilde{h}_{ab}$ as to the conformal metric or, if no
confusion can arise, simply as to the metric. The following equations
are derived from the ADM equations by using the standard rules for conformal
rescalings. The occurrence of some of the terms in the following equations
find their proper explanation in the general calculus for densities but we
shall not discuss this here any further.

The unknowns in our equations will be the fields
\begin{equation}
\label{1tunknowns}
\tilde{h}_{ab},\,\,\,\,\,
\eta = \log\,h,\,\,\,\,\,
\eta_a = \tilde{D}_a \log\,h,\,\,\,\,\,
\chi = h^{ab} \chi_{ab},\,\,\,\,\,
\end{equation}
\begin{equation}
\label{2tunknowns}
\hat{\gamma}_a = \tilde{\gamma}_a -
(\frac{1}{6} + \sigma)\,\tilde{D}_a\log\,h,\,\,\,\,\,
\tilde{\chi}_{ab},\,\,\,\,\,
\tilde{h}_{abc} = \tilde{h}_{ab,c},
\end{equation}
where $\tilde{\gamma}_a$ is defined for the metric $\tilde{h}_{ab}$ in
analogy to
(\ref{gaexpr}). Note that the power of the scaling factor $h$ has been chosen
such that we have $\tilde{h} = \det (\tilde{h}_{ab}) = 1$, whence
$0 = \tilde{h}_{,c} = \tilde{h}\,\tilde{h}^{ab}\,\tilde{h}_{ab,c} =
\tilde{h}^{ab}\,\tilde{h}_{abc}$. We denote by $\tilde{D}$ the
covariant
derivative, by $\tilde{\gamma}_a\,^b\,_c$ the connection coefficients,
and by
$\tilde{R}$ the Ricci scalar which are defined for the conformal metric
$\tilde{h}_{ab}$ by the standard rules. In
all expressions involving quantities carrying a tilde any index
operations are performed with the metric $\tilde{h}_{ab}$.

Our equations are obtained as follows
(cf. also \cite{baumgarte:shapiro}). {}From
(\ref{pbad}) we get by a direct calculation

\begin{equation}
\label{thevol}
\partial_t\,\tilde{h}_{ab} - \tilde{h}_{ab,c}\,\beta^c =
2\,\tilde{h}_{c(a}\,\beta^c\,_{,b)}
+ \frac{2}{3}\,\tilde{h}_{ab}\,\beta^c\,_{,c}
+ 2\,\alpha\,\tilde{\chi}_{ab},
\end{equation}

\begin{equation}
\label{2hevol}
\partial_t\,\eta - \eta_{,c}\,\beta^c = 2\,\beta^c\,_{,c}
+ 2\,\alpha\,\chi.
\end{equation}
Taking derivatives on both sides of the second equation gives

\begin{equation}
\label{d2hevol}
\partial_t\,\eta_a - \eta_{a,c}\,\beta^c =
2\,\alpha\,\tilde{D}_a\,\chi + 2\,\chi\,(\sigma\,\eta_a + \tilde{D}_a\,q)
+ \eta_c\,\beta^c\,_{,a} +  2\,\beta^c\,_{,ca}.
\end{equation}
Equation (\ref{pbag}) implies together with (\ref{pbad}) that
\[
\partial_t\,\chi - \chi_{,c}\,\beta^c
= D_c\,D^c\,\alpha - \alpha\,((1 - a)\,R + a\,R + \chi^2 - 3\,\lambda
+ \frac{\kappa}{2}\,(h^{ab}\,T_{ab} - 3\,\rho)),
\]
with an arbitrary real number $a$. To replace $R$ in the second term on the
right hand
side, we use the Hamiltonian constraint (\ref{hamconstr}), to replace $R$ in
the third term
we use the transformation law of the Ricci scalar under conformal
rescalings and the
expression of the Ricci scalar $\tilde{R}$ in terms of the conformal
quantities, which
give
\[
R = h^{-\frac{1}{3}}\,\left\{
\tilde{D}^c\,(\tilde{\gamma}_c - \frac{2}{3}\,\tilde{D}_c\,\log\,h)
- \tilde{\gamma}_c\,^c\,_b\,\tilde{\gamma}^b + \right.
\]
\[
\left. + \tilde{h}^{ab}\,\tilde{\gamma}_a\,^c\,_d\,\tilde{\gamma}_c\,^d\,_b
- \frac{1}{18}\,\tilde{D}_a\,\log\,h \,\tilde{D}^a\,\log\,h \right\}.
\]
Thus we get

\begin{equation}
\label{achevol}
h^{\frac{1}{3}}\,\frac{1}{\alpha}(\partial_t\,\chi - \chi_{,c}\,\beta^c)
= - a\,\tilde{D}^a\,\hat{\gamma}_a
+ (\sigma + a\,(\frac{1}{2} - \sigma))\,\tilde{D}^a\,\eta_a
\end{equation}
\[
+ \tilde{D}^a\,\tilde{D}_a q
+ a\,(\tilde{\gamma}_c\,^c\,_d\,\tilde{\gamma}^d
- \tilde{h}^{ab}\,\tilde{\gamma}_a\,^c\,_d\,\tilde{\gamma}_c\,^d\,_b)
\]
\[
+ (\frac{1}{18}\,a + \frac{1}{6}\,\sigma + \sigma^2)\,\eta_a\,\eta^a
+ (\frac{1}{6} + 2\,\sigma)\,\eta^a\,\tilde{D}_a q
+ \tilde{D}^a q\,\tilde{D}_a q
\]
\[
- h^{\frac{1}{3}}\,\{(1 - a)\,\tilde{\chi}_{ab}\,\tilde{\chi}^{ab}
+ \frac{1}{3}\,(1 + 2\,a)\,\chi^2 - (1 + 2\,a)\,\lambda
+ \frac{\kappa}{2}\,\{h^{ab}\,T_{ab} + (1 - 2\,a)\,\rho)\}.
\]
Using the equations above, the definition of $\hat{\gamma}_a$, and the
momentum
constraint (\ref{momconstr}) as well as its expression in terms of the
conformal quantities,
\[
\tilde{D}^c\,\tilde{\chi}_{ca} - \frac{2}{3}\,\tilde{D}_a\,\chi
+ \frac{1}{2}\,\tilde{\chi}_{ca}\,\tilde{D}^c\,\log\,h
+ \frac{1}{6}\,\chi\,\tilde{D}_a\log\,h
= \kappa\,T_{\mu \nu}\,n^{\mu}\,h^{\nu}\,_a,
\]
we obtain, with arbitrary real number $c$, by a direct calculation the
equation
\begin{equation}
\label{ggam}
\frac{1}{\alpha}\,\{\partial_t\,\hat{\gamma}_a
- \hat{\gamma}_{a,c}\,\beta^c\}
= - c\,\tilde{D}^c\,\tilde{\chi}_{ca}
+ 2\,(\frac{1}{2} + \frac{c}{3} - \sigma)\,\tilde{D}_a\,\chi
\end{equation}
\[
+ (2 + c)\,\kappa\,T_{\mu \nu}\,n^{\mu}\,h^{\nu}\,_a
+ 2\,\tilde{\chi}_a\,^c\,\{\hat{\gamma}_c
- (\frac{1}{2} + \frac{c}{4} - 2\,\sigma)\,\eta_c + \tilde{D}_c\,q\}
\]
\[
- \chi\,((\frac{1}{3} + 2\,\sigma)\,(\sigma\,\eta_a + \tilde{D}_a\,q)
+ \frac{c}{6}\,\eta_a)
- 2\,\tilde{\chi}^{cd}\,\tilde{\gamma}_c\,^b\,_d\,\tilde{h}_{ab}
\]
\[
+ \frac{1}{\alpha}\,\left\{h^{cd}\,h_{ac,b}\,\beta^b\,_{,d}
- D^{(c}\,\beta^{d)}\,(h_{ab}\,\gamma_c\,^b\,_d
+  h_{cb}\,\gamma_d\,^b\,_a)
- h^{cd}\,_{,b}\,h_{ac,d}\,\beta^b \right.
\]
\[
\left. - (\frac{1}{2} + \sigma)\,\eta_b\,\beta^b\,_{,a}
+ h^{cd}\,(h_{bc}\,\beta^b\,_{,a} + h_{ab}\,\beta^b\,_{,c})_{,d}
- (1 + 2\,\sigma)\,\beta^b\,_{,ba}\right\}.
\]
For the rescaled trace free part of the extrinsic curvature we get from
(\ref{pbag})
the equation
\[
\partial_t\,\tilde{\chi}_{ab} - \tilde{\chi}_{ab,c}\,\beta^c
- 2\,\tilde{\chi}_{c(a}\,\beta^c\,_{,b)} -
\frac{2}{3}\,\tilde{\chi}_{ab}\,\beta^c\,_{,c}
\]
\[
= - \alpha\,h^{- \frac{1}{3}}\,(R_{ab} - \frac{1}{3}\,R\,h_{ab})
+ h^{- \frac{1}{3}}\,(D_a\,D_b\,\alpha
- \frac{1}{3}\,D_c\,D^c\,\alpha\,h_{ab})
\]
\[
- \alpha\,(\chi\,\tilde{\chi}_{ab} -
2\,\tilde{h}^{cd}\,\tilde{\chi}_{ac}\,\tilde{\chi}_{bd}).
\]
To express the equation in terms of the conformal quantities we use the
transformation law of the Ricci tensor under conformal rescalings and the
expression
of the conformal Ricci tensor in terms of the connection coefficients to
get
\[
R_{ab} - \frac{1}{3}\,R\,h_{ab} =
- \frac{1}{2}\,\tilde{h}^{cd}\,\tilde{h}_{ab,cd} +
\tilde{D}_{(a}\,\tilde{\gamma}_{b)}
- \frac{1}{3}\,\tilde{h}_{ab}\,\tilde{D}_c\,\tilde{\gamma}^c
\]
\[
+ \tilde{\gamma}_c\,^d\,_a\,\tilde{h}_{ed}
\,\tilde{h}^{cf}\,\tilde{\gamma}_f\,^e\,_b
+ 2\,\tilde{\gamma}_d\,^e\,_c\,\tilde{h}^{df}
\,\tilde{h}_{e(a}\,\tilde{\gamma}_{b)}\,^c\,_f
+ \frac{1}{3}\,\tilde{h}_{ab}
\,(\tilde{\gamma}_c\,^c\,_d\,\tilde{\gamma}^d
- \tilde{h}^{ef}\,\tilde{\gamma}_e\,^c\,_d\,\tilde{\gamma}_c\,^d\,_f)
\]
\[
+ \frac{1}{36}\,(\tilde{D}_a\,\log\,h\,\tilde{D}_b\,\log\,h
- \frac{1}{3}\,\tilde{h}_{ab}\,\tilde{h}^{cd}\,
\tilde{D}_c\,\log\,h\,\tilde{D}_d\,\log\,h).
\]
Thus we obtain

\begin{equation}
\label{2chiprop}
h^{\frac{1}{3}}\,\frac{1}{\alpha}\,\{\partial_t\,\tilde{\chi}_{ab} -
\tilde{\chi}_{ab,c}\,\beta^c\}
= \frac{1}{2}\,\tilde{h}^{cd}\,\tilde{h}_{ab,cd}
- \tilde{D}_{(a}\hat{\gamma}_{b)}
+ \frac{1}{3}\,\tilde{h}_{ab}\,\tilde{D}^c\,\hat{\gamma}_c
\end{equation}
\[
+ \tilde{D}_a\,\tilde{D}_b q
- \frac{1}{3}\,\tilde{h}_{ab}\,\tilde{D}^c\,\tilde{D}_c q
+ \tilde{D}_a q\,\tilde{D}_b q - \frac{1}{3}\,h_{ab}\,
\tilde{D}^c q\tilde{D}_c q
\]
\[
+ (2\,\sigma - \frac{1}{3})\,(\eta_{(a}\,\tilde{D}_{b)} q
- \frac{1}{3}\,h_{ab}\,\eta^c\,\tilde{D}_c q)
+ (\sigma^2 - \frac{1}{3}\,\sigma - \frac{1}{36})\,
(\eta_a\,\eta_b - \frac{1}{3}\,h_{ab}\,\eta_c\,\eta^c)
\]
\[
- \tilde{\gamma}_c\,^d\,_a\,\tilde{h}_{ed}
\,\tilde{h}^{cf}\,\tilde{\gamma}_f\,^e\,_b
- 2\,\tilde{\gamma}_d\,^e\,_c\,\tilde{h}^{df}
\,\tilde{h}_{e(a}\,\tilde{\gamma}_{b)}\,^c\,_f
- \frac{1}{3}\,\tilde{h}_{ab}
\,(\tilde{\gamma}_c\,^c\,_e\,\tilde{\gamma}^e
- \tilde{h}^{cd}\,\tilde{\gamma}_c\,^e\,_f\,\tilde{\gamma}_e\,^f\,_d)
\]
\[
+ 2\,\tilde{\chi}_{c(a}\,S^c\,_{,b)}
+ \frac{2}{3}\,\tilde{\chi}_{ab}\,S^c\,_{,c}
- h^{\frac{1}{3}}\,(\chi\,\tilde{\chi}_{ab} -
2\,\tilde{h}^{cd}\,\tilde{\chi}_{ac}\,\tilde{\chi}_{bd}).
\]
In all the equations expressions like $\tilde{D}_a\tilde{\gamma}_b$ etc.
are defined by
the expressions which would hold if $\tilde{\gamma}_b$ denoted a tensor
field. Finally, we get from (\ref{thevol}) that
\begin{equation}
\label{dthevol}
\partial_t\,\tilde{h}_{abc} - \tilde{h}_{abc,d}\,\beta^d =
2\,\alpha\,\tilde{D}_c\,\tilde{\chi}_{ab}
\end{equation}
\[
+ 2\,\tilde{\chi}_{ab}\,\alpha\,(\sigma\,\eta_c + \tilde{D}_c q)
+ 4\,\alpha\,\tilde{\gamma}_c\,^d\,_{(a}\,\tilde{\chi}_{b)d}
+ \tilde{h}_{abd}\,\beta^d\,_{,c}
\]
\[
+ 2\,\beta^d\,_{(,a} \,\tilde{h}_{b)dc}
+ 2\,\tilde{h}_{d(b}\,\beta^d\,_{,a)c}
+ \frac{2}{3}\,(\tilde{h}_{abc}\,\beta^d\,_{,d}
+ \tilde{h}_{ab}\,\beta^d\,_{,dc}).
\]
If $c > 0$, the system (\ref{thevol}), (\ref{2hevol}), (\ref{d2hevol}),
(\ref{achevol}), (\ref{ggam}), (\ref{2chiprop}), (\ref{dthevol}) is symmetric
hyperbolic. This can be seen as follows. (i) We choose real numbers $e$, $f$
satisfying
\begin{equation}
\label{symequ}
e > 0,\,\,\,\,\,\,f > 0,\,\,\,\,\,\,
f\,a = - 2\,(\frac{1}{2} + \frac{c}{3} - \sigma),\,\,\,\,\,\,\,
e = \frac{\sigma}{2}\,f - (\frac{1}{2} - \sigma)\,
(\frac{1}{2} + \frac{c}{3} - \sigma),
\end{equation}
and multiply some of the equations by overall factors $c$, $e$, $f$,
$\alpha^{-1}$,
to obtain them in the form (writing out only their principal parts here)

\begin{equation}
\label{Zthevol}
\partial_t\,\tilde{h}_{ab} - \tilde{h}_{ab,c}\,\beta^c = \ldots,
\end{equation}

\begin{equation}
\label{Z2hevol}
\partial_t\,\eta - \eta_{,c}\,\beta^c = \ldots,
\end{equation}

\begin{equation}
\label{Zd2hevol}
\frac{e}{\alpha}\,\{\partial_t\,\eta_a - \eta_{a,c}\,\beta^c\} =
e\,\{2\,\tilde{D}_a\,\chi + \ldots\},
\end{equation}

\begin{equation}
\label{Zachevol}
h^{\frac{1}{3}}\,\frac{f}{\alpha}\,\{\partial_t\,\chi
- \chi_{,c}\,\beta^c\}
= f\,\{- a\,\tilde{D}^a\,\hat{\gamma}_a
+ 2\,\frac{e}{f}\,\tilde{D}^a\,\eta_a + \ldots\},
\end{equation}

\begin{equation}
\label{Zggam}
\frac{1}{\alpha}\,\{\partial_t\,\hat{\gamma}_a
- \hat{\gamma}_{a,c}\,\beta^c\}
= - c\,\tilde{D}^c\,\tilde{\chi}_{ca} - f\,a\,\tilde{D}_a\,\chi + \ldots,
\end{equation}

\begin{equation}
\label{Z2chiprop}
h^{\frac{1}{3}}\,\frac{c}{\alpha}\,\{\partial_t\,\tilde{\chi}_{ab} -
\tilde{\chi}_{ab,c}\,\beta^c\}
= c\,\{\frac{1}{2}\,\tilde{h}^{cd}\,\tilde{h}_{ab,cd}
- \tilde{D}_{(a}\hat{\gamma}_{b)}
+ \frac{1}{3}\,\tilde{h}_{ab}\,\tilde{D}^c\,\hat{\gamma}_c
+ \ldots\},
\end{equation}

\begin{equation}
\label{Zdthevol}
\frac{c}{4\,\alpha}\{\partial_t\,\tilde{h}_{abc}
- \tilde{h}_{abc,d}\,\beta^d\} =
c\,\{\frac{1}{2}\,\tilde{D}_c\,\tilde{\chi}_{ab} + \ldots\}.
\end{equation}
(ii) We contract both sides of (\ref{Zd2hevol}) and (\ref{Zggam}) with
$\tilde{h}^{ba}$,
both sides of (\ref{Z2chiprop}) with $\tilde{h}^{a(c}\,\tilde{h}^{d)b}$,
both sides of
(\ref{Zdthevol}) with  $\tilde{h}^{a(d}\,\tilde{h}^{e)b}\,\tilde{h}^{fc}$,
and add
on the right hand side of the equation obtained in this way from
(\ref{Zggam}) a term
of the form  $\frac{c}{3}\,\tilde{h}^{ba}\,\tilde{h}^{cd}\,\tilde{D}_a
\tilde{\chi}_{cd}$, which vanishes identically but whose formal occurence
makes the
symmetry manifest.

The range in which the coefficients $a$, $c$, $e$, $f$, $\sigma$ are
consistent with
(\ref{symequ}) can be seen by considering the following cases:
\[
i) \,\,\,\,\,\,\,\,\,a = 0 \leftrightarrow \sigma = \frac{1}{2}
+ \frac{c}{3}.
\]
In this case we have $e = (\frac{1}{4} + \frac{c}{6})\,f$ and we can choose
$c, f > 0$ arbitrarily.
If $a \neq 0$ we have
$f = - \frac{2}{a}\,(\frac{1}{2} + \frac{c}{3} - \sigma)$ and
$e = \{\sigma\,\frac{a - 1}{a} - \frac{1}{2}\}\,
(\frac{1}{2} + \frac{c}{3} - \sigma)$, which gives the following
restrictions
\[
ii)\,\,\,\,\,\,\,\,a < 0,\,\,\,\,\,\,\,
\frac{|a|}{2\,(|a| + 1)} < \sigma < \frac{1}{2} + \frac{c}{3},
\]
\[
iii)\,\,\,\,\,\,\,\,\,\,\,\,\,\,\,0 < a \le 1,\,\,\,\,\,\,\,
\frac{1}{2} + \frac{c}{3} < \sigma,
\]
\[
iv)\,\,\,\,\,\,\,1 < a < \frac{3 + 2\,c}{2\,c},\,\,\,\,\,\,\,
\frac{1}{2} + \frac{c}{3} < \sigma < \frac{a}{2\,(a - 1)}.
\]
We cannot have $e > 0$, $f > 0$ with $a \ge \frac{3 + 2\,c}{2\,c}$.

It is of interest to study the characteristics of the system. Equations
(\ref{Zthevol}) and
(\ref{Z2hevol}) contribute factors  $n(\xi) = n^{\mu}\,\xi_{\mu}$ to the
characteristic
polynomial. To find the other characteristics we analyse for which
$\xi_{\mu} \neq 0$
the following linear system of equations, defined by the principal symbol
map, admits
non-trivial solutions. We set $\xi^a = \tilde{h}^{ab}\,\xi_b$.
\[
n(\xi)\,\eta_a = 2\,\xi_a\,\chi,
\]
\[
h^{\frac{1}{3}}\,n(\xi)\,\chi = - a\,\xi^a\,\hat{\gamma}_a
+ 2\,\frac{e}{f}\,\xi^a\,\eta_a,
\]
\[
n(\xi)\,\hat{\gamma}_a = - c\,\xi^b\,\tilde{\chi}_{ba} - f\,a\,\xi_a\,\chi,
\]
\[
h^{\frac{1}{3}}\,n(\xi)\,\tilde{\chi}_{ab} =
\frac{1}{2}\,\tilde{h}_{abc}\,\xi^c
- \xi_{(a}\,\hat{\gamma}_{b)}
+ \frac{1}{3}\,\tilde{h}_{ab}\,\xi^c\,\hat{\gamma}_c,
\]
\[
n(\xi)\,\tilde{h}_{abc} =
2\,\xi_c\,\tilde{\chi}_{ab}.
\]
The condition $n(\xi) = 0$ implies $\tilde{\chi}_{ab} = 0$ and $\chi = 0$ but
there remains a fifteen-parameter freedom to choose the remaining unknowns.
If
\begin{equation}
\label{1charfac}
n(\xi) \neq 0,
\end{equation}
we derive from the equations above the further
equations (writing
$g(\xi, \xi) = g^{\mu \nu}\,\xi_{\mu}\,\xi_{\nu}$)
\[
h^{\frac{1}{3}}\,g(\xi, \xi)\,\tilde{\chi}_{ab} =
n(\xi)\,(\xi_{(a}\,\hat{\gamma}_{b)}
- \frac{1}{3}\,\tilde{h}_{ab}\,\xi^c\,\hat{\gamma}_c),
\]
whence
\[
\{g(\xi, \xi) + \frac{c}{2}\,h^{cd}\,\xi_c\,\xi_d\}\,\hat{\gamma}_a
= - (\frac{c}{6}\,h^{- \frac{1}{3}}\,\xi^c\,\hat{\gamma}_c
+ \frac{g(\xi, \xi)}{n(\xi)}\,f\,a\,\chi)\,\xi_a.
\]
The latter equation implies
\begin{equation}
\label{1charsyst}
\{g(\xi, \xi) + \frac{2\,c}{3}\,h^{cd}\,\xi_c\,\xi_d\}\,
\xi^a\,\hat{\gamma}_a
+ \frac{g(\xi, \xi)}{n(\xi)}\,\xi^c\,\xi_c\,f\,a\,\chi = 0,
\end{equation}
and, with $p^a$ such that $p^a\,\xi_a = 0$,
\[
\{g(\xi, \xi) + \frac{c}{2}\,h^{cd}\,\xi_c\,\xi_d\}\,
p^a\,\hat{\gamma}_a = 0,
\]
Finally we get
\begin{equation}
\label{2charsyst}
\{g(\xi, \xi) + (a - 1)\,(1 - 2\,\sigma)\,h^{cd}\,\xi_c\,\xi_d\}\,\chi
- a\,h^{- \frac{1}{3}}\,n(\xi)\,\xi^c\,\hat{\gamma}_c = 0.
\end{equation}
Equations (\ref{1charsyst}) and (\ref{2charsyst}), are of the form $A\,u = 0$
with the
unknown $u$ the transpose of $(\chi, \xi^i\,\hat{\gamma}_i)$ and the
matrix $A$ satisfying
\[
\det\,A = \{g(\xi, \xi)
+ \frac{2\,c}{3}\,(1 - a)\,h^{cd}\,\xi_c\,\xi_d\}\,
\{g(\xi, \xi) + (2\,\sigma - 1)\,h^{cd}\,\xi_c\,\xi_d\}.
\]
If
\begin{equation}
\label{2charfac}
g(\xi, \xi) = 0,
\end{equation}
it follows from the first of these equations that
$\hat{\gamma}_a = 0$. If $a \neq 1$, $\sigma \neq \frac{1}{2}$ there
remains the two-parameter freedom to choose $\tilde{\chi}_{ab}$ with
$\xi^a\,\tilde{\chi}_{ab} = 0$.
If $a = 1$ or $\sigma = \frac{1}{2}$ there remains the three-parameter
freedom to choose $\chi$, $\tilde{\chi}_{ab}$ satisfying
$\xi^a\,\tilde{\chi}_{ab} = - \frac{f\,a}{c}\,\xi_b\,\chi$.
If $g(\xi, \xi) \neq 0$ the field $\tilde{\chi}_{ij}$ is known once
$\hat{\gamma}_a$ has been determined. If
\begin{equation}
\label{3charfac}
g(\xi, \xi) + \frac{c}{2}\,h^{cd}\,\xi_c\,\xi_d = 0,
\end{equation}
there is a two-parameter freedom to choose $p^i\,\hat{\gamma}_i$ as
above. This is the
only freedom unless $a = \frac{1}{4}$ or $\sigma = \frac{1}{2}
+ \frac{c}{4}$,
conditions which exclude each other.
Further characteristics are given by the equations
\begin{equation}
\label{4charfac}
g(\xi, \xi) + \frac{2\,c}{3}\,(1 - a)\,h^{cd}\,\xi_c\,\xi_d = 0,
\end{equation}
\begin{equation}
\label{5charfac}
g(\xi, \xi) + (2\,\sigma - 1)\,h^{cd}\,\xi_c\,\xi_d = 0.
\end{equation}
The \lq physical\rq\ characteristics correspond of course to (\ref{2charfac}),
the two-parameter freedom pointed out in the case
$a \neq 1$, $\sigma \neq \frac{1}{2}$
corresponding to the two polarization states of gravitational waves and the
additional freedom in the other cases corresponding essentially to a
gauge freedom.
The timelike characteristics corresponding to (\ref{1charfac}) occur
because of the
transition from the system of second order to a system of first order.
These
characteristics are neither \lq physical\rq\ nor \lq harmful\rq.
Characteristics corresponding to (\ref{3charfac}) are spacelike, while the
nature of
the characteristics corresponding to (\ref{4charfac}) and (\ref{5charfac})
depends on the
constants $a$ and  $\sigma$. It can happen that one is spacelike while
the other is
timelike or both are spacelike.

We note here that it is possible to obtain by similar procedures
reduced equations, for somewhat more complicated unknowns, which have only
characteristics which are timelike or null \cite{fritelli:reula:II}.

Though the number $c > 0$ can be chosen arbitrarily (suitably adjusting
the
others) it is not possible to perform the limit $c \rightarrow 0$ while
keeping
the symmetric hyperbolicity of the system, however, the equations in
\cite{baumgarte:shapiro} can be considered as limit of equations which are
algebraically equivalent to our systems. We note also that it is not
possible to
perform a regular limit $\sigma \rightarrow 0$ which would make lapse and
shift
the gauge source functions. In the limit as $c \rightarrow 0$ (which we
can perform if
we do not insist on hyperbolicity) all characteristics, with the exception
of the gauge
dependent characteristics corresponding to (\ref{5charfac}), become
non-spacelike.

We finally remark on our gauge conditions. Under certain assumptions the
gauge
(\ref{qgauge}) coincides with the gauge with harmonic time coordinate and
prescribed
shift. {}From (\ref{gammadecomp}) follows the general relation
\[
\partial_t \alpha - \alpha_{,a}\,\beta^a
= \alpha^2\,\chi - \alpha^3\,\Gamma^0.
\]
Equation (\ref{2hevol}), written as an equations for $h$, entails together
with
(\ref{qgauge}) the equation
\[
\partial_t \alpha - \alpha_{,a}\,\beta^a = 2\,\sigma\,\alpha^2\,\chi
+ \alpha\,(2\,\sigma\,\beta^a\,_{,a} + \partial_t q - q_{,a} \beta^a).
\]
Thus the time coordinate $t$ is harmonic in our gauge, i.e.
$\Gamma^0 = 0$, if
$\sigma = \frac{1}{2}$ and $\partial_t q - q_{,a} \beta^a
= - \beta^a\,_{,a}$. In more
general situations the expression for $\Gamma^0$ implied by the equations
above
contains information about the solution and admits no direct conclusion
about
time-harmonicity in terms of $\alpha$ and $\beta^a$.

It follows from complete reductions based on the gauge source function
$q$ (cf. e.g
\cite{friedrich:hypred}, \cite{fritelli:reula:II}) that on solutions of
Einstein's
equations it is possible to achieve the corresponding gauge close to some
initial
surface. However, it has apparently never been shown that for prescribed
$\sigma > 0$,
$\beta^a$ and given spacelike hypersurface of some arbitrary
Lorentz manifold, coordinates can be constructed which realize these gauge
source
functions. It would be
useful to have a proof of the universal applicability of this type of
gauge and
information about its general behaviour.

\subsection{The Einstein--Euler System}\label{euler}

In the following we shall discuss the Einstein--Euler equations, i.e.
Einstein's
equation coupled to the Euler equation for a simple perfect fluid. Its
hyperbolicity has
been studied by various authors (cf. \cite{choquet58},
\cite{lichnerowicz94},
\cite{rendall:I}). Though the system is also important in the
cosmological context, our main concern in analysing the system here is to
control the evolution of compact perfect fluid
bodies, which are considered as models for \lq gaseous stars\rq. In this
situation
arises, besides the need to cast the equations into hyperbolic form, the
side
condition to control the evolution of the timelike boundary along which
the
Einstein--Euler equations go over into the Einstein vacuum field equations.

The analysis of the evolution of the fields in the neighbourhood of this
boundary poses
the basic difficulty in the discussion of compact fluid bodies. In the
case of
spherical symmetry, this problem was overcome in \cite{kind93}. If in
more general
situations the coordinates used in the hyperbolic reduction are governed
e.g. by wave
equations, as is the case in the harmonic gauge, there appears to be no
way to control
the motion of this boundary in these coordinates. In \cite{friedrich98}
a system of equations has been derived from the Einstein--Euler system
which combines hyperbolicity with the Lagrangian description of the flow,
so that the spatial
coordinates are constant along the flow lines. Thus the location of the
body is known
in these coordinates. In the following we shall discuss this system and
derive the
subsidiary system, which was not given in \cite{friedrich98}.

\subsubsection{The Basic Equations}\label{basiceq}

We shall use a frame formalism in which the information on the metric is
expressed
in terms of an orthonormal frame $\{e_k\}_{k = 0,\ldots,3}$ and all
fields, with the
possible exception of the frame itself, are given in this frame. To make
the formalism
easily comparable with the spin frame formalism which will be used later,
we shall use
a signature such that $g_{ik} \equiv g(e_i, e_k) =
{\rm diag}(1, -1, -1, -1)$.
Let $\nabla$
denote the the Levi--Civita connection of $g_{\mu \nu}$. The basic
unknowns of our
representation of the Einstein equations are given by
\[
e^{\mu}\,_k,\,\,\,\,\,\,\,\,
\Gamma_i\,^j\,_k,\,\,\,\,\,\,\,\,
C^i\,_{jkl}, \quad\mbox{matter variables,}\quad
\]
where $e^{\mu}\,_k = \,<e_k, x^{\mu}>$ are the coefficients of the frame
field in
some coordinates $x^{\mu}$, $\Gamma_i\,^j\,_k$  are the connection
coefficients,
defined by $\nabla_i\,e_k = \Gamma_i\,^j\,_k\,e_j$ and satisfying
$\Gamma_i\,^j\,_k\,g_{jl} + \Gamma_i\,^j\,_l\,g_{jk} = 0$, and
$C^i\,_{jkl}$ denotes
the conformal Weyl tensor in the frame $e_k$. The latter is obtained
from the
decomposition
\begin{equation}
\label{Rdecomp}
R_{ijkl} = C_{ijkl} + \{ g_{i[k}\,S_{l]j} - g_{j[k}\,S_{l]i} \},
\end{equation}
of the curvature tensor
\begin{equation}
\label{curv}
R^i\,_{jpq} = e_p(\Gamma_q\,^i\,_j) - e_{q}(\Gamma_p\,^i\,_j)
- \Gamma_k\,^i\,_j\,(\Gamma_p\,^k\,_q - \Gamma_q\,^k\,_p)
\end{equation}
\[
+ \Gamma_p\,^i\,_k\,\Gamma_q\,^k\,_j
- \Gamma_q\,^i\,_k\,\Gamma_p\,^k\,_j,
\]
where we also set $S_{ij} = R_{ij} - \frac{1}{6}\,g_{ij}\,R$, with
$R_{ij}$ and $R$ denoting the Ricci tensor and the Ricci scalar.
We shall need the notation
\[
T_i\,^k\,_j\,e_k =
- [e_i,e_j] + (\Gamma_i\,^l\,_j - \Gamma_j\,^l\,_i)\,e_l,
\]
\[
\Delta^i\,_{jkl} =
R^i\,_{jkl} - C^i\,_{jkl} - g^i\,_{[k}\,S_{l]j} + g_{j[k}\,S_{l]}\,^i,
\]
with $R^i\,_{jkl}$ understood as being given by (\ref{curv}). Furthermore
we set
\[
F_{jkl} = \nabla_i\,F^i\,_{jkl},
\quad\mbox{with}\quad
F^i\,_{jkl} = C^i\,_{jkl} - g^i\,_{[k}\,S_{l]j}.
\]

In the equations above we take account of the Einstein--Euler equations in
the form
\begin{equation}
\label{2einst}
S_{ik} = \kappa\,(T_{ik} - \frac{1}{3}\,g_{jk}\,T),
\end{equation}
with an energy-momentum tensor of a simple perfect fluid
\begin{equation}
\label{pfl}
T_{ik} = (\rho + p)\,U_i\,U_k - p\,g_{ik}.
\end{equation}
Here $\rho$ is the total energy density and $p$ the pressure, as measured
by an observer moving with the fluid, and $U$ denotes the
(future directed) flow vector
field, which satisfies $U_i\,U^i = 1$.

We shall need the decomposition
\begin{equation}
\label{Tdiv}
\nabla^j\,T_{jk} = q_k + q\,U_k,
\end{equation}
and the field
\begin{equation}
\label{Jex}
J_{jk} = \nabla_{[j}\,q_{k]},
\end{equation}
with
\begin{equation}
\label{qexpr}
q = U^i\,\nabla_i\,\rho + (\rho + p)\,\nabla_i\,U^i,
\end{equation}
\begin{equation}
\label{2qkexpr}
q_k = (\rho + p)\,U^i\,\nabla_i\,U_k
+ \{U_k\,U^i\,\nabla_i - \nabla_k\}\,p.
\end{equation}

We assume that the fluid is simple, i.e. it consists of only one class of
particles, and denote by $n$, $s$, $T$ the number density of
particles, the entropy per particle, and the absolute temperature as
measured by an observer moving with the fluid. We shall assume the first
law of equilibrium thermodynamics which has the familiar form
$d\,e = - p\,d\,v + T\,d\,s$  in terms of the volume $v = \frac{1}{n}$
and the energy  $e = \frac{\rho}{n}$ per particle. In terms of the
variables above, we have
\begin{equation}
\label{flth}
d\,\rho = \frac{\rho + p}{n}\,d\,n + n\,T\,d\,s.
\end{equation}

We assume an equation of state given in the form
\begin{equation}
\label{eqst}
\rho = f(n, s),
\end{equation}
with some suitable non-negative function $f$ of the number
density of particles and the entropy per particle. Using this in
(\ref{flth}), we obtain
\begin{equation}
\label{pT}
p = n\,\frac{\partial\,\rho}{\partial\,n} - \rho,\,\,\,\,\,\,
T = \frac{1}{n}\,\frac{\partial\,\rho}{\partial\,s},
\end{equation}
as well as the speed of sound $\nu$, given by
\begin{equation}
\label{sousp}
\nu^2 \equiv (\frac{\partial\,p}{\partial\,\rho})_s =
\frac{n}{\rho + p}\,\frac{\partial\,p}{\partial\,n},
\end{equation}
as known functions of $n$ and $s$. We require that the specific enthalpy
and the speed of sound are positive, i. e.
\begin{equation}
\label{fcond}
\frac{\rho + p}{n} > 0,\,\,\,\,\,\,
\frac{\partial\,p}{\partial\,n}
= n\,\frac{\partial^2\,\rho}{\partial\,n^2} > 0.
\end{equation}

We assume the law of particle conservation
\begin{equation}
\label{parcon}
U^i\,\nabla_i\,n + n\,\nabla_i\,U^i = 0.
\end{equation}
It implies together with $q = 0$ and (\ref{flth}) that the flow is
adiabatic, i.e. the entropy per particle is conserved along the flow
lines,
\begin{equation}
\label{isen}
U^i\,\nabla_i\,s = 0.
\end{equation}

The case of an isentropic flow, where the entropy is constant in
space and time, is of some interest. In this case the equation
of state can be given in the form
\begin{equation}
\label{homentropic}
p = h(\rho)
\end{equation}
with some suitable function $h$. As a special subcase we shall consider
pressure free
matter (\lq dust\rq\ ), where $h \equiv 0$.

We note that (\ref{eqst}), (\ref{pT}), (\ref{sousp}), (\ref{parcon}),
and (\ref{isen})
imply
\[
U^i\,\nabla_i\,p = - (\rho + p)\,\nu^2\,\nabla_i\,U^i,
\]
from which we get
\begin{equation}
\label{3qkexpr}
q_k = - \nabla_k\,p + (\rho + p)\,\{U^i\,\nabla_i\,U_k - \nu^2\,U_k\,
\nabla_i\,U^i\}.
\end{equation}
Finally, (\ref{isen}) implies the equation
\begin{equation}
\label{dsevol}
{\cal L}_U\,s_k = 0,
\end{equation}
for $s_k = \nabla_k\,s$, where ${\cal L}_U$ denotes the Lie derivative in
the
direction of $U$.

The Einstein--Euler equations are given in our representation by the
equations
$z = 0$, $q = 0$, (\ref{eqst}), (\ref{pT}), (\ref{sousp}), (\ref{parcon}),
and (\ref{isen}), where we
denote by $z$ the vector-valued quantity
\begin{equation}
\label{zeroqu}
z = (T_i\,^k\,_j,\,\,\Delta^i\,_{jkl},\,\,F_{jkl},\,\,q_k).
\end{equation}
which we shall refer to (as well as to each of its components)
as a \lq zero quantity\rq . These equations entail furthermore $J_{jk} = 0$
and (\ref{dsevol}).  After
making a suitable choice of gauge conditions we shall extract hyperbolic
evolution equations from this highly overdetermined system .

\subsubsection{Decomposition of Unknowns and Equations}\label{decomp}

{}From now on we shall assume that
\[
e_0 = U,
\]
so that we have $U^i = \delta^i\,_0$. For the further discussion of the
equations we
decompose the unknowns and the equations. With the vector field $U$ we
associate
\lq spatial\rq\ tensor fields, i.e. tensor fields $T_{i_1,\ldots,i_p}$
satisfying
\[
T_{i_1,\ldots,i_l,\ldots,i_p}\,U^{i_l} = 0,\,\,\,\,\,l = 1,\cdots,p.
\]
The subspaces orthogonal to $U$ inherit the metric
$h_{ij} = g_{ij} - U_i\,U_j$, and $h_i\,^j$ (indices being
raised and lowered with $g_{ij}$) is the orthogonal projector onto
these subspaces.

We shall have to consider the projections of various tensor fields
with respect to $U$ and its orthogonal subspaces. For a given tensor any
contraction with $U$ will be denoted by replacing the corresponding
index by $U$ and the projection with respect to $h_i\,^j$ will be
indicated by a prime, so that for a tensor field $T_{ijk}$ we write
e.g.
\[
T'_{iUk} = T_{mpq}\,h_i\,^m\,U^p\,h_k\,^q,
\]
etc. Denoting by $\epsilon_{ijkl}$ the totally antisymmetric tensor
field with $\epsilon_{0123} = 1$ and setting
$\epsilon_{jkl} = \epsilon'_{Ujkl}$, we have the decomposition
\[
\epsilon_{ijkl} = 2\,(U_{[i}\,\epsilon_{j]kl} - \epsilon_{ij[k}\,U_{l]}),
\]
and the relations
\[
\epsilon^{jkl}\,\epsilon_{jpq} = - 2\,\epsilon\,h^k\,_{[p}\,h^l\,_{q]},
\,\,\,\,\,\,
\epsilon^{jkl}\,\epsilon_{jkq} = - 2\,\epsilon\,h^l\,_q.
\]

Denoting by $C^{*}_{ijkl} = \frac{1}{2}\,C_{ijpq}\,\epsilon_{kl}\,^{pq}$
the dual of the conformal Weyl tensor, its $U$-electric and $U$-magnetic
parts
are given by by $E_{jl} = C'_{UjUl}$, $B_{jl} = C^{*'}_{UjUl}$
respectively.
With the notation
$l_{jk} = h_{jk} - U_j\,U_k$, we get the decompositions
\begin{equation}
\label{Cdecomp}
C_{ijkl} = 2\,(l_{j[k}\,E_{l]i} - l_{i[k}\,E_{l]j})
- 2\,(U_{[k}\,B_{l]p}\,\epsilon^p\,_{ij}
+ U_{[i}\,B_{j]p}\,\epsilon^p\,_{kl}),
\end{equation}
\begin{equation}
\label{C*decomp}
C^{*}_{ijkl} = 2\,U_{[i}\,E_{j]p}\,\epsilon^p\,_{kl}
- 4\,E_{p[i}\,\epsilon_{j]}\,^p\,_{[k}\,U_{l]}
- 4\,U_{[i}\,B_{j][k}\,U_{l]}
- B_{pq}\,\epsilon^p\,_{ij}\,\epsilon^q\,_{kl}.
\end{equation}
We set
\[
a^i = U^k\,\nabla_k\,U^i,\,\,\,\,\,
\chi_{ij} = h_i\,^k\,\nabla_k\,U_j,\,\,\,\,\,
\chi = h^{ij}\,\chi_{ij},
\]
so that we have
\[
\nabla_j\,U^i = U_j\,a^i + \chi_j\,^i,\,\,\,\,\,\,
a^i = h_j\,^i\,\Gamma_0\,^j\,_0,\,\,\,\,\,\,
\chi_{ij} = - h_i\,^k\,h_j\,^l\,\Gamma_k\,^0\,_l.
\]
Since $U$ is not required to be hypersurface orthogonal, the field
$\chi_{ij}$ will in general not be symmetric.
If the tensor field $T$ is spatial, i.e. $T = T'$, we define its spatial
covariant derivative by ${\cal D}\,T = (\nabla\,T)'$. i.e.
\[
{\cal D}_i\,T_{i_1,\ldots,i_p} =
\nabla_j\,T_{j_1,\ldots,j_p}
\,h_i\,^j\,h_{i_1}\,^{j_1} \ldots h_{i_p}\,^{j_p}.
\]
It follows that
\[
{\cal D}_i\,h_{jk} = 0,\,\,\,\,\,{\cal D}_i\,\epsilon_{jkl} = 0.
\]
Equation (\ref{isen}) then implies $s_k = {\cal D}_k\,s$.

To decompose the equations, we observe the relations
\[
q = \frac{2}{\kappa}\,h^{ij}\,F'_{iUj},
\,\,\,\,\,
q_k = - \frac{2}{\kappa}\,(h^{ij}\,F'_{ijk} - F'_{UkU}),
\]
and set
\[
P_i = F'_{UiU}
\]
\[
= {\cal D}^j \{E_{ji} - \kappa\,(\frac{1}{3}\,\rho
+ \frac{1}{2}\,p)\,h_{ji}\}
+ \frac{1}{2}\,\kappa\,(\rho + p)\,a_i
+ 2\,\chi^{kl}\,\epsilon^j\,_{l(i}\,B_{k)j},
\]

\[
Q_j = - \frac{1}{2}\,\epsilon_j\,^{kl}\,F'_{Ukl}
\]
\[
= {\cal D}^k\,B_{kj}
+ \epsilon_j\,^{kl}\,(2\,\chi^i\,_k - \chi_k\,^i)\,E_{il}
+ \kappa\,(\rho + p)\chi_{kl}\,\epsilon_j\,^{kl},
\]

\[
P_{ij} = \{ F'_{(i|U|j)}
- \frac{1}{3}\,h_{ij}\,h^{kl}\, F'_{kUl}\}
= {\cal L}_U\,E_{ij} + {\cal D}_k\,B_{l(i}\,\epsilon_{j)}\,^{kl}
\]
\[
- 2\,a_k\,\epsilon^{kl}\,_{(i}\,B_{j)l}
- 3\,\chi_{(i}\,^l\,E_{j)l} - 2\,\chi^l\,_{(i}\,E_{j)l}
+ h_{ij}\,\chi^{kl}\,E_{kl} + 2\,\chi\,E_{ij}
\]
\[
+ \frac{\kappa}{2}\,(\rho + p)\,(\chi_{(ij)}
- \frac{1}{3}\,\chi\,h_{ij}),
\]

\[
Q_{kl} = \frac{1}{2}\,\epsilon_{(k}\,^{ij}\,F'_{l)ij}
= {\cal L}_U\,B_{kl} - {\cal D}_i\,E_{j(k}\,\epsilon_{l)}\,^{ij}
\]
\[
+ 2\,a_i\,\epsilon^{ij}\,_{(k}\,E_{l)j}
- \chi^i\,_{(k}\,B_{l)i}
- 2\,\chi_{(k}\,^i\,B_{l)i}
+ \chi\,B_{kl}
- \chi_{ij}\,B_{pq}\,\epsilon^{pi}\,_{(k}\,\epsilon^{jq}\,_{l)}.
\]
Then we find the splitting

\begin{equation}
\label{2Fdeco}
F_{jkl} = 2\,U_j\,P_{[k}\,U_{l]} + h_{j[k}\,P_{l]}
+ Q_i\,(U_j\,\epsilon^i\,_{kl} - \epsilon^i\,_{j[k}\,U_{l]})
\end{equation}
\[
- 2\,P_{j[k}\,U_{l]} - Q_{ji} \epsilon^i\,_{kl}
- \frac{1}{2}\,\kappa\,h_{j[k}\,q_{l]}
- \frac{1}{3}\,\kappa\,q\,h_{j[k}\,U_{l]}.
\]

\subsubsection{The Reduced System}\label{redsys}

i) In the case of pressure free matter the equation $q_k = 0$ tells us
that
$e_0$ is geodesic and we can assume the frame field to be parallelly
transported in
the direction of $e_0$. Furthermore, we can assume the coordinates
$x^{\alpha}$,
$\alpha = 1, 2, 3$ to be constant on the flow lines of $e_0$ and the
coordinate
$t = x^0$ to be a parameter on the integral curves of $e_0$. These
conditions are
equivalent to
\[
\Gamma_0\,^j\,_k = 0,\,\,\,\,\,\,\,\,e^{\mu}\,_0 = U^{\mu}
= \delta^{\mu}\,_0,
\]
which, in turn, imply together with the requirement $p = 0$ the relation
$q_k = 0$ if $\rho + p > 0$. The unknowns to be determined are given by
\[
u = (e^{\mu}\,_a,\,\,\Gamma_b\,^i\,_j,\,\,E_{ij},\,\,B_{ij},\,\,\rho),
\]
where $a, b = 1, 2, 3$. The reduced system is given by
\begin{equation}
\label{dredequ}
T_0\,^j\,_a = 0,\,\,\,\,\,\Delta^i_{j0a} = 0,\,\,\,\,\,
P_{ij} = 0,\,\,\,\,\,Q_{kl} = 0,\,\,\,\,\,\,q = 0.
\end{equation}

ii) In the general case we cannot assume the frame to be parallelly
transported
because the evolution of $U$ is governed by the Euler equations. We
assume the vector fields $e_a$, $a = 1, 2, 3$, to be Fermi transported in
the direction of $U$ and the coordinates to be chosen as before such that
\[
\Gamma_0\,^a\,_b = 0,\,\,\,\,\,\,\,\,e^{\mu}\,_0 = U^{\mu}
= \delta^{\mu}\,_0.
\]
Our unknowns are given by
\[
u = (e^{\mu}\,_a,\,\,\Gamma_0\,^0\,_a,\,\,\Gamma_a\,^k\,_l,
\,\,E_{ij},\,\,B_{kl},\,\,\rho,\,\,n,\,\,s,\,\,s_k),
\]
and the reduced system by
\begin{equation}
\label{1flredequ}
T_0\,^j\,_a = 0,\,\,\,\,\,\Delta^c\,_{b0a} = 0,
\end{equation}
\[
\nu^2\,\Delta^c\,_{0ac} - \frac{1}{\rho + p}\,J'_{Ua} = 0,\,\,\,\,\,\,
\nu^2\,(\Delta^0\,_{a0b} + \frac{1}{\rho + p}\,J'_{ab}) = 0,
\]
\begin{equation}
\label{2flredequ}
P_{ij} = 0,\,\,\,\,\,Q_{kl} = 0,
\,\,\,\,\,\,q = 0,
\end{equation}
\[
{\cal L}_U\,n = - n\,\chi,\,\,\,\,\,\,
{\cal L}_U\,s = 0,\,\,\,\,\,{\cal L}_U\,s_k = 0,
\]
where it is assumed that the functions $p$, $\nu^2$, $\alpha$, $\beta$ are
determined from the equation of state (\ref{eqst}) according to (\ref{pT}),
(\ref{sousp}),
etc. and $\rho + p > 0$.

Since the functions $\rho$, $s$, $n$ are then determined as functions of
the
coordinates $x^{\mu}$, we remark that the relation
$\rho(x^{\mu}) = f(n(x^{\mu}), s(x^{\mu}))$ is satisfied as a consequence
of the
equations for $\rho$, $s$, and $n$, and the relation (\ref{pT}), since
$q = 0$ implies
$\partial_t\,\rho = \frac{d}{d\,t}\,f(n, s)$. Furthermore, we note that
in our
formalism ${\cal L}_U\,s_k = \partial_t\,s_k - (\Gamma_0\,^j\,_k -
\Gamma_k\,^j\,_0)\,s_j = e^{\mu}\,_k\,\partial_t\,s_{\mu}$ etc.

\vspace{.7cm}

When we solve the equations $P_{ij} = 0$ and $Q_{kl} = 0$, the symmetry of
the fields
$E_{ij}$, $B_{kl}$ has to be made explicit. The trace-free condition then
follows as a
consequence of the equations and the fact that the initial data are
trace-free.
With this understanding it is easy to see that the system (\ref{dredequ})
is symmetric
hyperbolic, the remaining equations only containing derivatives in the
direction of
$U$. To see that the system consisting of
(\ref{1flredequ}) and (\ref{2flredequ}) is also symmetric
hyperbolic, we write out some of the equations explicitly (taking the
opportunity to
correct some misprints in \cite{friedrich98}). It follows directly
from the
definition that
\begin{equation}
\label{intcond}
J_{kj} = (\rho + p)\,\left\{ U^i\,
(\nabla_k\,\nabla_i\,U_j - \nabla_j\,\nabla_i\,U_k)
- \nu^2\,U_j\,\nabla_k\,\nabla_i\,U^i\right.
\end{equation}
\[
\left. + \nu^2\,U_k\,\nabla_j\,\nabla_i\,U^i
- \nu^2\,\nabla_l\,U^l\,(\nabla_k\,U_j - \nabla_j\,U_k)
+ \nabla_k\,U^i\,\nabla_i\,U_j
- \nabla_j\,U^i\,\nabla_i\,U_k \right.
\]
\[
\left. + \epsilon\,\frac{\rho + p}{\nu^2}\,
(\frac{\partial^2\,p}{\partial\,\rho^2})_s\,\nabla_l\,U^l\,
(U_k\,U^i\,\nabla_i\,U_j - U_j\,U^i\,\nabla_i\,U_k) \right\}
\]
\[
+ (\alpha\,U_k\,\nabla_i\,U^i - \beta\,U^i\,\nabla_i\,U_k)\,s_j
- (\alpha\,U_j\,\nabla_i\,U^i - \beta\,U^i\,\nabla_i\,U_j)\,s_k,
\]
where we set
\[
\alpha = (\rho + p)\,\frac{\partial\,\nu^2}{\partial\,s}
- (1 + \frac{n}{\nu^2}\,\frac{\partial\,\nu^2}{\partial\,n})\,
\frac{\partial\,p}{\partial\,s} + \nu^2\,n\,T,\,\,\,\,\,\,
\beta = n\,T - \frac{1}{\nu^2}\,\frac{\partial\,p}{\partial\,s}.
\]
In particular,
\begin{equation}
\label{1intcond}
 - \frac{1}{\rho + p}\,J'_{Ua} = e_0(\Gamma_0\,^0\,_a)
- \nu^2\,e_a(\Gamma_c\,^c\,_0)
+ \Gamma_0\,^0\,_c\,(\Gamma_a\,^c\,_0 - \Gamma_0\,^c\,_a)
\end{equation}
\[
+ \left(\frac{\rho + p}{\nu^2}\,
(\frac{\partial^2\,p}{\partial\,\rho^2})_s -
\nu^2 \right)
\,\Gamma_c\,^c\,_0\,\Gamma_0\,^0\,_a
- \frac{\alpha}{\rho + p}\,\Gamma_c\,^c\,_0\,s_a,
\]
and
\begin{equation}
\label{2intcond}
- \frac{1}{\rho + p}\,J'_{ab}
= e_a(\Gamma_0\,^0\,_b) - \,e_b(\Gamma_0\,^0\,_a)
- \Gamma_0\,^0\,_c\,(\Gamma_a\,^c\,_b - \Gamma_b\,^c\,_a)
\end{equation}
\[
- \nu^2\,\Gamma_c\,^c\,_0\,(\Gamma_a\,^0\,_b - \Gamma_b\,^0\,_a)\}
- \frac{\beta}{\rho + p}\,
(\Gamma_0\,^0\,_a\,s_b - \Gamma_0\,^0\,_b\,s_a).
\]
{}From this follows that the last two equations of (\ref{1flredequ}) are
given by
\begin{equation}
\label{2Gred}
\partial_t\,\Gamma_0\,^0\,_a - \nu^2\,e_{c}(\Gamma_a\,^c\,_0) =
- \Gamma_0\,^0\,_c\,\Gamma_a\,^c\,_0
\end{equation}
\[
- \left(\frac{\rho + p}{\nu^2}
\,(\frac{\partial^2\,p}{\partial\,\rho^2})_s - \nu^2 \right)
\,\Gamma_c\,^c\,_0\,\Gamma_0\,^0\,_a
+ \frac{\alpha}{\rho + p}\,\Gamma_c\,^c\,_0\,s_a
\]
\[
+ \nu^2\,\left(\Gamma_k\,^c\,_0\,(\Gamma_a\,^k\,_c - \Gamma_c\,^k\,_a)
- \Gamma_a\,^c\,_k\,\Gamma_c\,^k\,_0
+ \Gamma_c\,^c\,_k\,\Gamma_a\,^k\,_0 + R^c\,_{0ac}\right),
\]

\begin{equation}
\label{3Gred}
\nu^2\,\partial_t\,\Gamma_a\,^0\,_b - \nu^2\,e_b(\Gamma_0\,^0\,_a) =
\nu^2\,\left(\Gamma_k\,^0\,_b\,(\Gamma_0\,^k\,_a
- \Gamma_a\,^k\,_0)\right.
\end{equation}
\[
\left.- \Gamma_0\,^0\,_c\,\Gamma_a\,^c\,_b + R^0\,_{b0a}
+ \Gamma_0\,^0\,_c\,(\Gamma_a\,^c\,_b - \Gamma_b\,^c\,_a)
\right.
\]
\[
\left. + \nu^2\,\Gamma_c\,^c\,_0\,(\Gamma_a\,^0\,_b - \Gamma_b\,^0\,_a)
+ \frac{\beta}{\rho + p}\,
(\Gamma_0\,^0\,_a\,s_b - \Gamma_0\,^0\,_b\,s_a)\right).
\]
The remaining equations of (\ref{1flredequ}) and (\ref{2flredequ}) (besides
$P_{ij} = 0$,
$Q_{kl} = 0$) again only contain derivatives in the direction of $U$.

\subsubsection{The Derivation of the Subsidiary
Equations}\label{subsidy}

We have to show that any solution to the reduced equations which
satisfies the
constraints on an initial hypersurface, i.e. for which $z = 0$ on the
initial
hypersurface, will satisfy $z = 0$ in the domain of dependence of the
initial
hypersurface with respect to the metric supplied by the solution. For
this purpose we
will derive a system of partial differential equations for those
components of $z$
which do not vanish already because of the reduced equations and the
gauge conditions.

We begin by deriving equations for $F_{jkl}$. There exist two different
expressions for
$F_{kl} \equiv \nabla^j\,F_{jkl}$. {}From the definition of $F_{jkl}$ and
from the
symmetries of the tensor field involved follows

\[
F_{kl} = \nabla^j\,\nabla^i\,F_{ijkl}
= \nabla^j\,\nabla^i\,C_{ijkl} - \nabla^j\,\nabla_{[k}\,S_{l]j}
\]
\[
=
- R^p\,_{[i}\,^{ij}\,C_{j]pkl}
+ C_{ijp[l}\, R^p\,_{k]}\,^{ij}
+ \frac{1}{2}\,T_i\,^p\,_j\,\nabla_p\,C^{ij}\,_{kl}
\]
\[
+ \nabla_{[k}\,\nabla^j\,S_{l]j}
- R^p\,_{[l}\,^j\,_{k]}\,S_{pj}
- R^p\,_j\,^j\,_{[k}\,S_{l]p}
+ \nabla_p\,S^j\,_{[l}\,\,T_{k]}\,^p\,_j,
\]
where we took into account that we do not know at this stage whether the
connection
coefficients $\Gamma_i\,^j\,_k$ supplied by the solution define a torsion
free
connection. {}From the reduced field equations, the definition of the zero
quantities,
and the symmetries of $C_{ijkl}$, it follows that
\begin{equation}
\label{Fexpr}
F_{kl} =
- \Delta^p\,_{[i}\,^{ij}\,C_{j]pkl}
+ C_{ijp[l}\,\Delta^p\,_{k]}\,^{ij}
+ \frac{1}{2}\,T_i\,^p\,_j\,\nabla_p\,C^{ij}\,_{kl}
\end{equation}
\[
+ \kappa\,J_{kl}
+ \Delta^p\,_{[l}\,^j\,_{k]}\,S_{pj}
- \Delta^p\,_j\,^j\,_{[k}\,S_{l]p} = N(z),
\]
where $N(z)$ is, as in the following, a generic symbol for a smooth
function (which
may change from equation to equation) of the zero quantities which
satisfies $N(0) =
0$.

On the other hand, because of the reduced equations, equation
(\ref{2Fdeco}) takes  the form
\begin{equation}
\label{3Fdeco}
F_{jkl} = 2\,U_j\,P_{[k}\,U_{l]}
+ h_{j[k}\,P_{l]}
+ Q_i\,(U_j\,\epsilon^i\,_{kl}
- \epsilon^i\,_{j[k}\,U_{l]})
- \frac{1}{2}\,\kappa\,h_{j[k}\,q_{l]}.
\end{equation}
Contracting with $\nabla^j$, decomposing the resulting expression into
$F'_{kU}$, $F'_{kl}$ and equating with the corresponding expressions
obtained from (\ref{Fexpr}), we arrive at equations of the form
\begin{equation}
\label{1Fsubeq}
{\cal L}_U\,P_k + \frac{1}{2}\,\epsilon_k\,^{ij}\,{\cal D}_i\,Q_j = N(z)
\end{equation}
\begin{equation}
\label{2Fsubeq}
{\cal L}_U\,Q_k - \frac{1}{2}\,\epsilon_k\,^{ij}\,{\cal D}_i\,P_j = N(z).
\end{equation}

The connection defined by the $\Gamma_i\,^j\,_k$ and the associated
torsion and curvature
tensors satisfy the first Bianchi identity
\[
\sum_{(jkl)} \nabla_j\,T_k\,^i\,_l =
\sum_{(jkl)} (R^i\,_{jkl} + T_j\,^m\,_k\,T_l\,^i\,_m),
\]
where $\sum_{(jkl)}$ denotes the sum over the cyclic permutation of the
indices $jkl$.
Setting here $j = 0$, observing that the symmetries of $C^i\,_{jkl}$,
$S_{kl}$ imply
$\sum_{(jkl)} R^i\,_{jkl} = \sum_{(jkl)} \Delta^i\,_{jkl}$, and taking
into account the
reduced equations, we get from this an equation of the form
\begin{equation}
\label{Gsubeq} e_0(T_a\,^i\,_b) = N(z).
\end{equation}

To derive equations for $\Delta^i\,_{jkl}$, we use the second Bianchi
identity
\begin{equation}
\label{2Bianchi}
\sum_{(jkl)} \nabla_j\,R^i\,_{mkl} =
- \sum_{(jkl)} R^i\,_{mnj}\,T_k\,^n\,_l.
\end{equation}
We write
\[
R^i\,_{jkl} =
\Delta^i\,_{jkl} + C^i\,_{jkl} + E^i\,_{jkl} + G^i\,_{jkl},
\]
with $E^i\,_{jkl} = g^i\,_{[k}\,S^{*}_{l]j} - g_{j[k}\,S^{*}_{l]}\,^i$,
$G^i\,_{jkl} = \frac{1}{2}\,S\,g^i\,_{[k}\,g_{l]j}$ and
$S^{*}_{jk} = S_{jk} - \frac{1}{4}\,S\,g_{jk}$. Using the well known
facts that
the left and right duals of $C_{ijkl}$ and $G_{ijkl}$ are equal
while the left dual of $E_{ijkl}$ differs from its right dual by a sign,
and using
the reduced equations, we get
\[
\sum_{(jkl)} \nabla_j\,R^i\,_{mkl} =
\sum_{(jkl)} \nabla_j\,\Delta^i\,_{mkl}
\]
\[
+ \frac{1}{2}\,\epsilon_{jkl}\,^p\,(\nabla_q\,C^q\,_{prs}
- \nabla_q\,E^q\,_{prs} + \nabla_q\,G^q\,_{prs})\,\epsilon_m\,^{irs}
\]
\[
= \sum_{(jkl)} \nabla_j\,\Delta^i\,_{mkl}
+ \frac{1}{2}\,\epsilon_{jkl}\,^p\,(F_{prs} +
\kappa\,g_{pr}\,q_s )\,\epsilon_m\,^{irs},
\]
whence, by (\ref{3Fdeco}) and (\ref{2Bianchi}),
\begin{equation}
\label{1Deq}
\sum_{(jkl)} \nabla_j\,\Delta^i\,_{mkl} = N(z).
\end{equation}
i) In the case of pressure free matter we get from (\ref{1Deq}), by
setting $j = 0$ and
using (\ref{dredequ}), an equation of the form
\begin{equation}
\label{dustDsubseq}
e_0(\Delta^i\,_{kab}) = N(z).
\end{equation}
Since, as remarked earlier, $q_k = 0$ by our gauge conditions, the system
of equations consisting of
(\ref{1Fsubeq}), (\ref{2Fsubeq}), (\ref{Gsubeq}), and (\ref{dustDsubseq})
constitutes the desired \lq subsidiary system\rq\ for the zero quantities
in the pressure free case.

ii) Using (\ref{1flredequ}), we get in the general case by the analogous
procedure only an equation of the form
\begin{equation}
\label{1Dsubseq}
e_0(\Delta^a\,_{bcd}) = N(z).
\end{equation}
By the antisymmetry of $J'_{ab}$ we know already that
\[
\Delta^0\,_{a0b} + \Delta^0\,_{b0a} = 0.
\]
We can express the equations for $\Delta^0\,_{[b|0|c]}$,
$\Delta^0\,_{abc}$.
in terms of the quantities
\begin{equation}
\label{Ddecomps}
\Delta_a = \frac{1}{2}\,\epsilon_a\,^{bc}\,\Delta^0\,_{b0c},\,\,\,\,\,\,
\Delta^{*}_{ab} = \frac{1}{2}\,\epsilon_{(a}\,^{cd}\,
\Delta^0\,_{b)cd},\,\,\,\,\,\,
\Delta^{*}_a = \frac{1}{2}\,\Delta^c\,_{0ac},
\end{equation}
because
\[
\Delta^0\,_{[b|0|c]} = - \Delta_a\,\epsilon^a\,_{bc},\,\,\,\,\,\,
\Delta^0\,_{abc} = - \Delta^{*}_{ad}\,\epsilon_{bc}\,^d + 2\,h_{a[b}\,
\Delta^{*}_{c]}.
\]
{}From (\ref{1Deq}) we get
\[
e_0\,(\Delta^0\,_{abc}) + e_c\,(\Delta^0\,_{a0b})
+ e_b\,(\Delta^0\,_{ac0})
= N(z),
\]
which implies for the quantities (\ref{Ddecomps}) equations
\[
2\,e_0\,(\Delta^{*}_a) - \epsilon_a\,^{bc}\,e_b\,(\Delta_c) = N(z),
\]
and
\[
e_0\,(\Delta^{*}_{ab}) - e_{(a}\,(\Delta_{b)})
+ h_{ab}\,h^{cd}\,e_c\,(\Delta_d)
= N(z).
\]
Since we have
\[
h^{ab}\,e_a\,(\Delta_b)
= - \frac{1}{2}\,\epsilon^{abc}\,e_a\,(\frac{1}{\rho + p}\,J'_{bc})
= - \frac{1}{2}\,\frac{1}{\rho + p}\,\epsilon^{ijk}\,\nabla_i\,
\nabla_j\,q_k + N(z)
= N(z),
\]
we can write the second equation in the form
\[
e_0\,(\Delta^{*}_{ab}) - e_{(a}\,(\Delta_{b)}) = N(z).
\]
{}From (\ref{1Deq}) we get furthermore
\[
e_a\,(\Delta^0\,_{bcd}) + e_d\,(\Delta^0\,_{bac})
+ e_c\,(\Delta^0\,_{bda})
= N(z),
\]
which implies in terms of the quantities (\ref{Ddecomps}) an equation of
the form
\[
h^{ab}\,e_a\,(\Delta^{*}_{bc}) - \epsilon_c\,^{ab}\,e_a\,(\Delta^{*}_b).
\]
By a direct calculation we derive from (\ref{Jex}) the equation
\[
2\,{\cal L}_U\,J'_{ij} = 4\,{\cal D}_{[i}\,J'_{Uj]}
- h_i\,^p\,h_j\,^q\,(\Delta^l\,_{npq} + \Delta^l\,_{qnp}
+ \Delta^l\,_{pqn})\,U^n\,q_l
\]
\[
- h_i\,^p\,h_j\,^q\,(T_l\,^n\,_p\,\nabla_n\,q_q
+ T_q\,^n\,_l\,\nabla_n\,q_p + T_p\,^n\,_q\,\nabla_n\,q_l)\,U^l
\]
\[
- 2\,a_i\,J'_{Uj} + 2\,a_j\,J'_{Ui},
\]
which can be rewritten by (\ref{1flredequ}) in the form
\begin{equation}
\label{(r+p)D}
(\rho + p)\,\{e_0\,(\Delta_a) + 2\,\nu^2\,\epsilon_a\,^{bc}\,
e_b\,(\Delta^{*}_c)\}
= N(z).
\end{equation}
{}From the equations above we obtain the system
\begin{equation}
\label{2Dsubseq}
(\rho + p)\,\{2\,\nu^2\,e_0\,(\Delta^{*}_a)
- \nu^2\,\epsilon_a\,^{bc}\,e_b\,(\Delta_c)\} = N(z),
\end{equation}
\begin{equation}
\label{3Dsubseq}
(\rho + p)\,\{e_0\,(\Delta_a) + \nu^2\,\epsilon_a\,^{bc}\,e_b\,
(\Delta^{*}_c)
+ \nu^2\,h^{bc}\,e_b\,(\Delta^{*}_{ca})\} = N(z),
\end{equation}
\begin{equation}
\label{4Dsubseq}
c_{ab}\,(\rho + p)\,\{\nu^2\,e_0\,(\Delta^{*}_{ab})
- \nu^2\,e_{(a}\,(\Delta_{b)})\} = N(z),
\end{equation}
where $c_{ab} = 1$ if $a = b$ and $c_{ab} = 2$ if $a \neq b$.
Finally, we obtain from (\ref{Jex}), (\ref{1flredequ}) the equation
\begin{equation}
\label{qksusseq}
{\cal L}_U\,q_a = 2\,\nu^2\,(\rho + p)\,\Delta^c\,_{0ac}.
\end{equation}
Equations (\ref{1Fsubeq}), (\ref{2Fsubeq}), (\ref{Gsubeq}), (\ref{1Dsubseq}),
(\ref{2Dsubseq}), (\ref{3Dsubseq}), (\ref{4Dsubseq}), and (\ref{qksusseq})
constitute the
subsidiary equations for the zero quantities in the general case.

We note here that in the reduced system it has not been built in explicitly
that the energy-momentum tensor has vanishing divergence (cf. equation
(\ref{Tdiv})). While we assumed the equation $q = 0$ as part of the reduced
equations, we verify the vanishing of the quantity $q_k$ by deriving
the subsidiary equations and using the uniqueness property for these equations.

In the present formalism the gauge conditions are taken care of by the
explicit form
of some of the unknowns, however the list of constraints is much longer
than in the
previous discussions.  We shall not try to demonstrate that the
constraints are
preserved in the specific case of \lq floating fluid balls\rq.. Though the
construction of
data for fluid balls of compact support which are embedded in
asymptotically vacuum
data has been shown and their smoothness properties near the boundary
have been
discussed \cite{nagy}, the evolution in time of these data and the
precise smoothness
properties of the fields near and possible jumps travelling along the
boundary have
not been worked out yet. However, without a precise understanding of the
behaviour of
the solution near the boundary the conservation of the constraints cannot
be
demonstrated.

Our reduced system also found applications in cosmological context where
the fluid is
spread out, with $\rho + p > 0$, over the time slices
(cf. \cite{elst:ellis},
\cite{reula}). In this case the desired conclusion follows from the fact
that the
subsidiary systems are symmetric hyperbolic, have right hand sides of the
form
$N(z)$, and the characteristics of the reduced system and the subsidiary
system are
as follows (where we use only the frame components
$\xi_k = \xi_{\mu}\,e^{\mu}\,_k$
of the covector $\xi$).

(i) In the case of pressure free matter the characteristic polynomial of
the reduced
system is of the form
\[
c\,(\xi_0)^K\,(\xi^2_0 + \frac{1}{4}\,h^{ab}\,\xi_a\,\xi_b)^L
\,(g^{\mu \nu}\,\xi_{\mu}\,\xi_{\nu})^N,
\]
with positive integers $K$, $L$, $N$ and constant factor $c$, while the
characteristic
polynomial of the subsidiary system only contains the first two factors.
Thus the
characteristics of the subsidiary system are timelike with respect to
$g_{\mu \nu}$
(cf. also the remarks in Sect. \ref{basics}).

(ii) In the general case the characteristic polynomial of the reduced
equations is of
the form
\begin{equation}
\label{flcharcone}
c\,(\xi_0)^K\,(\xi^2_0 + \frac{1}{4}\,h^{ab}\,\xi_a\,\xi_b)^L
\,(\xi^2_0 + \nu^2\,h^{cd}\,\xi_c\,\xi_d)^M\,(g^{\mu \nu}\,
\xi_{\mu}\,\xi_{\nu})^N,
\end{equation}
with positive integers $K$, $L$, $M$, $N$ and constant factor $c$, while
the
characteristic polynomial of the subsidiary system is generated by powers
of the
first three factors.

We note that the equations (\ref{2Gred}) and (\ref{3Gred}) contribute to the
principal symbol the
third factor which corresponds to the sound cone pertaining to the fluid.
If one wants to ensure that the sound does not travel with superluminal
speed
one has to require the equation of state (\ref{eqst}) to be such that
$\nu \le 1$.
For the question of existence and uniqueness of solutions there is no
need to impose
such a condition, but if $\nu > 1$, the domain of dependence with respect
to $g_{\mu \nu}$, as it has been defined in Sect. \ref{basics}, cannot be
shown any longer,
by the arguments given in Sect. \ref{basics}, to be also a domain of
uniqueness. In any case, it
follows from the characteristic polynomials that in a domain where the
solution of
the reduced system is unique according to those arguments, the
constraints will be
satisfied if they hold on the initial hypersurface.

We note that the system simplifies considerably in the isentropic case.
In the
reduced system the function $\rho + p$ then neither occurs in the
principal part
nor in a denominator. In the case of the subsidiary system a more
detailed discussion
is required to understand the consequences of the occurrence of the
various factors
$\rho + p$.

In our procedure the fluid equations serve two purposes, they determine
the motion of
the fluid as well as the evolution of the frame. If we set $\kappa = 0$
in all
equations the fluid equations decouple from the geometric equations and
we obtain a
new hyperbolic reduction of the vacuum field equations. In this procedure
any exotic \lq equation of state\rq\ may be prescribed as long as it ensures
a useful,
long-lived gauge.

If the initial data for the Einstein--Euler equations are such that $U$,
$\rho$, $p$,
$n$, $s$, and the equation of state can be smoothly extended through
the boundaries
of the fluid balls, this suggests using the \lq extended fluid\rq\ to control
the
evolution of the gauge in the vacuum part of the solution near the
boundary.

\subsection{The Initial Boundary Value Problem}\label{ibvp}

In the previous section we studied a problem involving a distinguished
timelike
hypersurface. Its evolution in time was determined by a physical process.
There are also
important problems where the Einstein equations are solved near timelike
hypersurfaces
which are prescribed for practical reasons, e.g. to perform numerical
calculations on finite
grids. The underlying initial boundary value problem for Einstein's field
equations, where
initial data are prescribed on a (spacelike) hypersurface $S$ and
boundary data a
(timelike) boundary $T$ which intersect at a $2$-surface
$\Sigma = T \cap S$, has been
analysed in detail in the article \cite{friedrich:nagy}. The solution to
this problem
requires a hyperbolic reduction which needs to satisfy, beyond the
conditions discussed at
the beginning of this section, certain side conditions. In the following
we want to comment
on those aspects of the work in \cite{friedrich:nagy} which illustrate
the flexibility of the
field equations in performing reductions and on certain characteristics
of the reduced
system. For the full analysis of the initial boundary value problem we
refer to
\cite{friedrich:nagy}.

Since we are dealing with a problem for equations which are essentially
hyperbolic, the
problem can be localized. In suitably adapted coordinates $x^{\mu}$,
defined on some
neighbourhood of a point $p \in \Sigma$, the manifold $M$ on which the
solution is to be
determined will then be given in the form
$M = \{ x \in {\R}^4|x^0 \ge 0, x^3 \ge 0 \}$, the
initial hypersurface by $S = \{x \in M | x^0 = 0\}$ and the boundary by
$T = \{x \in M | x^3 = 0\}$. Clearly, we will have to prescribe Cauchy
data on $S$ as
before, we will have to analyse which kind of boundary data
are admitted by the
equations, and on the edge $\Sigma = \{x \in M | x^0 = 0,\,\,x^3 = 0\}$
the data will have
to satisfy some consistency conditions, as is always the case in
initial boundary value
problems.

\subsubsection{Maximally Dissipative Initial Boundary Value Problems}
\label{maxdis}

We have seen in Sect. \ref{symhyp} that energy estimates provide
a basic tool for obtaining
results about the existence and uniqueness of solutions to symmetric
hyperbolic
systems. To explain the side conditions which have to be satisfied in a
hyperbolic
reduction of an initial boundary value problem for Einstein's field
equations, we
consider what will happen if we try to obtain energy estimates in the
present situation.
Assume that we are given on $M$ in the coordinates $x^{\mu}$ a linear
symmetric hyperbolic
system of the form
\begin{equation}
\label{lsyhy}
A^{\mu}\partial_{\mu}\,u = B\,u + f(x),
\end{equation}
for an ${\R}^N$-valued unknown $u$. The matrices
$A^{\mu} = A^{\mu}(x)$, $\mu = 0, 1, 2, 3$, are smooth functions on $M$
which take values in the set of symmetric $N \times N$-matrices, there
exists a
1-form $\xi_{\mu}$ such that $A^{\mu}\,\xi_{\mu}$ is positive definite,
$B = B(x)$ is a smooth matrix-valued function and $f(x)$ a smooth
${\R}^N$-valued function on $M$. For convenience we assume that the
positivity condition is satisfied with $\xi_{\mu} = \delta^0\,_{\mu}$.

If we assume that $u$ vanishes for large positive values of $x^{\alpha}$,
$\alpha = 1, 2, 3$, and if the relation
\[
\partial_{\mu}( ^tu \, A^{\mu} \, u) = \,^tu\,K\,u + 2\,^tu\,f
\quad\mbox{with}\quad
K = B + \,^tB + \partial_{\mu}\,A^{\mu},
\]
implied by (\ref{lsyhy}), is integrated over a set
$M_{\tau} = \{ x \in M | 0 \le x^0 \le \tau \}$, defined by some number
$\tau \ge 0$, we
obtain the relation
\[
\int_{S_{\tau}}\,^tu\,A^0\,u\,dS =
\int_S \,^tu\,A^0\,u\,dS +
\int_{M_{\tau}} \{^tu\,K\,u + 2\,^tu\,f\}\,dV
+ \int_{T_{\tau}}\,^tu\,A^3\,u\,dS,
\]
involving boundary integrals over
$S_{\tau} = \{ x \in M | x^0 = \tau \}$ and
$T_{\tau} = \{ x \in M | 0 \le x^0 \le \tau, x^3 = 0 \}$.
Obviously, the structure of the {\it normal matrix} $A^3$ plays a
prominent role here. If the
last term on the right hand side is non-positive, we can use the
equation above to obtain
energy estimates for proving the existence and uniqueness of solutions.

By this (and certain considerations which will become clear when we have
set up our reduced
system) we are led to consider the following {\it maximally dissipative
boundary value
problem}.

We choose $g \in C^{\infty}(S, {\R}^N)$ and require as initial condition
$u(x) = g(x)$ for $x \in S$. We choose a smooth map $Q$ of $T$ into the
set of linear
subspaces of ${\R}^N$ and require as boundary condition
$u(x) \in Q(x)$ for $x \in T$. The type of map $Q$ admitted here is
restricted by the
following assumptions.

(i) The set $T$ is a characteristic of (\ref{lsyhy}) of constant
multiplicity, i.e.
\[
{\rm dim}({\rm ker}\,A^3(x)) = {\rm const.} > 0,\,\,\,\,\, x \in T.
\]

(ii) The map $Q$ is chosen such as to ensure the desired non-positivity
\[
^tu\,A^3(x)\,u \le 0,\,\,\,\,\,u \in Q(x),\,\,\,\,\, x \in T.
\]

(iii) The subspace $Q(x)$, $x \in T$, is a maximal with (ii), i.e.
the dimension of $Q(x)$ is equal to
number of non-positive eigenvalues of $A^3$ counting multiplicity.

The specification of $Q$ can be expressed in terms of linear equations.
Since $A^3$ is
symmetric, we can assume, possibly after a transformation of the
dependent variable,
that at a given point $x \in T$
\[
A^3 = \kappa\,\left[ \begin{array}{ccc}
- I_j & 0 & 0 \\
0 & 0_k & 0\\
0 & 0 & I_l
\end{array} \right],\,\,\,\,\,\,\kappa > 0,
\]
where $I_j$ is a $j \times j$ unit matrix, $0_k$ is a $k \times k$ zero
matrix etc.
and $j + k + l = N$. Writing
$u = \,^t(a,b,c) \in {\R}^j \times {\R}^k \times {\R}^l$ we find
that at $x$ the
linear subspaces admitted as values of $Q$ are neccessarily given by
equations of
the form $0 = c - H\,a$ where $H = H(x)$ is a $l \times j$ matrix
satisfying
\[
- \,^ta \, a + \,^ta \, ^tH \,H \,a \le 0, \,\,\,\,\, a \in {\R}^j,
\quad\mbox{i.e.}\quad ^tH\,H \le I_j.
\]
We note that there is no freedom to prescribe data for the component $b$
of $u$
associated with the kernel of $A^3$. More specifically, if $A^3 \equiv 0$
on $T$,
energy estimates are obtained without imposing conditions on $T$ and the
solutions are determined uniquely by the initial condition on $S$. By
subtracting a suitable
smooth function from $u$ and redefining the function $f$, we can convert
the homogeneous
problem above to an inhomogeneous  problem and vice versa. Inhomogeneous
maximal dissipative
boundary conditions are of the form
\begin{equation}
\label{bdrycond}
q = c - H \, a,
\end{equation}
with $q = q(x)$, $x \in T$, a given ${\R}^l$-valued function representing
the free boundary data on $T$.

Maximally dissipative boundary value problems as outlined above have been
worked out in
detail in \cite{rauch}, \cite{secchi:I} for the linear case and in
\cite{gues}
\cite{secchi:II} for quasi-linear problems (see also these articles for
further references).
If we want to make use of this theory to analyse the initial boundary
value problem for
Einstein's field equations we will have to solve two problems which go
beyond what is known from the standard Cauchy problem.

(i) We will have to find a reduction, involving a symmetric hyperbolic
system, which gives us
sufficient information on the normal matrix so that we can control the
conditions above.

(ii) To demonstrate the preservation of the constraints we will have to
discuss an
initial boundary value problem for the subsidiary system. This should be
such as to admit a
uniqueness proof. Moreover, there is the problem of getting sufficient
control on the
solution near $T$. While the initial data for the reduced system will of
course be
arranged such that the constraints are satisfied on the initial
hypersurface $S$, it will a
priori not be clear that we will have sufficient information on the
behaviour of the
solution to the reduced equations and on the data on $T$ in order to
conclude that the constraints will be satisfied on $T$.

The choice of representation of the field equations, of the gauge
conditions and the gauge
source functions, and, in particular, the choice of the reduced
equations will largely be
dominated by the second problem.

\subsubsection{The Representation of the Einstein Equations}\label{rep}

In \cite{friedrich:nagy} the initial boundary value problem for
Einstein's vacuum field
equations was analysed in terms of the equations
\begin{equation}
\label{einstrepr}
T_i\,^k\,_j = 0,\,\,\,\,\,\Delta^i\,_{jkl} = 0,\,\,\,\,\,F_{jkl} = 0,
\end{equation}
of the previous section with everywhere vanishing energy-momentum tensor.
We shall use these equations together with the conventions and
notation introduced in the previous section.

\subsubsection{The Gauge Conditions}\label{gauge}

The gauge, which we assume here for simplicity extends to all
of $M$, has been
chosen as follows. On the initial hypersurface $x^0 = 0$ and $x^{\alpha}$,
$\alpha = 1, 2, 3$, are coordinates with $x^3 = 0$ on $\Sigma$ and
$x^3 > 0$ elsewhere. The
timelike unit vector field $e_0$ on $M$ is tangent to $T$,  orthogonal
to the 2-surfaces
$S_c = \{x^3 = c = {\rm const.} > 0\}$ in  $S$, and it points towards $M$ on
$S \cap U$. The
coordinates $x^{\mu}$ satisfy  $e^{\mu}\,_0 = e_0(x^{\mu})
= \delta^{\mu}\,_0$ on $M$ and
the sets $T_c = \{x^3 = c\}$ are smooth timelike  hypersurfaces of $M$
with $T_0 = T$. The
unit vector field $e_3$ is normal to the hypersurfaces $T_c$ and points
towards $M$ on $T$.
The vector fields $e_A$, $A = 1, 2$, are tangent to $T_c \cap S$ and such
that they form
with $e_0$, $e_3$ a smooth orthonormal frame field on $S$. On the
hypersurfaces $T_c$ these fields
are Fermi transported in the direction of $e_0$ with respect to the
Levi--Civita connection
$D$ defined by the metric induced on $T_c$. The $e_k$ form a smooth
orthonormal frame field
on $U$. We refer this type of gauge as an \lq adapted gauge\rq.
Notice that it leaves a freedom to choose the timelike vector field
$e_0$ on $M \setminus S$.

In analysing the initial boundary value problem it will be necessary
to distinguish between
interior equations on the submanifolds
$S$, $T$, $T_c$, $\Sigma$, $S_c$. Since our frame is adapted to these
submanifolds, this
can be done by distinguishing four groups of indices. They are given,
together with the
values they take, as follows
\[
a, c, d, e, f = 0, 1, 2;\,\,\,\,\,\,i,j,k,l,m,n = 0, 1, 2, 3;
\]
\[
p, q, r, s, t = 1, 2, 3;\,\,\,\,\,\,A, B, C, D = 1, 2.
\]
We assume the summation convention for each group.

By our conditions the frame coefficients $e^{\mu}\,_k$ satisfy
\begin{equation}
\label{egauge}
e^{\mu}\,_0 = \delta^{\mu}\,_0,\,\,\,\,\,\,e^3\,_a = 0,\,\,\,\,\,\,
e^3\,_3 > 0 \quad\mbox{on}\quad M,
\end{equation}
while the frame connection coefficients satisfy
\begin{equation}
\label{Fgauge}
\Gamma_0\,^A\,_B = 0.
\end{equation}
The fields $e_a$ satisfy on $T_c$ the equations
\begin{equation}
\label{frame}
D_{e_0}\,e_0 = \Gamma_0\,^A\,_0\,e_A,\,\,\,\,\,
D_{e_0}\,e_A = - g_{AB}\,\Gamma_0\,^B\,_0\,e_0.
\end{equation}
Thus, given the hypersurfaces $T_c$, the evolution of the coordinates
$x^{\alpha}$,
$\alpha = 0, 1, 2$, and the frame vector fields $e_a$ off $S$ is governed
by the coefficients
$\Gamma_0\,^A\,_0$.

Another part of the connection coefficients defines the intrinsic connection
$D$ on $T_c$, since
$D_a\,e_c = D_{e_a}\,e_c = \Gamma_a\,^b\,_c\,e_b$. The remaining
connection coefficients,
given by
\begin{equation}
\label{ftrgauge}
\chi_{ab} = g(\nabla_{e_a}\,e_3, e_b) = \Gamma_a\,^j\,_3\,g_{jb}
= \Gamma_a\,^3\,_b = \Gamma_{(a}\,^3\,_{b)},
\end{equation}
define the second fundamental form of the hypersurfaces $T_c$ in the
frame $e_a$.
In the reduced equations, the symmetry of $\chi_{ab}$ has to be taken
into account
explicitly. A special role is played by mean extrinsic curvature
\begin{equation}
\label{mec}
\chi \equiv g^{ab}\,\chi_{ab} = g^{jk}\,\Gamma_j\,^3\,_k =
\nabla_{\mu}\,e^{\mu}\,_3,
\end{equation}
since it can be regarded as the quantity controlling the evolution of the
hypersurfaces
$T_c$ and thus of the coordinate $x^3$.

We now choose two smooth functions $F^A \in C^{\infty}(M)$ as gauge
source functions. These will occur explicitly in the reduced equations and
will play the role of connection coefficients for the solution, namely
$F^A = \Gamma_0\,^A\,_0$. Furthermore we will choose a
function $f \in C^{\infty}(M)$ which will play the role of the mean
extrinsic curvature
on the hypersurfaces $T_c$. Here the interpretation is somewhat more
complicated.
On $T$ the function $\chi = f|_T$ must be regarded as the free datum
which, together with
certain data on $\Sigma$, indirectly specifies the boundary $T$. However,
for $x^3 > 0$ the
function $f$ plays the role of a gauge source function which determines
the gauge dependent
hypersurface $T_c$, $c > 0$. It is a remarkable feature of the Codazzi
equations
that they admit this freedom while at the same time implying hyperbolic
equations.

This example clearly shows the importance of the freedom to dispose of
the gauge source
functions. While we could choose $F^A = 0$ locally (cf. the remarks in
\cite{friedrich:nagy}
about certain subtleties arising here), we need the full freedom to make
use of the gauge
source functions $f$, since otherwise we could only handle restricted
types of boundaries.

\subsubsection{The Reduced Equations}\label{redeq}

Using the gauge conditions above, we extract from (\ref{einstrepr}) the
following reduced
system for those components of the unknowns $e^{\mu}\,_k$,
$\Gamma_i\,^j\,_k$,
$E_{ij}$, $B_{kl}$ which are not determined already by the gauge conditions
and the chosen
gauge source functions. Where it has not already been done  explicitly, it is
understood that in
the following equations the connection coefficients $\Gamma_0\,^A\,_0$
and $\chi$ are replaced
by the gauge source functions $F^A$ and $f$ respectively.
The torsion free condition gives
\begin{equation}
\label{eprop}
0 = - T_0\,^k\,_p\,e^{\mu}\,_k =  \partial_t\,e^{\mu}\,_p
- (\Gamma_0\,^q\,_p - \Gamma_p\,^q\,_0)\,e^{\mu}\,_q
- \Gamma_0\,^0\,_p\,\delta^{\mu}\,_0.
\end{equation}
The Gauss equations with respect to $T_c$ provide the
equations
\begin{equation}
\label{G00Aprop}
0 = \Delta^B\,_{00A} =
e_0(\Gamma_A\,^B\,_0) - e_A(F^B)
+ \Gamma_C\,^B\,_0\,\Gamma_A\,^C\,_0
\end{equation}
\[
- \Gamma_A\,^B\,_C\,F^C + F^B\,F^C\,g_{AC}
+ \chi_0\,^B\,\chi_{A0} - \chi_A\,^B\,\chi_{00}
- C^B\,_{00A},
\]

\begin{equation}
\label{GC0Aprop}
0 = \Delta^B\,_{C0A} = e_0(\Gamma_A\,^B\,_C)
+ F^B\,\Gamma_A\,^0\,_C
+ \Gamma_A\,^B\,_0\,F^D\,g_{CD}
\end{equation}
\[
+ \Gamma_D\,^B\,_C\,\Gamma_A\,^D\,_0
+ \chi_0\,^B\,\chi_{AC} - \chi_A\,^B\,\chi_{0C}
- C^B\,_{C0A}.
\]
Codazzi's equations with respect to $T_c$ imply

\begin{equation}
\label{chi01prop}
0 = g^{ab}\,\Delta^3\,_{ab1}  =
D_0\,\chi_{01} - D_1\,\chi_{11} - D_2\,\chi_{12}
- D_1(f),
\end{equation}

\begin{equation}
\label{chi02prop}
0 = g^{ab}\,\Delta^3\,_{ab2} =
D_0\, \chi_{02} - D_1\,\chi_{12} - D_2\,\chi_{22}
- D_2(f),
\end{equation}

\begin{equation}
\label{chi11prop}
0 = \Delta^3\,_{101} =
D_0\,\chi_{11} - D_1\,\chi_{01} - C^3\,_{101},
\end{equation}

\begin{equation}
\label{chi12prop}
0 = \Delta^3\,_{201} + \Delta^3\,_{102} =
2\,D_0\,\chi_{12} - D_1\,\chi_{02} - D_2\,\chi_{01}
- C^3\,_{201} - C^3\,_{102},
\end{equation}

\begin{equation}
\label{chi22prop}
0 = \Delta^3\,_{202} =
D_0\,\chi_{22} - D_2\,\chi_{02} - C^3\,_{202},
\end{equation}
where it is understood that the component $\chi_{00}$, which
appears only in undifferentiated form, is given by
$\chi_{00} = \chi_{11} +\chi_{22} + f$.
The remaining Ricci identities give

\begin{equation}
\label{G3ABprop}
0 = \Delta^A\,_{B03} = e_0(\Gamma_3\,^A\,_B)
+ F^A\,\Gamma_3\,^0\,_B + \Gamma_3\,^A\,_0\,F^C\,g_{BC}
+ \Gamma_C\,^A\,_B\,\Gamma_3\,^C\,_0
\end{equation}
\[
+ \Gamma_3\,^A\,_B\,\Gamma_3\,^3\,_0
+ \chi_0\,^A\,\Gamma_3\,^3\,_B
- \Gamma_3\,^A\,_3\,\chi_{0B}
- \Gamma_C\,^A\,_B\,\chi_0\,^C
- C^A\,_{B03},
\]

\begin{equation}
\label{G3A0prop}
0 = \Delta^A\,_{003} = e_0(\Gamma_3\,^A\,_0) - e_3(F^A)
+ \chi_0\,^A\,\Gamma_3\,^3\,_0
- \Gamma_3\,^A\,_B\,F^B
+ \Gamma_B\,^A\,_0\,\Gamma_3\,^B\,_0
\end{equation}
\[
+ \Gamma_3\,^A\,_0\,\Gamma_3\,^3\,_0
- \Gamma_B\,^A\,_0\,\chi_0\,^B
- \Gamma_3\,^3\,_B\,g^{BA}\,\chi_{00}
- F^A\, \chi_{00}
- C^A\,_{003},
\]

\begin{equation}
\label{G33Bprop}
0 = \Delta^3\,_{A03} + \Delta^3\,_{03A}
= e_0(\Gamma_3\,^3\,_A) - e_A(\Gamma_3\,^3\,_0)
\end{equation}
\[
+ \Gamma_3\,^3\,_0\,F^B\,g_{BA}
+ \Gamma_3\,^3\,_C\,\Gamma_A\,^C\,_0,
\]

\begin{equation}
\label{G330prop}
0 = g^{ab}\,\Delta^3\,_{ab3} =
e_0(\Gamma_3\,^3\,_0) + g^{AB}\,e_A(\Gamma_3\,^3\,_B) - e_3(f)
\end{equation}
\[
- g^{ab}\,\Gamma_3\,^3\,_k \Gamma_b\,^k\,_a
+ g^{ab}\,\Gamma_b\,^3\,_k\Gamma_3\,^k\,_a
+ g^{ab}\,\Gamma_m\,^3\,_a (\Gamma_3\,^m\,_b - \Gamma_b\,^m\,_3).
\]

In the previous section we saw how to extract a symmetric hyperbolic
system from the
Bianchi identities. However, for reasons given below, we shall not choose
that system here.
Instead we choose the \lq boundary adapted system\rq\
\begin{equation}
\label{basyst}
\begin{array}{rclcrcl}
P_{11}- P_{22} &=& 0 & \hspace{1cm} & Q_{11}- Q_{22} &=& 0 \\
2\,P_{12} &=& 0 &\hspace{1cm} & 2\,Q_{12} &=& 0 \\
P_{11} + P_{22} &=& 0 & \hspace{1cm} & Q_{11} + Q_{22} &=& 0 \\
P_{13} &=& \frac{1}{2} Q_2& \hspace{1cm} &Q_{13} &=&-\frac{1}{2} P_2 \\
P_{23} &=&-\frac{1}{2} Q_1& \hspace{1cm}& Q_{23} &=& \frac{1}{2} P_1,
\end{array}
\end{equation}
written as a system for the unknown vector $u$ which is the transpose
of
\[
((E_{-}, 2E_{12}, E_{+}, E_{13}, E_{23}),
(B_{-}, 2B_{12}, B_{+}, B_{13}, B_{23})).
\]
Here $E_{\pm} = E_{11} \pm E_{22}$, and $B_{\pm} = B_{11} \pm B_{22}$ and
it
is understood that the relations $g^{ij}\,E_{ij} = 0$ and
$g^{ij}\,B_{ij} = 0$ are used everywhere to replace the fields $E_{33}$
and $B_{33}$ by our
unknowns. Written out explicitly, this system takes the form
$\left({\bf I}^{\mu} + {\bf A}^{\mu}\,\right)\,\partial_{\mu}\,u = b$,
with
$$
{\bf I}^{\mu} = \left[
\begin{array}{cc}
I^{\mu} & 0  \\
0 & I^{\mu}
\end{array} \right],
\;\;\;\;\;\;
{\bf A}^{\mu} = \left[
\begin{array}{cc}
 0 & A^{\mu} \\
^TA^{\mu} & 0
\end{array} \right],
$$
where
\begin{equation}
\label{matr}
I^{\mu} = \delta^{\mu}\,_0\left[
\begin{array}{ccccc}
1 & 0 & 0 & 0 & 0 \\
0 & 1 & 0 & 0 & 0 \\
0 & 0 & 1 & 0 & 0 \\
0 & 0 & 0 & 1 & 0 \\
0 & 0 & 0 & 0 & 1
\end{array} \right],
\;\;\;\;\;\;
A^{\mu} = \left[
\begin{array}{ccccc}
0 & - e^{\mu}\,_3 & 0 &  e^{\mu}\,_2 & e^{\mu}\,_1 \\
e^{\mu}\,_3 & 0 & 0 & - e^{\mu}\,_1 & e^{\mu}\,_2 \\
0 & 0 & 0 & e^{\mu}\,_2 & - e^{\mu}\,_1 \\
- e^{\mu}\,_2 & e^{\mu}\,_1 & - e^{\mu}\,_2 & 0 & 0 \\
- e^{\mu}\,_1 &- e^{\mu}\,_2 & e^{\mu}\,_1 & 0 &  0
\end{array} \right].
\end{equation}

The reduced system consisting of (\ref{eprop}) to (\ref{basyst}), is
symmetric hyperbolic. However, beyond that the choice of this particular
system was motivated by the following
specific features.

(i) The theory of maximally dissipative initial value problems applies to
our reduced
equations. In equations (\ref{eprop}) to (\ref{G330prop}) the derivative
$\partial_{x^3}$, which
by our gauge conditions occurs only with the directional derivative
$e_3$, is applied to the
gauge source functions but not to the unknowns, while (\ref{matr}) shows
that we have perfect
control on the non-trivial part of the normal matrix arising from
(\ref{basyst}). Our
discussion of maximally dissipative initial value problems, which led to
(\ref{bdrycond}), and
the form of the matrices (\ref{matr}) suggest that we can prescribe besides
the datum $\chi$,
which characterizes the boundary, precisely two free functions as
boundary data on $T$. This
is confirmed by the detailed discussion in \cite{friedrich:nagy}, though
in general a number of technical details have to be taken care of .

(ii) If instead of (\ref{basyst}) we had chosen the system $P_{ij} = 0$,
$Q_{kl} = 0$ as
equations for the electric and magnetic part of the conformal Weyl
tensor, the theory of
maximally  dissipative initial value problems would also have applied.
We would, however, have
come to the conclusion that besides the mean extrinsic curvature four
functions could be
prescribed freely on $T$. This apparent contradiction is resolved when
one tries to show the
preservation of the constraints, i.e. that those equations contained in
(\ref{einstrepr}) are
satisfied which are not already solved because of the gauge conditions
and the reduced
equations. In the case of the reduced equations above this can be shown
for the following
reason. The subsidiary system splits in this case into a hierachy of
symmetric hyperbolic
subsystems with the following property. The first subsystem has vanishing
normal matrix on
$T$. This implies under suitable assumptions on the domain of the
solution to the reduced
equations that all  unknows in this subsystem must vanish, because the
data on $S$ are of
course arranged such that all constraints are satisfied. Furthermore, it
follows for any
subsystem in the hierarchy that its normal matrix vanishes if the
unknowns of
all previous subsystems in the hierarchy vanish. {}From this the
desired conclusion follows in
a finite number of steps. If we had considered instead the system
$P_{ij} = 0$, $Q_{kl} =
0$, the discussion whether the constraints are preserved would have
become quite complicated
and would have led us in the end to the conclusion that only two
functions are really free on
$T$ while the others are subject to restrictions determined by the
evolution properties
of the reduced system.

We end our discussion of the initial boundary value problem with an
observation
about the characteristics of the reduced system.
Equations (\ref{eprop}) to (\ref{G330prop})
contribute a factor of the form
\[
\xi_0^K\,(\xi_0^2 - \xi_1^2
- \xi_2^2)^L\,(2\,\xi_0^2 - \xi_1^2 - \xi_2^2)^M,
\]
to the characteristic polynomial (using again only the frame components
of the covector
$\xi$). The corresponding characteristics are timelike or null with
respect to $g_{\mu
\nu}$. However, the subsystem (\ref{basyst}) contributes a factor
\[
\xi_0^2\,(\xi_0^2 - \xi_1^2 - \xi_2^2)^2\,
(\xi_0^2 - 2\,\xi_1^2 - 2\,\xi_2^2 - \xi_3^2)^2.
\]
If we denote by $\sigma^j$ the 1-forms dual to the vector fields $e_k$,
so that
$<\sigma^j, e_k>\,= \delta^j\,_k$ and
$g_{\mu \nu} = \sigma^0_{\mu}\,\sigma^0_{\nu} - \sigma^1_{\mu}\,
\sigma^1_{\nu}
- \sigma^2_{\mu}\,\sigma^2_{\nu} - \sigma^3_{\mu}\,\sigma^3_{\nu}$,
the characteristics associated with the third factor in the
polynomial above can be described as the null hypersurfaces with respect
to the metric
\begin{equation}
\label{extcharibvp}
k_{\mu \nu} = \sigma^0_{\mu}\,\sigma^0_{\nu}
- \frac{1}{2}\,\sigma^1_{\mu}\,\sigma^1_{\nu}
- \frac{1}{2}\,\sigma^2_{\mu}\,\sigma^2_{\nu}
- \sigma^3_{\mu}\,\sigma^3_{\nu}.
\end{equation}

\subsection{The Einstein--Dirac System}\label{edirac}

Apparently, not much has been shown so far about the existence of
solutions to the Einstein--Dirac system for general data. The
initial value problem for this system was considered in \cite{bao85a},
but no existence theorem for the evolution equations was proved there.
J. Isenberg has suggested to us that it should be possible to
show the well-posedness of the equations by formulating the equations
as a system of wave equations in a way similar to what was done for
the Cauchy problem for classical supergravity in \cite{bao85b}. However,
this idea has not been worked out in the literature.

We shall indicate here how to obtain symmetric hyperbolic evolution
equations from the Einstein--Dirac system. To avoid lengthy calculations,
we shall not
discuss the complete reduction procedure but only use this system to
illustrate certain
questions arising in the reduction.

We shall write the equations in terms of the $2$-component spin frame
formalism, which may
be thought of as the spinor version of the frame formalism used in the
previous sections.
The fields and the equations will be expressed in terms of a spin frame
$\{\iota_a\}_{a = 0,1}$, which is normalized with respect to the
antisymmetric bilinear form
$\epsilon$ in the sense that it satisfies  $\epsilon_{ab}
= \epsilon(\iota_a, \iota_b)$
with $\epsilon_{01} = 1$. The associated double null frame is given by
$e_{aa'} = \iota_a\,\bar{\iota}_{a'}$, it satisfies $\bar{e}_{aa'}
= e_{aa'}$ and
$g(e_{aa'}, e_{bb'}) = \epsilon_{ab}\,\epsilon_{a'b'}$

All spinor fields (with the possible exception of the basic spin frame
itself and the
vector fields $e_{aa'}$) will be given with respect to the spin frame
above and we
shall use $\epsilon_{ab}$ and $\epsilon^{ab}$, defined by the requirement
$\epsilon_{ab}\,\epsilon^{cb} = \delta_a\,^c$ (the Kronecker symbol), to
move indices
according to the rule $\omega^a = \epsilon^{ab}\,\omega_b$, $\omega_a =
\omega^b\,\epsilon_{ba}$.

We use the covariant derivative operator $\nabla$ acting on spinors,
which is derived from
the Levi--Civita connection of $g$ and satisfies $\nabla\,\epsilon_{ab}
= 0$, to define
connection coefficients $\Gamma_{aa'bc} = \Gamma_{aa'(bc)}$ by
\[
\nabla_{aa'}\,\iota_b = \nabla_{e_{aa'}}\,\iota_b
= \Gamma_{aa'}\,^c\,_b\,\iota_c.
\]
For any spinor field $\omega^a$ we have
\[
(\nabla_{cc'}\,\nabla_{dd'} - \nabla_{dd'}\,\nabla_{cc'})\,\omega^a =
- R^a\,_{bcc'dd'}\,\omega^b -  T_{cc'}\,^{ee'}\,_{dd'}\,
\nabla_{ee'}\omega^a,
\]
with vanishing torsion
\begin{equation}
\label{tors}
0 = T_{bb'}\,^{dd'}\,_{cc'}\,e_{dd'} = \nabla_{bb'}\,e_{cc'}
- \nabla_{cc'}\,e_{bb'}
- [e_{bb'}, e_{cc'}]
\end{equation}
\[
= \Gamma_{bb'}\,^d\,_c\,e_{dc'}
+ \bar{\Gamma}_{bb'}\,^{d'}\,_{c'}\,e_{cd'}
- \Gamma_{cc'}\,^d\,_b\,e_{db'}
- \bar{\Gamma}_{cc'}\,^{d'}\,_{b'}\,e_{bd'}
- [e_{bb'}, e_{cc'}],
\]
and curvature spinor field
\begin{equation}
\label{spincurv}
R_{abcc'dd'} = e_{dd'}(\Gamma_{cc'ab}) - e_{cc'}(\Gamma_{dd'ab})
+ \Gamma_{dd'ae}\,\Gamma_{cc'}\,^e\,_b
\end{equation}
\[
+ \Gamma_{ed'ab}\,\Gamma_{cc'}\,^e\,_d
- \Gamma_{cc'ae}\,\Gamma_{dd'}\,^e\,_b
- \Gamma_{ec'ab}\,\Gamma_{dd'}\,^e\,_c
\]
\[
+ \Gamma_{de'ab}\,\bar{\Gamma}_{cc'}\,^{e'}\,_{d'}
- \Gamma_{ce'ab}\,\bar{\Gamma}_{dd'}\,^{e'}\,_{c'}
- T_{cc'}\,^{ee'}\,_{dd'}\,\Gamma_{ee'ab}.
\]
The latter has the decomposition
\begin{equation}
\label{spincurvdecomp}
R_{abcc'dd'} = - \Psi_{abcd}\,\epsilon_{c'd'} - \Phi_{abc'd'}\,
\epsilon_{cd}
+ \Lambda\,\epsilon_{c'd'}\,(\epsilon_{bd}\,\epsilon_{ac} +
\epsilon_{ad}\,\epsilon_{bc}),
\end{equation}
into the conformal Weyl spinor field $\Psi_{abcd} = \Psi_{(abcd)}$ as
well as the Ricci
spinor
$\Phi_{aba'b'} = \Phi_{(ab)(a'b')} = \bar{\Phi}_{aba'b'}$ and the scalar
$\Lambda$, which
allow us to represent the Ricci tensor in the form
\[
R_{aa'bb'} = 2\,\Phi_{aba'b'} + 6\,\Lambda\,\epsilon_{ab}\,\epsilon_{a'b'}.
\]
The Bianchi identity reads
\begin{equation}
\label{spinBian}
\nabla^f\,_{a'}\,\Psi_{abcf} = \nabla_{(a}\,^{f'}\,\Phi_{bc)a'f'}.
\end{equation}

\subsubsection{The Field Equations}\label{fieldeq}

The Einstein--Dirac system is specified (cf. \cite{penrose:rindler:I}) by
a pair of
$2$-spinor fields $\phi_a$, $\chi_{a'}$ satisying the Dirac equations
\begin{equation}
\label{dirac}
\nabla^a\,_{a'}\phi_a = \mu\,\chi_{a'},\,\,\,\,\,\,\,
\nabla_a\,^{a'}\chi_{a'} = \mu\,\phi_a,
\end{equation}
with a real constant $\mu$, and the Einstein equations with
energy-momentum tensor
\begin{equation}
\label{energymomentum}
T_{aa'bb'} = \frac{i\,k}{2}\,\{
\phi_a\,\nabla_{bb'}\bar{\phi}_{a'} - \bar{\phi}_{a'}\,\nabla_{bb'}
\phi_{a'}
+ \phi_b\,\nabla_{aa'}\bar{\phi}_{b'} - \bar{\phi}_{b'}\,\nabla_{aa'}
\phi_{b}
\end{equation}
\[
- \bar{\chi}_a\,\nabla_{bb'}\chi_{a'} + \chi_{a'}\,\nabla_{bb'}
\bar{\chi}_{a}
- \bar{\chi}_b\,\nabla_{aa'}\chi_{b'} + \chi_{b'}\,\nabla_{aa'}
\bar{\chi}_{b}\}.
\]
The Einstein equations then take the form

\begin{equation}
\label{treinst}
\Lambda = - \frac{i\,k\,\kappa\,\mu}{3}\,(\phi_a\bar{\chi}^a
- \bar{\phi}_{a'}\chi^{a'}),
\end{equation}
\begin{equation}
\label{trfreinst}
\Phi_{aba'b'} = \frac{i\,k\,\kappa}{2}\,\{
\phi_{(a}\,\nabla_{b)(a'}\bar{\phi}_{b')}
- \bar{\phi}_{(a'}\,\nabla_{b')(a}\phi_{b)}
\end{equation}
\[
- \bar{\chi}_{(a}\,\nabla_{b)(a'}\chi_{b')}
+ \chi_{(a'}\,\nabla_{b')(a}\bar{\chi}_{b)}\}.
\]

The discussion of this system is complicated by the fact that the Dirac
equations are
of first order while derivatives of the spinor fields also appear on the
right hand side of
(\ref{trfreinst}). Consequently, the right hand side of the Bianchi
identity (\ref{spinBian})
is given by an expression involving the derivatives of the spinor fields
from zeroth to
second order. Therefore we need to derive equations for these quantities
as well.

By taking derivatives of the Dirac equations and commuting derivatives,
we obtain
\begin{equation}
\label{1dphi}
\nabla^a\,_{a'}\nabla_{bb'}\phi_a
= - R^h\,_a\,^a\,_{a'bb'}\,\phi_h + \mu\,\nabla_{bb'}\chi_{a'}
\end{equation}
\begin{equation}
\label{2dphi}
\nabla^a\,_{a'}\nabla_{cc'}\nabla_{bb'}\phi_a
= - \nabla_{cc'}R^h\,_a\,^a\,_{a'bb'}\,\,\phi_h
\end{equation}
\[
- R^h\,_a\,^a\,_{a'bb'}\,\nabla_{cc'}\phi_h
+ \mu\,\nabla_{bb'}\nabla_{cc'}\chi_{a'}
- R^h\,_b\,^a\,_{a'cc'}\,\nabla_{hb'}\phi_a
\]
\[
- \bar{R}^{h'}\,_{b'}\,^a\,_{a'}\,_{cc'}\,\nabla_{bh'}\phi_a
- R^h\,_a\,^a\,_{a'cc'}\,\nabla_{bb'}\phi_h
\]
and similar equations for the derivatives of $\chi_{a'}$. It is important
here that the
curvature quantity
\[
R^h\,_a\,^a\,_{a'bb'} = - \Phi^h\,_{da'b'} -
3\,\Lambda\,\epsilon_{a'b'}\epsilon_d\,^h
\]
which occurs in these equations does not contain the conformal Weyl
spinor. We
can use (\ref{treinst}) and (\ref{trfreinst}) to express
$R^h\,_a\,^a\,_{a'bb'}$
and its derivative
in (\ref{1dphi}) and (\ref{2dphi}) in terms of the spinor fields and their
derivatives to obtain
a complete system of equations for $\phi_a$, $\nabla_{bb'}\phi_a$,
$\nabla_{cc'}\nabla_{bb'}\phi_a$ and the corresponding fields derived
from $\chi_{a'}$.

These fields are not quite independent of each other. If we define
symmetric fields
$\phi_{aca'}$, $\phi_{abca'b'}$, $\chi_{aa'c'}$, $\chi_{aba'b'c'}$ by
setting
\[
\phi_{ac}\,^{a'} = \nabla_{(a}\,^{a'}\phi_{c)},\,\,\,\,\,\,\,
\phi_{abc}\,^{a'b'} = \nabla_{(a}\,^{(a'}\nabla_{b}\,^{b')}\phi_{c)}
\]
\[
\chi_a\,^{a'c'} = \nabla_a\,^{(a'}\chi^{c')},\,\,\,\,\,\,\,
\chi_{ab}\,^{a'b'c'} = \nabla_{(a}\,^{(a'}\nabla_{b)}\,^{b'}\chi^{c')},
\]
we get from the Dirac equations
\[
\nabla_{aa'}\phi_b = \phi_{aba'} - \frac{\mu}{2}\,\epsilon_{ab}\,\chi_{a'},
\]
\[
\nabla_{cc'}\nabla_{bb'}\phi_a = \phi_{abcb'c'}
- \frac{1}{2}\,\epsilon_{b'c'}\,\Psi_{abch}\,\phi^h
+ \frac{2}{3}\,\epsilon_{c(a}\,\Phi_{b)hb'c'}\,\phi^h
\]
\[
+ 2\,\Lambda\,\phi_{(a}\epsilon_{b)c}\,\epsilon_{b'c'}
+ \frac{2}{3}\,\mu\,\epsilon_{a(b}\,\chi_{c)b'c'}
- \frac{1}{2}\,\mu^2\,\phi_a\,\epsilon_{bc}\,\epsilon_{b'c'},
\]
and similar relations for the derivatives of $\chi_{a'}$. {}From these, the
equations above,
and (\ref{spinBian}), (\ref{treinst}) and (\ref{trfreinst}) we can derive
equations of the form
\begin{equation}
\label{dddphi}
\nabla^a\,_{a'}\phi_{abb'} = M^1_{ba'b'},\,\,\,\,\,\,
\nabla^a\,_{a'}\phi_{abcb'c'} = M^2_{bca'b'c'},
\end{equation}
\begin{equation}
\label{dddchi}
\nabla_a\,^{a'}\chi_{ba'b'} = N^1_{abb'},\,\,\,\,\,\,
\nabla_a\,^{a'}\chi_{bca'b'c'} = N^2_{abcb'c'},
\end{equation}
where $M^1_{ba'b'}$, $N^1_{abb'}$ denote functions of $\phi_a$,
$\chi_{a'}$, $\phi_{aca'}$,
$\chi_{aa'c'}$, while $ M^2_{bca'b'c'}$, $N^2_{abcb'c'}$ depend in
addition on
$\phi_{abca'b'}$, $\chi_{aba'b'c'}$, and $\Psi_{abcd}$. Note that this
introduces (or rather
makes explicit) further non-linearities.

\subsubsection{Hyperbolic Equations from the Einstein--Dirac
System}\label{hypdir}

Our field equations for the unknowns
\[
e^{\mu}\,_k,\,\,\,\,\Gamma_{aa'bc},\,\,\,\,\Psi_{abcd},\,\,\,\,
\phi_a,\,\,\,\,\phi_{abb'},\,\,\,\,\phi_{abcb'c'},\,\,\,\,
\chi_{a'},\,\,\,\,\chi_{ba'b'},\,\,\,\,\chi_{bca'b'c'},
\]
are now given by (\ref{tors}) and (\ref{spincurvdecomp}) (with the left hand
side understood as being given by (\ref{spincurv})), (\ref{spinBian}),
(\ref{dirac}), (\ref{dddphi}) and (\ref{dddchi}). Here Einstein's equations
(\ref{treinst}) and (\ref{trfreinst})
are used to express   quantities derived from the Ricci tensor in terms
of the spinor
fields and their derivatives.

When we try to deduce a hyperbolic reduced system from these equations,
the first two
equations, which determine the gauge dependent quantities, will require the
choice of a gauge, while we expect the remaining equations, which are
tensorial, to
contain subsystems which are hyperbolic irrespective of any gauge. This
is indeed the
case and there are, due to the fact that most of the equations are
overdetermined, various possibilites to extract such systems.

In \cite{friedrich:1985} sytems of spinor equations have been considered
which are built from systems of the type
\[
\nabla^b\,_{a'}\,\psi_{b \beta} = F_{a' \beta}(x^{\mu}, \psi_{c \gamma}),
\]
or their complex conjugates, where $\beta$ denotes a multi-index of some
sort. If the
components corresponding to different values of the indices $b$, $\beta$
are independent of
each other the equations
\[
- \nabla^b\,_{0'}\,\psi_{b \beta} = - F_{0' \beta}
\]
\[
\nabla^b\,_{1'}\,\psi_{b \beta} = F_{1' \beta}
\]
form a symmetric hyperbolic system. Equations (\ref{dirac}) are thus
symmetric hyperbolic as
they stand. If symmetries are present which relate the index $b$ to a
group of unprimed spinor indices comprised by $\beta$, the equations to
be extracted are
slightly different. For instance, we obtain from (\ref{spinBian}) a
symmetric hyperbolic
system (regarding all fields besides the Weyl spinor field as given) of
form
\[
\nabla^f\,_{1'}\,\Psi_{000f} = \ldots,
\]
\[
\nabla^f\,_{1'}\,\Psi_{ab1f} - \nabla^f\,_{0'}\,\Psi_{ab0f} = \dots,
\]
\[
- \nabla^f\,_{0'}\,\Psi_{111f} = \dots
\]
where the symmetry $\Psi_{abcd} = \Psi_{(abcd)}$ is assumed explicitly
so that there are
five complex unknown functions. Equations (\ref{dddphi}) and (\ref{dddchi})
can be dealt with similarly.

To compare the characteristics of the system above with previous
hyperbolic systems
extracted from the Bianchi identity, we set
\[
e_0 = \frac{1}{\sqrt{2}}(e_{00'} + e_{11'}),\,\,\,\,\,
e_1 = \frac{1}{\sqrt{2}}(e_{01'} + e_{10'}),
\]
\[
e_2 = \frac{-i}{\sqrt{2}}(e_{01'} - e_{10'}),\,\,\,\,\,
e_3 = \frac{1}{\sqrt{2}}(e_{00'} - e_{11'}).
\]
Then the characteristic polynomial is given up to a positive constant
factor by
\begin{equation}
\label{1spinredchar}
\xi_{\mu}\,e^{\mu}\,_0\,\,k^{\rho \nu}\,\xi_{\rho}\,\xi_{\nu}
\,\,g^{\lambda \sigma}\,\xi_{\lambda}\,\xi_{\sigma},
\end{equation}
which contains the degenerate quadratic form
\[
k^{\mu \nu} = 2\,e^{\mu}\,_0\,e^{\nu}\,_0 - e^{\mu}\,_1\,e^{\nu}\,_1 -
e^{\mu}\,_2\,e^{\nu}\,_2.
\]
The cone $\{k^{\rho \nu}\,\xi_{\rho}\,\xi_{\nu} = 0\}$ is the product of
a $2$-dimensional
cone in the plane $\{\xi_{\mu}\,e^{\mu}\,_3 = 0\}$ with the real line
so that its set of
generators is diffeomorphic to $S^1 \times {\R}$. The associated
characteristics are
timelike. The special role played here by the vector field $e_3$ allows
us to adapt the
system to situations containing a distinguished direction.

Another method to extract symmetric hyperbolic equations from spinor
equations has been
discussed in \cite{friedrich:global}. It is based on the space-spinor
formalism in
which an arbitrary normalized timelike vector field is used to express
all spinor
fields and spinor equations in terms of fields and equations containing
only unprimed
indices. If the fields and equations are then decomposed into their
irreducible parts
(a direct, though somewhat lengthy algebraic procedure), the equations
almost
automatically decompose into symmetric hyperbolic propagation equations and
constraints.

For simplicity we choose the timelike vector field to be
\[
\sqrt{2}\,e_0 = \tau^{aa'}\,e_{aa'}
\quad\mbox{with}\quad
\tau_{aa'} = \epsilon_0\,^a\,\epsilon_{0'}\,^{a'} +
\epsilon_1\,^a\,\epsilon_{1'}\,^{a'}.
\]
Since $\tau_{aa'}\,\tau^{ba'} = \epsilon_a\,^b$ etc., maps generalizing
the map
$\omega_{a'} \rightarrow \tau_a\,^{a'}\,\omega_{a'}$ to spinors of
arbitrary
valence are bijective and allow us to obtain faithful representations of
all spinor
relations in terms of unprimed spinors.  Writing $\nabla_{ab} =
\tau_b\,^{a'}\,\nabla_{aa'} = \frac{1}{2}\,\epsilon_{ab}\,P
+ {\cal D}_{ab}$, we
obtain a representation of the covariant derivative operator in terms of
the
directional derivative operators $P = \tau^{aa'}\,\nabla_{aa'}$,
${\cal D}_{ab} =
\tau_{(b}\,^{a'}\,\nabla_{a)a'}$ acting in the direction of $e_0$ and in
directions
orthogonal to $e_0$ respectively. In particular, (\ref{spinBian}) splits
under the
operations indicated above into \lq constraints\rq\
\[ {\cal D}^{fg}\,\Psi_{abfg} = \ldots,
\]
and \lq evolution equations\rq\
\[
P\,\Psi_{abcd} - 2\,{\cal D}_{(a}\,^f\,\Psi_{bcd)f} = \ldots
\]
If the latter are multiplied by the binomial coefficients
${4 \choose a + b + c + d}$, they are seen to be symmetric hyperbolic.
The characteristic
poynomial of this system is again of the form
\[
\xi_{\mu}\,e^{\mu}\,_0\,\,k^{\rho \nu}\,\xi_{\rho}\,\xi_{\nu}
\,\,g^{\lambda \sigma}\,\xi_{\lambda}\,\xi_{\sigma},
\]
however, since there is no privileged spacelike direction singled out
here, we have
a non-degenerate quadratic form
\begin{equation}
\label{innercone}
k^{\mu \nu} = (1 + c)\,e^{\mu}\,_0\,e^{\nu}\,_0
- e^{\mu}\,_1\,e^{\nu}\,_1 -
e^{\mu}\,_2\,e^{\nu}\,_2 - e^{\mu}\,_3\,e^{\nu}\,_3,
\end{equation}
with some constant $c > 0$, as in the case of the system used in the case
of the
Einstein--Euler equations.

By the method outlined above symmetric hyperbolic systems are also obtained
for equations
(\ref{dddphi}) and (\ref{dddchi}). For instance, from the first of equations
(\ref{dddphi}) we obtain
for $\phi_{abc} = \tau_c\,^{b'}\,\phi_{abb'} = \phi_{(ab)c}$ an equation
of the form
$\nabla^a\,_d\,\phi_{abc} = \ldots$, where we only indicate the principal
part.
Using the decomposition
\[
\phi_{abc} = \phi^*_{abc} - \frac{2}{3}\,\epsilon_{c(a}\,\phi^*_{b)}
\quad\mbox{with}\quad \phi^*_{abc} = \phi_{(abc)},\,\,\,\,\,\,\phi^*_b
= \phi_{fb}\,^f,
\]
and the decomposition of $\nabla_{ab}$, we get a
system of the form
\[
P\,\phi^*_a - \frac{2}{3}\,{\cal D}^b\,_a\,\phi^*_b
+ 2\,{\cal D}^{bc}\,\phi^*_{abc}
= 2\,\nabla^{bc}\,\phi_{bac} = \ldots,
\]
\[
{3 \choose a + b + c}\left\{
P\,\phi^*_{abc} - 2\,{\cal D}^d\,_{(a}\,\phi^*_{bc)d}
- \frac{2}{3}\,{\cal D}_{(ab}\,\phi^*_{c)}\right\}
\]
\[
= - 2\,{3 \choose a + b + c}\,\nabla^f\,_{(a}\,\phi_{|f|bc)} = \ldots
\]
which is symmetric hyperbolic.

It is well known that equations for spinor fields of spin $\frac{m}{2}$,
$m > 2$
give rise to consistency conditions (cf. \cite{penrose:rindler:I}). For
instance,
the equation
\begin{equation}
\label{spequ}
\nabla^a\,_{a'}\,\phi_{abcb'c'} = H_{bca'b'c'},
\end{equation}
where we consider the right hand side as given, implies the relation
\[
\phi_{abcb'c'}\,\Psi^{abc}\,_d + 4\,\phi_{abcd'(b'}\,\Phi^{abd'}\,_{c')} =
\nabla^{aa'}\,H_{ada'b'c'},
\]
which reduces e.g. in the case of vanishing right hand side to a particular
relation between the
background curvature and the unknown spinor field. Depending on the type
of equation
and the background space-time, such consistency conditions may forbid the
existence of
any solution at all. Nevertheless, equation (\ref{spequ}) implies a
symmetric hyperbolic
system for which the existence of solutions is no problem. Difficulties
will arise if one wants to show that the constraints implied by
(\ref{spequ}) are preserved.

This example emphasizes the need to show the preservation of the
constraints. Because
in our case the right hand sides of the equations are given by very
specific functions
of the unknowns themselves, we can expect to obtain useful subsidiary
equations.

There are again various methods to obtain hyperbolic equations for the
gauge-dependent
frame and connection coefficients, which depend in particular on the
choice of gauge
conditions. In \cite{friedrich:1985}, \cite{friedrich:global} the
coordinates and the frame
field have been subject to wave equations (nonlinear in the case of the
frame field). Here
we shall indicate a gauge considered in \cite{friedrich:hypred}
(which can, of course, also be implemented in the frame formalism
considered in the previous sections).

We shall denote by $T$ a \lq time flow vector field\rq\  and by $x^{\mu}$
coordinates on some
neighbourhood of an initial hypersurface $S$. We assume $T$ to be
transverse to $S$, the
\lq time coordinate\rq\ $t \equiv x^0$ to vanish on $S$, and the relation
$<d\,x^{\mu}, T>\, = \delta^{\mu}\,_0$ to hold on the neighbourhood such
that we can write
$T = \partial_t$.

The frame $e_{aa'}$ is chosen such that the timelike vector field
$\tau^{aa'}\,e_{aa'}$ is orthogonal to $S$. Using the expansion
\[
e_{aa'} = \frac{1}{2}\,\tau_{aa'}\,\tau^{cc'}\,e_{cc'}
- \tau^b\,_{a'}\,e_{ab}
\quad\mbox{with}\quad e_{ab} = \tau_{(b}\,^{a'}\,e_{a)a'},
\]
we can write
\[
T = \alpha\,\tau^{cc'}\,e_{cc'} + \beta^{cc'}\,e_{cc'}
= \alpha\,\tau^{cc'}\,e_{cc'} + \beta^{ab}\,e_{ab},
\]
with
\begin{equation}
\label{bcond}
\tau^{cc'}\,\beta_{cc'} = 0, \,\,\,\,\,\,\,\beta_{ab}
= \tau_{(a}\,^{a'}\,\beta_{b)a'}.
\end{equation}
Thus the evolution of the coordinates off $S$ is determined by the fields
$\alpha \neq 0$ and
$\beta^{aa'}$ and we can write
\begin{equation}
\label{cgau}
\tau^{cc'}\,e_{cc'} = \frac{1}{\alpha}\,(\partial_t - \beta^{ab}\,e_{ab}).
\end{equation}

Since we have $\nabla_T\,\iota_c = \Gamma^b\,_c\,\iota_b$, the evolution
of the frame is
determined by the functions
\[
\Gamma_{bc} = T^{aa'}\,\Gamma_{aa'bc}.
\]
and we can write
\begin{equation}
\label{fgau}
\tau^{aa'}\,\Gamma_{aa'bc} = \frac{1}{\alpha}\,(\Gamma_{bc}
- \beta^{ae}\,\Gamma_{aebc}),
\end{equation}
with $\Gamma_{aebc} = \tau_{(e}\,^{a'}\,\Gamma_{a)a'bc}$.

We now consider the fields $\alpha = \alpha(x^{\mu}) > 0$,
$\beta^{ab} = \beta^{ab}(x^{\mu})$ (together four real functions) as
\lq coordinate gauge source functions\rq\ and the field
$\Gamma_{bc} = \Gamma_{bc}(x^{\mu})$ (six real functions) as
\lq frame gauge source functions\rq. This is feasible, because given these
functions, we can find  smooth coordinates and a frame fields close to an
initial hypersurface such that the given functions assume the meaning given
to them above.

The gauge conditions are then expressed by the requirement that the right
hand sides of
(\ref{cgau}) and (\ref{fgau}) are given in terms of the gauge source functions
and
$e_{ab}$ and $\Gamma_{aebc}$. Thus it remains to obtain evolution equations
for
$e^{\mu}\,_{ab}$ and $\Gamma_{aebc}$.

Reading the quantity $e^{\mu}\,_{aa'}$ for fixed index $\mu$  as the
expression of the
differential of $x^{\mu}$ in the frame $e_{aa'}$, we can write (\ref{tors})
in the form
\[
\nabla_{aa'}e^{\mu}\,_{bb'} - \nabla_{bb'}e^{\mu}\,_{aa'} = 0.
\]
Contracting this equation with $T^{aa'}$ and $\tau_c\,^{b'}$ and
symmetrizing, we
obtain the equation
\[
0 = \nabla_T\,e^{\mu}\,_{cb} - e^{\mu}\,_{b'(b}\nabla_T\,\tau_{c)}\,^{b'}
- e^{\mu}\,_{aa'}\nabla_{bc}T^{aa'},
\]
which can be rewritten in the form
\[
\partial_t\,e^{\mu}\,_{ab} = \ldots,
\]
where the right hand side can be expressed in terms of the gauge source
functions and
their derivatives and the unknowns. By using (\ref{spincurvdecomp}) with
(\ref{spincurv}) on the left hand side, we can derive in a similar way an
equation
\[
\partial_t\,\Gamma_{aebc} = \ldots,
\]
with the right hand side again being given in terms of the gauge source
functions and
their derivatives and the unknowns.

Thus we obtain symmetric hyperbolic reduced equations for all unknowns
except those
given by the left hand sides of the gauge conditions (\ref{cgau}) and
(\ref{fgau}).
Our procedure applies of course to various other sytems. Our choice of
gauge is of
interest because of the direct relation between the gauge source
functions and
the evolution of the gauge. The causal nature of the evolution can be
controlled
explicitly because the formalism allows us to calculate the value of the
norm
$g(T, T) = 2\,\alpha^2 + \beta_{ab}\,\beta^{ab}$. This may prove useful
if it is desired to control the effect of the choice of gauge source
functions on the long time evolution of the gauge in numerical calculations
of space-times.

\subsection{Remarks on the Structure of the Characteristic
Set}\label{remchar}

We have seen that for certain reduced systems there occur besides the
\lq physical\rq\
characteristics, given by null hypersurfaces, also characteristics which
are timelike
or spacelike with respect to the metric $g_{\mu \nu}$. Timelike
characteristics, which usually occur if a system of first order is deduced
from a system of second
order, are usually harmless and of no physical significance. The
spacelike
characteristics, which are partly due to the choice of gauge condition
and partly due
to the use made of the constraints, have no physical significance either.
Though they
are associated with non-causal propagation, there is a priori nothing bad
about them
and it rather depends on the applications one wants to make whether they
are harmful or
not.

In the characteristic polynomial (\ref{flcharcone}) of the reduced
equations for the Einstein--Euler system there appears a factor $\xi^2_0
+ \frac{1}{4}\,h^{ab}\,\xi_a\,\xi_b$
which corresponds to timelike characteristics. In vacuum the corresponding
cone
has of course no
physical meaning, since in general there is no preferred timelike vector field
available. In the
perfect fluid case there is a distinguished timelike vector field present.
This, and
perhaps the symmetry of the inner cone with respect to the fluid vector,
has led some
people to speculate on the physical significance of that cone
\cite{elst:ellis}.
However, in some of the later examples of hyperbolic equations deduced
from the
Bianchi identity, which could also be used in the fluid case, the
structure of the
characteristics is drastically different from the one observed in the
fluid case
(cf. (\ref{extcharibvp}) and also the degenerate cone arising in
(\ref{1spinredchar})).
In particular, the factor above does not occur in their characteristic
polynomials. The
large arbitrariness in extracting hyperbolic equations, which arises from
different
use made of the constraints implied by the Bianchi identity, suggests
that in the case
of the Einstein--Euler system the only \lq physical characteristics\rq\ are
those associated
with the fluid vector, the null cone of $g_{\mu\nu}$, and the sound cone.

The null cone of the metric (\ref{extcharibvp}) touches the null cone of
the metric $g$
in the directions of $\pm\,e_3$ but it is spacelike in all other
directions. Thus all
null hypersurfaces of it are spacelike or null for $g_{\mu \nu}$. Such a
cone has
the effect that the \lq domain of uniqueness\rq\ defined by the techniques
discussed in Sect. \ref{symhyp} may decrease. However, as we have seen,
is does not prevent us from proving useful results.

There is also no reason to assume that the additional characteristics
necessarily
create problems in numerical calculations. In situations where the
maximal slicing
condition can be used, the occurrence of spacelike characteristics which
are related,
as in our examples, in a rigid way with the metric should be innocuous.
Also,
numerical calculations based on equations with inner characteristic cones
as
observed above have been performed without difficulties
(\cite{frauendiener}, \cite{huebner}).

\section{Local Evolution}\label{locevol}

\subsection{Local Existence Theorems for the Einstein
Equations}\label{locexist}

The purpose of this section is to present a local existence theorem for
the Einstein vacuum equations. By (abstract) vacuum initial data we mean a
three-dimensional manifold $S$ together with a Riemannian metric $h_{ab}$
and a symmetric tensor $\chi_{ab}$ on $S$ which satisfy the vacuum constraints
(see Sect. \ref{basics}). A corresponding solution of the vacuum Einstein
equations is a
Lorentzian metric $g_{\alpha\beta}$ on a four-dimensional manifold $M$ and
an embedding $\phi$ of $S$ into $M$ such that $h_{ab}$ and $\chi_{ab}$
coincide with the pull-backs via $\phi$ of the induced metric on $\phi(S)$
and the second fundamental form of that manifold respectively and the Einstein
tensor of $g_{\alpha\beta}$ vanishes. If $\phi(S)$ is a Cauchy surface for
the space-time $(M,g_{\alpha\beta})$ then this space-time is said to be a
Cauchy
development of the data $(S,h_{ab},\chi_{ab})$. The basic local existence
theorem for the vacuum Einstein equations says that every vacuum initial data
set has at least one Cauchy development. In fact to make this precise it is
necessary to fix the differentiability properties which are assumed for
the data and demanded of the solution. For instance, the result holds
if the differentiability class for both data and solutions is taken to
be $C^\infty$. Note that there is no need to require any further
conditions on the spatial dependence of the data.

The proof of local existence will now be outlined. We follow essentially
the original method of \cite{bruhat52} except for the fact that we reduce
second order equations to first order symmetric hyperbolic systems and that
we use harmonic mappings rather than harmonic coordinates. The use of
harmonic mappings, as discussed in Sect. \ref{reductions},
allows us to work globally in space even if the manifold
$S$ cannot be covered by a single chart. Using harmonic coordinates it
would be necessary to construct solutions local in space and time and then
piece them together. Choose a fixed Lorentz metric on
$S\times\R$, for instance the metric product of the metric $h_{ab}$ with
$-dt^2$. This comparison metric will be denoted by $\bar g_{\alpha\beta}$.
The idea is to look for a solution $g_{\alpha\beta}$ on an open subset $U$
of $\R\times S$ such that the identity is a harmonic map from
$(U, g_{\alpha\beta})$ to
$(U, \bar g_{\alpha\beta})$. This is a condition which is defined
in a global invariant way. Its expression in local coordinates is
$g^{\beta\gamma}(\Gamma^\alpha_{\beta\gamma}
-\bar\Gamma^\alpha_{\beta\gamma})=0$
where $\Gamma^\alpha_{\beta\gamma}$ and $\bar\Gamma^\alpha_{\beta\gamma}$
are the Christoffel symbols of $g_{\alpha\beta}$ and $\bar g_{\alpha\beta}$
respectively. In the terminology of Sect. \ref{evole} this means that we
choose $g^{\beta\gamma}\bar\Gamma^\alpha_{\beta\gamma}$ as a gauge source
function.

Next consider the question of reduction of nonlinear wave equations to
symmetric hyperbolic form. This is done as follows. Let
$g^{\alpha\beta}(t,x,u)$ be functions of $(t,x,u)$
which for each fixed value of $(t,x,u)$ make up a symmetric matrix of Lorentz
signature and consider an equation of the form:
\begin{equation}\label{genwave}
g^{\alpha\beta}\d_\alpha\d_\beta u+F(t,x,u,Du)=0
\end{equation}
This has been formulated in a local way but a corresponding class of
equations can be defined in the case that the unknown $u$ is a section of a
fibre bundle. As in the above treatment of symmetric hyperbolic equations
on a manifold, consideration will be restricted to the case of sections of
a vector bundle $V$. Choose a fixed connection on $V$. Then the class of
equations to be considered is obtained by replacing the partial derivatives
in the above equation by covariant derivatives defined by the given
connection. Let $u_\alpha=\nabla_\alpha u$. Then the equation (\ref{genwave})
can be written as:
\begin{eqnarray*}
-g^{00}\nabla_0 u_0-2g^{0a}\nabla_a u_0
&=&g^{ab}\nabla_a u_b+F(t,x,u_0,u_a)                         \\
g^{ab}\nabla_0 u_a&=&g^{ab}\nabla_a u_0+K^b                  \\
\nabla_0 u&=&u_0
\end{eqnarray*}
Here $K^b$ is a term involving the curvature of the connection which is of
order zero in the unknowns of the system. This is a symmetric hyperbolic
system for the unknowns $u$ and $u_\alpha$. Since $u$ is allowed to be a
section of a vector bundle we are dealing
with a system of equations. However the functions $g^{\alpha\beta}$ must
be scalars. The appropriate initial data for the second order equation
consists of the values of $u$ and $\d_t u$ on the initial hypersurface.
{}From these the values of the functions $u_\alpha$ on the initial
hypersurface may be determined. Thus an initial data set for the symmetric
hyperbolic system is obtained. It satisfies the additional constraint
equation $\nabla_\alpha u=u_\alpha$. Applying the existence theory for
symmetric hyperbolic systems gives a solution $(u,u_\alpha)$. To show
that the function $u$ obtained in this way is a solution of the original
second order equation it is necessary to show that the constraint equation
is satisfied everwhere. That $\nabla_0 u=u_0$ follows directly from the first
order system. It also follows from the first order system that
$\nabla_0 (u_a-\nabla_a u)=0$. Since this is a first order homogeneous ODE for
$u_a-\nabla_a u$, the vanishing of the latter quantity for $t=0$ implies its
vanishing everywhere.

In the case of the vacuum Einstein equations the bundle $V$ can be taken
to be the bundle of symmetric covariant second rank tensors. The connection
can be chosen to be the Levi--Civita connection defined by
$\bar g_{\alpha\beta}$. This is only one possible choice but note that it is
important that this connection does not depend on the unknown in the equations,
in this case the metric $g_{\alpha\beta}$.

Now a proof of local in time existence for the vacuum Einstein equations will
be presented. Let $h_{ab}$ and $\chi_{ab}$ be the initial data. In
Sect.
\ref{basics} it was shown that the vacuum Einstein equations reduce to a
system of nonlinear wave equations when harmonic coordinates, or the
generalization involving gauge source functions, are used. As was already
indicated in that section, there is no loss of generality in imposing this
condition locally in time. If there exists a development of particular initial
data then there exists a diffeomorphism $\phi$ of a neighbourhood of the
initial hypersurface such that the pull-back of the metric with the given
diffeomorphism satisfies the harmonic condition with respect to $\bar
g_{\alpha\beta}$. In fact $\phi$ can be chosen to satisfy some additional
conditions. The harmonic condition is equivalent to a nonlinear wave equation
for $\phi$. Solving the local in time Cauchy problem for this wave equation
provides the desired diffeomorphism. The existence theory for this Cauchy
problem follows from that for symmetric hyperbolic systems by the reduction
to first order already presented. The initial data for $\phi$ will be
specified as follows. It is the identity on the initial hypersurface
and the contraction of the derivative of $\phi$ with the normal vector with 
respect to $\bar g_{\alpha\beta}$ should agree on $S$ with the corresponding 
quantity constructed from the identity on $M$.
Since the vector $\d/\d t$ is the unit normal
vector to $S$ with respect to $\bar g_{\alpha\beta}$ it will also have this
property with respect to $g_{\alpha\beta}$.

A local solution of the Einstein equations corresponding to prescribed
initial data can be obtained as follows. Let $h_{ab}$ and $\chi_{ab}$ denote
the components of the tensors making up the initial data in a local chart
as above. A set of initial data for the harmonically reduced vacuum Einstein
equations consists of values for the whole metric $g_{\alpha\beta}$ and its
time derivative on the initial hypersurface. A data set of this kind can
be constructed from $h_{ab}$ and $\chi_{ab}$ as follows. (The following
equations are expressed in local coordinates, but their invariant meaning
should be clear.)
\begin{eqnarray*}
g_{ab}=h_{ab},\ \ g_{0a}=0,\ \ g_{00}=-1       \\
\d_t g_{ab}=2\chi_{ab},\ \ \d_t g_{0a}=h^{bc}(h_{ab,c}-(1/2)h_{bc,a})
-h_{ab}\bar\Gamma^b_{cd}h^{cd},\\
\d_t g_{00}=-2h^{ab}\chi_{ab}
\end{eqnarray*}
These data are chosen in such a way that the harmonic condition is
satisfied on the initial hypersurface.

Corresponding to the initial data for the reduced equations there is a
unique local solution of these equations. It remains to show that it is
actually a solution of the Einstein equations provided the initial data
$h_{ab}$ and $\chi_{ab}$ satisfy the Einstein constraint equations. In
order to do this it suffices to show that the harmonic conditions are
satisfied everywhere since under those conditions the reduced
equations are equivalent to the Einstein equations. Let $\Delta^\alpha
=g^{\beta\gamma}(\Gamma^\alpha_{\beta\gamma}
-\bar\Gamma^\alpha_{\beta\gamma})$.
That the harmonic conditions are satisfied can be verified using the fact
that the quantities $\Delta^\alpha$ satisfy a linear homogeneous
system of wave equations. By uniqueness for this system the
$\Delta^\alpha$ vanish provided the initial data $\Delta^\alpha$ and
$\d_t\Delta^\alpha$
vanish on the initial hypersurface. The first of these was built into
the construction of the data for the reduced equations. The second is a
consequence of the combination of the reduced equations with the Einstein
constraints.

\subsection{Uniqueness}\label{unic}

The argument of the last section gives an existence proof for solutions
of the vacuum Einstein equations, local in time. It does not immediately say
anything about uniqueness of the space-time constructed. The solution of
the reduced equations is unique, as a consequence of the uniqueness
theorem for solutions of symmetric hyperbolic systems. However the
freedom to do diffeomorphisms has not yet been explored. In fact it
is straightforward to obtain a statement of uniqueness of the solution
corresponding to given abstract initial data, up to diffeomorphism.
Suppose two solutions $g_1$ and $g_2$ with the same initial data are
given. Choose a reference metric $\bar g$ as before and determine
diffeomorphisms $\phi_1$ and $\phi_2$ such that the identity is a
harmonic map with respect to the pairs $(\bar g,g_1)$ and $(\bar g,g_2)$
respectively. The transformed metrics satisfy the same system of reduced
equations with the same initial data and thus must coincide on a
neighbourhood of the initial hypersurface. Thus $g_1$ and
$(\phi_2\circ\phi_1^{-1})^*g_2$ coincide on a neighbourhood of the
initial hypersurface. This statement is often referred to as `geometric
uniqueness'.

For hyperbolic equations it is in general hard to prove theorems about
global existence of solutions due to the possibility of the formation
of singularities. On the other hand, it is possible to prove global
uniqueness theorems. Proving global uniqueness for the Einstein equations
is more difficult due to difficulties with controlling the freedom to do
diffeomorphisms. In this context we use the word 'global' to mean
not just applying to a subset of a given solution, but to
whole solutions having suitable intrinsic properties.
The strategy which has just been used to prove local
existence using harmonic maps cannot be applied directly. For doing so
would require a global existence theorem for harmonic maps and that we do
not have in general.

A global uniqeness theorem was proved by abstract means by Choquet-Bruhat
and Geroch \cite{choquet69}, who introduced the notion of the maximal
Cauchy development. If initial data for the Einstein equations coupled
to some matter fields on a manifold $S$ are prescribed, a {\it development}
of the data is a solution of the Einstein-matter system on a manifold $M$
together with an embedding $\phi$ of $S$ into a $M$ which induces the correct
initial data and for which the image of $S$ is a Cauchy surface. Another
development with a solution of the Einstein-matter equations on a manifold
$M'$ and an embedding $\phi'$ is called an {\it extension} of the first if
there is a diffeomorphism $\psi$ from $M$ to an open subset $U$ of $M'$
which maps the given metric on $M$ onto the restriction of the metric on $M'$
to $U$ and also maps the matter fields on $M$ to those on $U$ obtained by
restriction from $M'$. If we are dealing with the vacuum Einstein equations
then the requirement on the matter fields is absent. In \cite{choquet69} the
following theorem was proved for the vacuum case:

\noindent
{\bf Theorem} Let $S$ be an initial data set. Then there exists a development
$M$ of $S$ which is an extension of every other development of $S$. This
development is unique up to isometry.

\vskip 10pt\noindent
The development whose existence and uniqueness is asserted by this theorem
is called the {\it maximal Cauchy development} of the initial data set.
Uniqueness up to isometry means the following. If we have two developments
of the same data given by embeddings $\phi$ and $\phi'$ of $S$ into
manifolds $M$ and $M'$ respectively then there exists a diffeomorphism
$\psi:M\to M'$ which is an isometry and satisfies $\phi'=\phi\circ\psi$. The
proof of this theorem does not depend strongly on the vacuum assumption. One
potential problem in extending it to certain matter fields is gauge freedom
in those fields. This requires the concept of extension to be defined in
a slightly different way. For instance in the case of gauge fields it is
necessary to consider not only diffeomorphisms of the base manifold, but
also automorphisms of the principal bundle entering into the definition of
the theory. We do not expect that this leads to any essential difficulty, but
in any concrete example one should pay attention to this point.

The proof of this theorem applies directly to Zorn's lemma and it is an
open question
whether it is possible to remove the use of the axiom of choice from the
argument. The maximal Cauchy development is often very useful in formulating
certain arguments. However it remains very abstract and gives the subjective
impression of being difficult to pin down.

It should be emphasized that, despite its global aspects, it would be
misleading to consider the above theorem as a global existence theorem for
the Einstein equations in any sense.
A comparison with ordinary differential equations
may help to make this clear. If an ordinary differential equation for a
function $u(t)$ is given (with smooth coefficients) then the standard local
existence theorem for ordinary differential equations says that given an
initial value $u_0$ there exists a $T>0$ and a unique solution $u(t)$ on
the interval $(-T,T)$ with $u(0)=u_0$. Now one can ask for the longest
time interval $(-T_1,T_2)$ on which a solution of this kind exists for
a fixed initial value $u_0$. This is called the maximal interval of
existence and has a similar status to that of the maximal Cauchy development.
The fact that the maximal interval of existence is well-defined says nothing
about the question whether the solution exists globally or not, which is
the question whether $T_1$ and $T_2$ are infinite or not. The existence
of the maximal Cauchy development says nothing about the global properties
of the solution obtained in this way, whereas global existence of the
solution of an ordinary differential equation does mean that the solution
has a certain global property, namely that it exists for an infinitely
long time.

\subsection{Cauchy Stability}\label{caustab}

Cauchy stability of the initial value problem for the Einstein equations
is the statement that, in an appropriate sense, the solution of the Einstein
equations depends continuously on the initial data. This continuity
statement has two parts, which can be stated intuitively in the following
way. Firstly, if a solution corresponding to one initial data set is
defined on a suitable closed region, then the solution corresponding to
any initial data set close enough to the original one will be defined on
the same region. Closeness is defined in terms of Sobolev norms. Care is
needed with the interpretation of the phrase \lq the same region\rq\ due
to the diffeomorphism invariance. To make it precise, something has to be
said about how regions of the different spacetimes involved are to be
compared with each other. Secondly, the solution defined on this
common region depends continuously on the initial data, where continuity
is again defined in terms of Sobolev norms. In non-compact situations it
is appropriate to use local Sobolev norms for this, i.e. the Sobolev norms
of restrictions of a function to compact sets.

Rather than make this precise in general we will restrict to one case where
the formulation of the statement is relatively simple, but which is still
general enough to give a good idea of the basic concepts. Consider initial
data sets for the Einstein equations on a compact manifold $S$. For
definiteness let us restrict to the vacuum case. As has been discussed above,
solutions can be constructed by using the harmonically reduced equations.
One step in this process is to associate to geometric data $(h_{ab},\chi_{ab})$
full data $(g_{\alpha\beta},\d_t g_{\alpha\beta})$. For each of these pairs
let us choose the topology of the Sobolev space $H^s(S)$ for the first
member and that of $H^{s-1}(S)$ for the second. Then standard properties of
Sobolev spaces show that if $s$ is sufficiently large the mapping from
geometric data to full data is continuous.  Suppose now that we have one
particular solution of the Einstein vacuum equations with data on $S$.
The corresponding solution of the harmonically reduced equations exists
on some region of the form $S\times [-T,T]$. If $s$ is sufficiently large
then there exists an open neighbourhood of the given data in
$H^s(S)\times H^{s-1}(S)$ such that for any data in this neighbourhood
there exists a corresponding solution in $H^s(S\times [-T,T])$. Moreover
the mapping from data to solutions defined on this neighbourhood is
continuous (in fact differentiable). This has been proved by
Choquet-Bruhat \cite{choquet74}.

The theorem concerning a compact initial hypersurface can also be modified
to give a local statement of the following type. Let initial data for the
Einstein equations be given on some manifold $S$ and suppose that a
corresponding solution is given on a manifold $M$. There is a neighbourhood
$U$ of $S$ where harmonic reduction is possible globally. This identifies
$U$ with an open subset of $S\times\R$. This contains a set of the form
$V\times [-T,T]$ (for some open subset $V$ of $S$ and some $T>0$) which
contains any given point of the initial hypersurface. If we cut off the
the initial data for the harmonically reduced equations and use the domain
of dependence, we can use the above statement for a compact initial
hypersurface to get continuous dependence on initial data for a possibly
smaller set $V'\times [-T',T']$. Summing up, each point sufficiently close
to the initial hypersurface has a neighbourhood $W_1$ with compact closure
such that there is an open subset $W_2$ of the initial hypersurface with
compact closure such that the following properties hold. If the restriction
of an initial data set for the Einstein equations is sufficiently close to
that of the original data set in $H^s(W_2)$ then there exists a corresponding
solution of class $H^s$ on a neighbourhood of $W_1$. Moreover, the resulting
mapping from $H^s(W_2)$ to $H^s(W_1)$ is continuous.

\subsection{Matter Models}\label{matmod}

To specify a matter model in general relativity three elements are
required. The first is a set of tensors (or perhaps other geometrical
objects) on space-time which describe the matter fields. The second is
the equations of motion which are to be satisifed by these fields. The
third is the expression for the energy-momentum tensor in terms of the
matter fields which is to be used to couple the matter to the Einstein
equations. Note that in general both the matter field equations and
the expression for the energy-momentum tensor involve the space-time
metric. Thus it is impossible to consider matter in isolation from the
space-time metric. In solving the Cauchy problem it is necessary to deal
with the coupled system consisting of the Einstein equations and the
equations of motion for the matter fields.

There are two broad classes of matter models which are considered in
general relativity, the field theoretical and phenomenological matter
models. The distinction between these is not sharply defined but is
useful in order to structure the different models. The intuitive idea
is that the field theoretic matter models correspond to a fundamental
description while the phenomenological models represent an effective
description of matter which may be useful in certain situations.
Within the context of classical general relativity, which is the
context of this article, the pretension of the field theoretic matter
models to be more fundamental is not well founded since on a fundamental
level the quantum mechanical nature of matter should be taken into
account.

Before going further, a general remark on the Einstein-matter equations is
in order. Suppose that in any given coordinate system the matter equations
can be written in symmetric hyperbolic form in terms of a variable $u$.
Consider the system obtained by coupling the harmonically reduced Einstein
equations, written in first order symmetric hyperbolic form, to the
sytem for $u$. If the coupling is only by terms of order zero then the
combined system is symmetric hyperbolic and a local existence theorem
for the reduced Einstein-matter system is obtained. The condition for this
to happen is that the equations for the matter fields contain at most
first derivatives of the metric (in practice the Christoffel symbols) and
that the energy-momentum tensor contains no derivatives of $u$. When
these conditions are satisfied, local existence for the Einstein-matter
equations (not just the reduced equations) can be proved using the same
strategy as we presented in the vacuum case. The fact, which should hold
for any physically reasonable matter model, that the energy-momentum tensor
is divergence free as a consequence of the matter field equations, can be
is used derive the equation which allows it to be proved that the harmonic
condition propagates.

It would be unreasonable to try and describe here all the matter models
which have ever been used in general relativity. We will, however, attempt
to give a sufficiently wide variety of examples to illustrate most of the
important features to be expected in general. We start with the field
theoretic models.

The simplest case is where the matter field is a single real-valued
function $\phi$. The equations of motion are:
$$\nabla^\alpha\nabla_\alpha\phi=m^2\phi+V'(\phi)$$
Here $m$ is a constant and $V$ is a smooth function which is
$O(\phi^3)$ for $\phi$ close to zero. A typical example would be
$V(\phi)=\phi^4$. The energy-momentum tensor is:
$$T_{\alpha\beta}=\nabla_\alpha\phi\nabla_\beta \phi
-[(1/2)(\nabla^\gamma\phi\nabla_\gamma\phi)+m^2\phi^2+2V(\phi)]
g_{\alpha\beta}$$
The equation for $\phi$ is a nonlinear wave equation and so may be
reduced to a symmetric hyperbolic system. When it is coupled to the
harmonically reduced Einstein equations via the energy-momentum tensor
above and the whole system reduced to first order there is no coupling
in the principal part. As mentioned above this is enough to allow a
local existence theorem to be proved. Note that the
splitting off of the nonlinear term $V$ and the sign condition following
from the form $m^2$ of the coefficient in the linear term are irrelevant
for the local well-posedness of the equations. On the one hand they are
motivated by considerations of the physical interpretation of the
equations. On the other hand they have an important influence on the
global behaviour of solutions. This matter model is often referred to
as \lq the scalar field\rq\ , although when used without qualification
this often means the special case $m=0$, $V=0$.

The scalar field can be thought of as a mapping into the real line. It
can be generalized by considering mappings into a Riemannian manifold
$N$. An equivalent of the massless scalar field with vanishing potential
is the nonlinear $\sigma$-model or wave map as it is often known to
physicists and mathematicians respectively. It is a mapping from
space-time into a manifold $N$ with Riemannian metric $h$ called the
target manifold. The field
equations and energy-momentum tensor have a coordinate-invariant
meaning but we will content ourselves with giving the expressions in
coodinate systems on $M$ and $N$. The field equations are
$$\nabla_\alpha\nabla^\alpha\phi^A+\Gamma^A_{BC}(\phi)\nabla_\alpha\phi^B
\nabla^\alpha\phi^C=0$$
where $\Gamma^A_{BC}$ are the Christoffel symbols of $h$ in some
coordinate system. The energy-momentum tensor is:
$$T_{\alpha\beta}=h_{AB}[\nabla_\alpha\phi^A\nabla_\beta\phi^B-
(1/2)(\nabla_\gamma\phi^A\nabla^\gamma\phi^B)g_{\alpha\beta}]$$
The special case where $N$ is the complex plane with the flat Euclidean
metric corresponds to the complex scalar field. In contrast to the case of
the scalar field, the wave map does not allow the addition of a mass term
or a potential term in any obvious way. If $N$ has the structure of a
vector space there is an obvious way of defining a mass term and further
structure on $N$ may lead to natural ways of defining a potential.
These features occur in the case of Higgs fields. It may be noted that
the wave maps considered here, which are sometimes also called hyperbolic
harmonic maps, are related mathematically to the harmonic gauge discussed
in Sect. \ref{reductions}. The role of the connection $\bar\Gamma$ in
Sect. \ref{reductions} is played here by the Levi--Civita connection
of the target manifold.

One of the most familiar matter models in general relativity is the Maxwell
field. This is described by an antisymmetric tensor $F_{\alpha\beta}$.
The equations of motion for the source-free Maxwell field are
$\nabla_\alpha F^{\alpha\beta}=0$ and
$\nabla_\alpha F_{\beta\gamma}+\nabla_\beta F_{\gamma\alpha}
+\nabla_\gamma F_{\alpha\beta}=0$ and the energy-momentum tensor is
$$T_{\alpha\beta}=F_\alpha{}^\gamma F_{\beta\gamma}-(1/4) F^{\gamma\delta}
F_{\gamma\delta}g_{\alpha\beta}$$
The second set of Maxwell equations can be solved locally by writing
$F_{\alpha\beta}=\nabla_\alpha A_\beta-\nabla_\beta A_\alpha$ for a
potential $A_\alpha$. Then the
other equations can be regarded as second order equations for $A_\alpha$.
Note, however, that if space-time has a non-trivial topology then it may
be impossible to find a global potential which reproduces a given field
$F_{\alpha\beta}$. In the same way that the scalar field can be
generalized to get wave maps, the Maxwell field can be generalized to
get Yang--Mills fields. We will not give the global description of these
fields involving principal fibre bundles but only give expressions in
local coordinates and a local gauge. The model is defined by the choice
of a Lie algebra (which we describe via a basis) and a positive definite
quadratic form on the Lie algebra with components $h^{IJ}$ in this basis.
Let $C^I_{JK}$ be the structure constants in this basis. The basic matter
field is a one-form $A_\alpha^I$ with values in the Lie algebra and the
field strength is defined by
$$F^I_{\alpha\beta}=\nabla_\alpha A^I_\beta-\nabla_\beta A^I_\alpha+
C^I_{JK}A^J_\alpha A^K_\beta$$
The field equations are
$$\nabla_\alpha F^{I\alpha\beta}+C^I_{JK}A^J_\alpha F^{K\alpha\beta}=0$$
In the special case where the Lie algebra is one-dimensional (and hence
Abelian) the Yang--Mills equations reduce to the Maxwell equations. Note
however that the Yang--Mills field cannot be described by the field
strength alone. The description in terms of a potential is indispensible.

A complication which arises when studying the initial value problem for
the Yang--Mills or Einstein--Yang--Mills systems is that of gauge
invariance. Although the potential is required it is not uniquely
determined. Gauge transformations of the form:
$$A_\alpha^I\mapsto A^I_\alpha+(g^{-1}\nabla_\alpha g)^I$$
leave $F^I_{\alpha\beta}$ invariant. Here $g$ is a function taking values
in a Lie group with the given Lie algebra and the expression $g^{-1}
\nabla g$ can naturally be identified with a Lie-algebra-valued one-form.
Fields related by a gauge transformation describe the same physical system.
The ambiguity here is similar to that of the ambiguity of different
coordinate systems in the case of the Einstein equations. It can be solved
in an analogous way by the use of the Lorentz gauge. This is similar to the
harmonic coordinate condition and reduces the Yang--Mills equations on
any
background to a system of nonlinear wave equations which can, if desired,
be reduced to a symmetric hyperbolic system. Combining harmonic
coordinates and Lorentz gauge produces a reduced Einstein--Yang--Mills
system which can be handled by the same sort of techniques as the reduced
vacuum Einstein equations. Of course there are a number of steps which
have to be checked, such as the propagation of Lorentz gauge. All these
comments apply equally well to the Einstein--Yang--Mills--Higgs system
obtained by coupling the Yang--Mills field to a Higgs field. (Now gauge
transformations for the Higgs field must also be specified.)

A field theoretic matter model whose Cauchy problem does not fit easily into
the above framework is that given by the Dirac equation. It has been
discussed in Sect. \ref{edirac}.

Probably the best known phenomenological matter model in general relativity
is the perfect fluid. This has already been discussed at some length in
Sect. \ref{euler}. Here we mention some complementary aspects.
Recall that the basic matter fields are the energy density $\rho$,
a non-negative real-valued function, and the four-velocity $U^\alpha$, a unit
timelike vector field. The energy-momentum tensor is given by
$$T^{\alpha\beta}=(\rho+p)U^\alpha U^\beta+pg^{\alpha\beta}$$
in the signature used here, namely $(-,+,+,+)$. The pressure $p$ is given
in the
isentropic case in terms of an equation of state $p=h(\rho)$. The
usual assumptions on this equation of state is that it is a non-negative
continuous function defined on an interval $[\rho_0,\infty)$ which is
positive for $\rho>0$. Also it should be $C^1$ for $\rho>0$ with positive
derivative $h'(\rho)=dp/d\rho$ there. The equations of motion of the fluid,
the Euler equations, are given by the condition that the energy-momentum
tensor should be divergence-free. The Euler equations can be written as
a well-posed symmetric hyperbolic system in terms of the basic variables
provided we restrict to cases where $\rho\ge C>0$ for some $C>0$. Here the
condition $h'>0$ is crucial. In the study of spatially homogeneous
cosmological models the equation of state $p=k\rho$ with $k<0$ is sometimes
considered. We emphasize that a fluid with an equation of state of this kind
cannot be expected to have a well-posed initial value problem. This has to do
with the fact that the speed of sound, which is the square root of $h'$, is
imaginary in that case. In Sect. \ref{illposed} we prove a related but
simpler result, namely that in special relativity the Euler equations with
this equation of state, linearized about a constant state, have an ill-posed
initial value problem. The system obtained by coupling the Euler equations to
the harmonically reduced Einstein equations can be written as a symmetric
hyperbolic system in the case that the Euler equations can be written
symmetric hyperbolic in terms of the basic variables, as stated above.

In the case where the density is everywhere positive, writing the Euler
equations in symmetric hyperbolic form is not trivial. One approach is
to take $\rho$ and the spatial components $U^i$ of the velocity as
variables and to express $U^0$ in terms of the $U^i$ and the metric via
the normalization condition $U^\alpha U_\alpha=-1$ (see \cite{stewart90}).
Another possibility (see \cite{brauer95}) is to consider the Euler equations
as evolution equations for $\rho$ and $U^\alpha$ and treat the
normalization condition as a constraint, whose propagation must be
demonstrated. The treatments above are limited to the isentropic
Euler equations. For a fluid which is not isentropic the conservation
equation for the energy-momentum tensor must be supplemented by the
equation of conservation of entropy $u^\alpha\nabla_\alpha s=0$. The
equation of state can then be written in the form $p=h(\rho,s)$ or,
equivalently, in the form $p=f(n,s)$, where $n$ is the number density
of particles. We are not aware that a local existence theorem non-isentropic
Euler equations has been proved by a generalization of the method for the
isentropic case just outlined, although there is no reason to suppose that
it cannot be done, provided the condition $\partial h/\partial \rho>0$ is
satisfied. As we saw in Sect. \ref{euler}, the Einstein equations
coupled to the non-isentropic Euler equations can be brought into symmetric
hyperbolic form by introducing additional variables in a suitable way,
following \cite{friedrich98}. The Cauchy problem for the general (i.e. not
necessarily isentropic) Euler equations coupled to the Einstein equations
had much earlier been solved by other means by Choquet-Bruhat using
the theory of Leray hyperbolic systems \cite{choquet58}.

If it is desired to show the existence of solutions of the Einstein--Euler
system representing dynamical fluid bodies (such as oscillating stars) then
problems arise. Either the density must become zero somewhere, in which case
the Euler equations as written in the usual variables fail to be symmetric
hyperbolic there, or at least
the density must come arbitrarily close to zero at infinity, which is
almost as bad. Treating this situation as a pure initial value problem
is only possible under restrictive circumstances \cite{rendall:I} and the
general problem is still open. It would seem more promising to try to use the
theory of initial boundary value problems, explicitly taking account of
the boundary of the fluid. So far this has only been achieved in the
spherically symmetric case with $\rho_0>0$ \cite{kind93}.

One case of a fluid which is frequently considered in general relativity
is dust. This is defined by the condition that the presure should be
identically zero. Since in that case $h'=0$ the straightforward
method of writing the fluid equations as a symmetric hyperbolic system
does not work. The symmetric hyperbolic system of \cite{friedrich98}
discussed in Sect. \ref{euler} also covers the
dust case as does the existence theorem of Choquet-Bruhat \cite{choquet58}
using a Leray hyperbolic system.

The next phenomenological matter model we will consider comes from kinetic
theory. It does not quite fit into the framework we have used so far to
describe matter models since the fundamental matter field is a
non-negative function $f$ on the cotangent bundle of space-time. (Often the
case of particles with a fixed mass is considered in which case it is
defined on the subset of the cotangent bundle defined by the condition
$g^{\alpha\beta}p_\alpha p_\beta=-1$, known as the mass shell.) The idea is
that the matter consists of particles which are described statistically
with respect to their position and momentum. The function $f$ represents
the density of particles. The geodesic flow of the space-time metric defines
a vector field (Liouville vector field) on the cotangent bundle. Call it $L$.
This flow describes the evolution of individual test particles. The equation
of motion for the particles is $Lf=Q(f)$ where $Q(f)$ is an integral
expression which is quadratic in its argument. This is the Boltzmann equation.
It describes collisions
between the particles in a statistical way. The case $Q=0$ is the
collisionless case, where the equation $Lf=0$ obtained is often called the
Vlasov equation. The energy-momentum tensor is defined by:
$$T^{\alpha\beta}=\int f p^\alpha p^\beta d\omega(p)$$
where $d\omega(p)$ represents a natural measure on the cotangent space or
mass shell, depending on the case being considered.

The coupled Einstein--Boltzmann system in harmonic coodinates cannot be a
hyperbolic system is any usual sense for the simple reason that it is not
even a system of differential equations, due to the integrals occurring.
Nevertheless, the techniques used to prove existence and uniqueness for
hyperbolic equations can be adapted to prove local existence for the
Einstein--Boltzmann system \cite{bancel73}. In contrast to the case
of the perfect fluid nothing particular happens when the energy density
vanishes so that there is no problem in describing isolated concentrations
of matter. The Einstein--Vlasov system can be used to describe globular
clusters and galaxies. For the analogue of this in Newtonian theory
see \cite{binney87}. Literature on the relativistic case can be found in
\cite{shapiro85}.

Another kind of phenomenological matter model is elasticity theory. Apart
from its abstract interest, self-gravitating relativistic elasticity is of
interest for describing the solid crust of neutron stars. As in the case of
a perfect fluid the field equations are equivalent to the equation that the
divergence of the energy-momentum tensor is zero. What is different is the
nature of the matter variables and the way they enter into the definition
of the energy-momentum tensor. This is complicated and will not be treated
here. The Cauchy problem for the Einstein equations coupled to elasticity
theory has been discussed by Choquet-Bruhat and
Lamoureux--Brousse \cite{choquet73}. A local existence theorem for the
equations of relativistic elasticity has been proved by
Pichon \cite{pichon66} for data belonging to Gevrey classes. These are classes
of functions which are more special that $C^\infty$ functions in that the
growth of their Taylor coefficients is limited. However they do not have
the property of analytic functions, that fixing the function on a small open
set determines it everywhere.

In non-relativistic physics, the Euler equations are an approximation to
the Navier--Stokes equations where viscosity and heat conduction are
neglected. The Navier--Stokes equations are dissipative with no limit to
the speed at which effects can propagate. Mathematically this has the
effect that the equations are parabolic with no finite domain of dependence.
It is problematic to find an analogue of the Navier--Stokes equations in
general relativity which takes account of the effects of diffusion and heat
conduction. One possibility is to start from the Boltzmann equation, which
does have a finite domain of dependence and try to do an expansion in the
limit where the collision term is large. This is analogous to the Hilbert
and Chapman--Enskog expansions in non-relativistic physics. The first
attempts to do this led to equations which probably have no well-posed
Cauchy problem (Landau--Lifschitz and Eckart models). Hiscock and
Lindblom \cite{hiscock85} have shown that the linearization of these equations
about an equilibrium state have solutions which grow at arbitrarily large
exponential rates. In response to this
other classes of models were developed where the fluid equations are
symmetric hyperbolic. (For information on this see \cite{geroch91}).
These models do have a well-posed Cauchy problem and the main difficulty
seems to be to decide between the many possible models. Since the variables
used to formulate the symmetric hyperbolic system for the fluid are not
differentiated in forming the energy-momentum tensor, the system obtained
by coupling these fluids to the Einstein equations can be written in
symmetric hyperbolic form.

More general matter models can be obtained from those already mentioned
by combining different types of matter field. For instance there is the
charged scalar field which combines a scalar field and a Maxwell field.
This often produces no extra difficulties at all for the local in time
Cauchy problem due to the fact, mentioned above, that the system obtained
by taking two symmetric hyperbolic systems together is also symmetric
hyperbolic, provided the coupling between the two systems is via terms
of order zero. It is also routine to allow charged particles in a kinetic
model, obtaining the Einstein--Maxwell--Boltzmann system \cite{bancel73}.
Charged fluids are more complicated. The local in time Cauchy
problem has been treated in two cases, those of zero conductivity and
infinite conductivity. The latter model is also known as
magnetohydrodynamics. It has been shown to be have a well-posed initial
value problem only for data in Gevrey classes \cite{lichnerowicz67},
\cite{lichnerowicz94}.

\subsection{An Example of an Ill-Posed Initial Value
Problem}\label{illposed}

It may be hard to appreciate the significance of a system of equations
having a well-posed initial value problem since most examples which come
up in practice do have this property. In this section we present an
ill-posed example which is close to examples which relativists are familiar
with. Consider the special relativistic Euler equations with equation of
state $p=k\rho$ where $k<0$. A special solution is given by constant density
and zero spatial velocity. Now consider the equations obtained by linearizing
the Euler equations about this background solution. The unknowns in the
linearized system will be denoted by adding a tilde to the corresponding
unknowns in the nonlinear system. For simplicity we take the background
density to be unity. The linearized equations are:
\begin{eqnarray*}
\d_t\tilde\rho&=&-2(1+k)\d_a\tilde u^a   \\
\d_t\tilde u^a&=&-\frac{k}{2(1+k)}\delta^{ab}\d_b\tilde\rho
\end{eqnarray*}
It will be shown that given any $T>0$ there exist periodic initial data
$(\tilde\rho_0,\tilde u^a_0)$ of class $C^\infty$ such that it is not true
that there is a unique corresponding solution, periodic in the space
coordinates, on the time
interval $[0,T]$. The condition of periodicity here does not play an
essential role. It is adopted for convenience. Instead of thinking of
smooth periodic functions on $\R^3$ with can equally well think in terms
of smooth functions on a torus $T^3$. The space of smooth functions on
$T^3$ can be made into a topological vector space $X$ in a standard way.
Convergence of functions in the sense of this topology means uniform
convergence of the functions and their derivatives of all orders.
In a similar way the space of smooth functions on $[0,T]\times T^3$
can be made into a topological vector space $Y$. These are Fr\'echet spaces
\cite{rudin91}. The unknown in the linearized Euler equations can be thought
of as an element of $X^4$ and the data as an element of $Y^4$. Let $Z$ be the
closed linear subspace of $X^4$ consisting of solutions of the linearized
Euler equations. Consider the linear mapping $L:Z\to Y^4$ defined by
restricting solutions to $t=0$. It is continuous with respect to the
relevant topologies. If existence and uniqueness held for all initial
data then this linear map would be invertible. Hence, by the open mapping
theorem \cite{rudin91}, it would have continuous inverse. What this means
concretely is that given any sequence $(\tilde\rho_{0,n},\tilde u^a_{0,n})$
which is uniformly bounded together with each of its derivatives, the
corresponding sequence of solutions (which exists by assumption) must
also be bounded together with each of each of its derivatives. Thus to
prove the desired theorem it is enough to exhibit a uniformly bounded
sequence of initial data and a corresponding sequence of solutions which
is not uniformly bounded. This can be done explicitly as follows:
\begin{eqnarray*}
\tilde\rho_{0,n}&=&\sin nx                 \\
\tilde u^a_{0,n}&=&0                       \\
\tilde\rho_n&=&\sin nx\cosh (n\sqrt{-k}t)  \\
\tilde u^1_n&=&\frac{\sqrt{-k}n}{2(1+k)}\sin nx\sinh (n\sqrt{-k}t)  \\
\tilde u^2_n&=&\tilde u^3_n=0
\end{eqnarray*}
{}From these explicit formulae we get an idea what is going wrong. Fourier
modes of increasing frequencies of the initial data grow at increasing
exponential rates. In the case of a fluid where the equation of state has
a positive value of $k$ the hyperbolic functions in the above formulae
are replaced by trigonometric ones and the problem does not arise.

In fact in the above example the density perturbation satisfies a second
order equation which is elliptic. After a rescaling of the time coordinate
it reduces to the Laplace equation. The computation which has just been
done should be compared with the remarks on the Cauchy problem for the
Laplace equation on p. 229 of \cite{courant62}, Vol. 2. The corresponding
example for the Laplace equation goes back to Hadamard \cite{hadamard52}.

Solutions of the Einstein equations coupled to a fluid with an equation of
state of the type considered in this section have been considered in the
context of inflationary models \cite{barrow87}, \cite{wainwright97}. While
this is unproblematic for spatially homogeneous models, the above
ill-posedness result suggests strongly that this kind of model cannot give
reasonable results in the inhomogeneous case.

Some general comments on well-posedness and stability will now be made.
Suppose a solution of a system of evolution equations is given.
Assume for simplicity that this solution is time-independent, although
a similar discussion could be carried out more generally. The solution
is called stable if in order to ensure that a solution stays close to
the original solution to any desired accuracy, it is enough to require
it to be sufficiently close at one time. Closeness is measured in some
appropriate norm. Well-posedness has no influence on stability in this
sense. Already for ordinary differential equations with smooth coefficients,
which always have a well-posed initial value problem, stability does not
in general hold. Solutions of the linearized sytem about the given solution
can grow exponentially. However there is a constant $k$ such that no
linearized  solution can grow faster than $Ce^{kt}$. This is a rather general
feature of well-posed evolution equations. For instance any linear
symmetric hyperbolic system allows an exponential bound with some
constant $k$ independent of the solution. If no such bound is possible,
then it is said that there is a violent instability. (Cf. \cite{majda84},
Sect. 4.4 for this terminology.) The presence of a violent
instability
is closely related to ill-posedness, as can be seen in the above example.

\subsection{Symmetries}\label{syms}

If an initial data set for the Einstein-matter equations possesses
symmetries, then we can expect these to be inherited by the corresponding
solutions. This will be discussed here in the case of the vacuum Einstein
equations. There is nothing in the argument which obviously makes
essential use of the vacuum condition and it should extend to reasonable
types of matter. It makes use of the maximal Cauchy development and so
any restrictions on the matter model which might come up there would
appear again in the present context.

The following only covers symmetries of spacetime which
leave a given Cauchy surface invariant. It is based on group actions rather
than Killing vectors. A more extensive discussion of symmetries of
spacetime and their relations to the Cauchy problem can be found in
Sect. 2.1 of \cite{chrusciel91}.

By a symmetry of an initial data set $(S,h_{ab},\chi_{ab})$ for the vacuum
Einstein equations we mean a diffeomorphism $\psi:S\to S$ which leaves
$h_{ab}$ and $\chi_{ab}$ invariant. Let $\phi$ be the embedding of $S$
into its maximal Cauchy development. Then $\bar\phi=\phi\circ\psi$ also
satisfies the properties of the embedding in the definition of the
maximal Cauchy devlopment. Hence, by uniqueness up to isometry, there
exists an isometry $\bar\psi$ of the maximal Cauchy development onto itself
such that $\bar\psi\circ\phi=\bar\phi$. This means that $\bar\psi\circ\phi
=\phi\circ\psi$. Thus $\bar\psi$ is an isometry of $M$ whose restriction to
$\phi(S)$ is equal to $\psi$. We see that a symmetry of the initial data
extends to a symmetry of the solution. Next we wish to show that this
extension is unique. Since a general theorem of Lorentzian (or Riemannian)
geometry says that two isometries which agree on a open set agree everywhere
it suffices to show that any two isometries $\bar\psi$ with the properties
described agree on a neighbourhood of $\phi(S)$. Let $p$ be a point of
$\phi(S)$. A neighbourhood of $p$ can be covered with Gauss coordinates
based on $\phi(S)$. An isometry preserves geodesics and orthogonality.
Hence if, when expressed in Gauss coordinates, it is the identity for
$t=0$ it must be the identity everywhere. This completes the proof of
the uniqueness of $\bar\psi$.

Now consider the situation where a Lie group $G$ acts on $S$ in such a
way that each transformation $\psi_g$ of $S$ corresponding to an element
of the group is a symmetry of the initial data. Let $H$ be the isometry
group of the maximal Cauchy development and $H_S$ the group of
all isometries of the maximal Cauchy development which leave $\phi(S)$
invariant. The group $H_S$ is a closed subgroup of the Lie group $H$
and thus is itself a Lie group. Each $\psi_g$ is the restriction of a
unique element $\bar\psi_g$ of $H_S$. Using uniqueness again we must have
$\bar\psi_{gh}=\bar\psi_g\bar\psi_h$ for all elements $g$ and $h$ of $G$.
Thus we obtain a homomorphism from $G$ to $H_S$. This shows that there exists
an action of the group $G$ on $M$ by isometries which extends the action
on $\phi(S)$ arising from the original action on $S$ by means of the
identification using $\phi$. However this argument does not show that the
resulting action of $G$ is smooth. To show this consider first the group
$H_I$ of all symmetries of the initial data. It is a closed subgroup
of the isometry group of $h$ and therefore has the structure of a Lie
group. The above considerations show that restriction defines an isomorphism
of groups from $H_S$ to $H_I$. (We identify $S$ with $\phi(S)$ here.) If we
knew that this mapping was continuous a general theorem on Lie
groups \cite{varadarajan84} would show that it is also an isomorphism of Lie
groups. The continuity can be seen by noting that the topology of an
isometry group coming from its Lie group structure coincides with the
compact open topology \cite{kobayashi72}. The continuity of the restriction
mapping in the compact open topology follows immediately from the definitions.
We conclude that as Lie groups $H_S$ and $H_I$ can be identified.

Now we come back to the action of $G$. The action of $G$ on initial data
is a smooth mapping $G\times S\to S$. It is the composition of a smooth
homomorphism from $G$ to $H_I$ with the action of $H_I$ on $S$. By the
comments of the last paragraph this can be identified with the composition
of a smooth homomorphism from $G$ to $H_S$ with the action of $H_S$. In
this way we obtain a smooth action of $G$ on $M$ which leaves $S$ invariant
and restricts to the original action on the initial data. It is the
action of $G$ which we previously considered.


\section{Outlook}\label{outlook}

This article is intended to be an informative tour through its subject,
rather than an exhaustive account. The latter would in any case be
impossible in an article of this length, given the amount
of literature which now exists. The aim of this section is to mention
a few of the important things which have been left out, and to direct the
reader to useful sources of information concerning these. A good starting
point is the review article of Choquet-Bruhat and York \cite{choquet80}
which is still very useful. (See also \cite{fischer79}.) There is an
extensive treatment of the constraints in reference \cite{choquet80}. For
a selection of newer results on the constraints, see e.g.
\cite{andersson92}, \cite{andersson94}, \cite{isenberg95},
\cite{isenberg96}, \cite{kannar97} and \cite{choquet99}.

The most obvious omission of the present article is the lack of
statements on the global Cauchy problem. A review with pointers
to further sources can be found in \cite{rendall98}. This material
is too recent to be discussed in \cite{choquet80} and a lot has happened
since then. It is natural that once some of the basic local questions
had been solved attention turned to global issues. The latter are
now central to present research on the Cauchy problem. Most of the
existing results concern spacetimes with high symmetry, although in
the meantime there are also a few theorems on space-times without
symmetries which are small but finite perturbations of space-times of
special types. This is an area which is developing vigorously at the
moment. Often the statements obtained about global existence of solutions
are accompanied by information on the global qualitative properties of
the solutions. A discussion of the asymptotic behaviour of a particular
class of solutions and its relevance for the modelling of certain
physical systems can be found in \cite{friedrich:rp} and
\cite{friedrich:grg15}.

In the introduction we mentioned the possible applications of ideas
connected with the Cauchy problem to analytical and numerical
approximations. Up to now progress on establishing an effective
interaction between theoretical developments and the applications
of approximate methods to concrete physical problems such as the
generation of gravitational waves has been limited. For instance, little
has been done on the question of proving theorems on analytical
approximations in general relativity since \cite{damour90},
\cite{rendall92}, and \cite{rendall:II}. There is no shortage of things to
be done. For instance one tempting goal would be a precise formulation
and justification of the quadrupole formula.

As for the link to numerical relativity, the discussions in Sect.
\ref{reductions}, apart from their interest for purely analytical reasons,
could potentially be exploited for improving numerical codes. Many new
hyperbolic reductions have been suggested recently with the aim of
providing equations which would ensure a stable time evolution. We have
added a few more. We also pointed out various different gauge conditions.
Their usefulness for stable long-time numerical calculations still has
to be explored. We have seen that different representations of the field
equations and different formalisms allow us to employ different gauge
conditions. The possibilities of controlling the lifetime of a gauge
by a judicious choice of gauge source functions have neither been
investigated analytically nor numerically in a systematic way.

One of the main interests in the analysis of the initial boundary
problem lies in the fact that many approaches to numerical relativity
require the introduction of timelike boundary hypersurfaces which
reduce the calculations to spatially finite grids. A good analytical
understanding of the initial boundary value problem does not guarantee
the stability of long-time evolutions for problems with timelike
boundaries, but the latter are likely to fail without a proper
understanding of the analytical background. In this respect we also
consider the analytical investigations of the conformal Einstein
equations (cf. \cite{friedrich:rp}, \cite{friedrich:grg15}) as an
opportunity for numerical
relativity. They allow us in principle to calculate entire spacetimes,
including their asymptotic behaviour, on finite grids without the need
to introduce artificial boundaries in the physical spacetime.

Another place where the theory of the Cauchy problem could have something
to contribute is the application of numerical methods developed for
systems of conservation laws (or, more generally, systems of balance laws)
to the Einstein equations, a procedure
which has been popular recently. In a certain sense this means
accepting an analogy between the Einstein equations and the Euler equations.
At the moment this amounts to no more than the fact that both systems are
quasi-linear hyperbolic and can be formulated as systems of balance
laws. Unlike the Euler equations, the Einstein equations do not appear
to admit a preferred formulation of this kind. There are different
alternatives and no known criterion for choosing between them. It is very
tempting to import the vast amount of knowledge which has been
accumulated concerning the numerical solution of the Euler equations
into general relativity. On the other hand it is not at all clear how
many of these techniques are really advantageous for, say, the vacuum
Einstein equations. One might hope that analytical theory
could throw some light on these questions.

There can be little doubt that increased cooperation between people
working on analytical and numerical aspects of the evolution of
solutions of the Einstein equations would lead to many new insights.
As both fields progress cases where a rewarding collaboration would be
possible are bound to present themselves. It suffices for someone to
show the initiative required to profit from this situation.

An open problem which has been touched on already is that of the
existence of solutions of the Einstein--Euler system describing fluid
bodies. This kind of free boundary problem is poorly understood even
in classical physics, although there has been significant progress
recently \cite{wu99}. The central importance of this becomes clear when
it is borne in mind that the usual applications of gravitational
theory in astrophysics (except for cosmology) concern self-gravitating
{\it bodies}.

We hope to have succeeded in showing in this article that the study
of the Cauchy problem for the Einstein equations is a
field which presents a variety of fascinating challenges. Perhaps, with
luck, it will stimulate others to help tackle them.

\vskip 10pt\noindent
{\it Acknowledgements} During the preparation of this article we have
benefitted from discussions with many people. We are grateful to all
of them. We thank in particular Yvonne Choquet-Bruhat, Oliver Henkel,
Jim Isenberg, Satyanad Kichenassamy and Bernd Schmidt.


\makeatletter                  
\@takefromreset{equation}{section}
\renewcommand{\theequation}{\arabic{equation}}%
\renewcommand{\thesubequation}{\arabic{equation}\alph{eqsubcnt}}
\makeatother                   


\end{document}